\documentclass{IEEE-Con-Sys-mag}
\pdfoutput=1

\title{Pick-to-Learn for Systems and Control \stitle{data-driven synthesis with state-of-the-art safety guarantees}}
\author{Dario Paccagnan, Daniel Marks, Marco C. Campi, Simone Garatti\\
}
\affil{%
D. Paccagnan (d.paccagnan@imperial.ac.uk) is with the Department of Computing, Imperial College London, UK;\\
D. Marks (daniel.marks@cs.ox.ac.uk) is with the Department of Computer Science, University of Oxford, UK; work conducted while affiliated with Department of Computing, Imperial College London, UK;\\
M.C. Campi (marco.campi@unibs.it) is with the Department of Information Engineering, University of Brescia, Italy;\\
S. Garatti (simone.garatti@polimi.it) is with the Dipartimento di Elettronica, Informazione e Bioingegneria, Politecnico di Milano, Italy.
}

\let\labelindent\relax
\usepackage{enumitem}

\usepackage{booktabs}
\usepackage{amsthm}
\usepackage{amssymb}
\usepackage{amsmath}
\usepackage{mathrsfs}
\usepackage{hyperref}
\usepackage{makecell}
\usepackage{mathtools}
\usepackage{slashbox}

\usepackage{algorithm}
\usepackage{algorithmicx}
\usepackage{algpseudocode}
\algrenewcommand{\algorithmiccomment}[1]{\hfill\textit{// #1}}
\allowdisplaybreaks

\usetikzlibrary{arrows.meta, positioning, shapes.geometric}

\usepackage[most]{tcolorbox}
\usepackage{graphicx}
\usepackage{pgfplots}
\hypersetup{
    unicode=false,          %
    pdftoolbar=true,        %
    pdfmenubar=true,        %
    pdffitwindow=false,     %
    pdfstartview={FitH},    %
    pdfnewwindow=true,      %
    colorlinks=true,       %
    linkcolor=blue,          %
    citecolor=blue,        %
    filecolor=magenta,      %
    urlcolor=red,           %
    breaklinks=true
}
\usepackage{cleveref}
\usepackage{xcolor}
\usepackage[sort, numbers, compress]{natbib}
\usepackage{here}
\usepackage{caption}
\usepackage{subcaption}
\usepackage{graphicx}

\usepackage{tikz}
\usetikzlibrary{shapes.geometric, arrows}
\tikzstyle{arrow} = [thick,->,>=stealth]
\tikzstyle{arrow_nohead} = [thick,-,>=stealth]
\tikzstyle{startstop} = [rectangle, rounded corners, minimum width=1cm, minimum height=2.5cm,text centered, text width=3cm, draw=black, fill=red!30]
\usetikzlibrary{automata, positioning, arrows}

\newcommand{\spacing}{\vspace{0.1 cm}}

\newcommand{\be}{\begin{equation}}
\newcommand{\ee}{\end{equation}}
\newcommand{\bz}{{\boldsymbol{z}}}

\newtheorem{definition}{\bf Definition}
\newtheorem{theorem}{Theorem}

\newtheorem{corollary}{Corollary}
\newtheorem{remark}{Remark}

\newtheorem{problem}{Sidebar}

\newcommand{\mc}{\mathcal}
\newcommand{\mb}{\mathbb}

\newcommand{\bh}{{\boldsymbol{h}}}
\newcommand{\bT}{{\boldsymbol{T}}}
\newcommand{\bD}{{\boldsymbol{D}}}

\renewcommand{\k}{k}

\newlength\figureheighteps 
\newlength\figurewidtheps 

\usepackage{marginnote} 
\usepackage{ifthen}
\makeatletter
\newcommand{\DP}[1]{%
  \ifthenelse{\isodd{\value{page}}}
    {\setlength{\marginparwidth}{1cm}\normalmarginpar\marginpar{\textcolor{red}{\scriptsize DP: #1}}}
    {\setlength{\marginparwidth}{2cm}\reversemarginpar\marginpar{\textcolor{red}{\scriptsize DP: #1}}}
}
\makeatother

\newcommand{\T}{H}

\usepackage[normalem]{ulem}
\usepgfplotslibrary{groupplots}
\usepackage{float}
\usepackage{hyperref}

\begin{document}

\maketitle

\vspace*{-10mm}

\begin{summary}
Data-driven methods have become paramount in modern systems and control problems characterized by growing levels of complexity. In safety-critical environments, deploying these methods requires rigorous guarantees, a need that has motivated much recent work at the interface of statistical learning and control. 
However, many  existing approaches achieve this goal at the cost of sacrificing valuable data for testing and calibration, or by constraining the choice of learning algorithm, thus leading to suboptimal performances.

In this paper, we  describe \emph{Pick-to-Learn (P2L) for Systems and Control}, a framework that allows \emph{any} data-driven control method to be equipped with state-of-the-art safety and performance guarantees. P2L enables the use of \emph{all} available data to \emph{jointly} synthesize and certify the design, eliminating the need to set aside data for calibration or validation purposes. 

In presenting a comprehensive version of P2L for systems and control, this paper demonstrates its effectiveness across a range of core problems, including optimal control, reachability analysis, safe synthesis, and robust control. In many of these applications, P2L delivers designs and certificates that outperform commonly employed methods, and shows strong potential for broad applicability in diverse practical settings.
\end{summary}

\section{Introduction}
\medskip

\noindent
The increasing availability of data and the rapid advancements in learning-based methods have, over the past years, blurred the boundaries between the fields of \emph{Systems \& Control} and \emph{Machine Learning}. Indeed, data-availability and learning-based techniques have not only shown great potential for enhanced control synthesis, but they have also driven fundamental research in new applications such as autonomous driving \cite{salzmann2020trajectron}, intelligent robotics \cite{murphy2019introduction}, and assisted clinical treatment \cite{haidar2016artificial}.

Due to the critical nature of many control tasks, the deployment of data-driven policies in real-world settings requires equipping them with \emph{safety} and \emph{performance guarantees}. For example, autonomous driving, robotics, and industrial automation require controllers that satisfy strict state constraints, often in the face of high levels of uncertainty. In  learning-based approaches, safety and performance certificates are themselves \emph{based on data}. This need has motivated much recent work at the interface of statistical learning and control, where guarantees must integrate seamlessly with data-driven designs,  \cite{CampiCareGaratti2021,Dorfler2023-A,Dorfler2023-B}.

Many existing approaches to provide safety and performance guarantees either require setting aside valuable data for testing and calibration, or constrain the choice of algorithms during design, which ultimately leads to suboptimal performances. As a result, there remains a strong need to develop broadly applicable, theoretically sound methods that exploit the information in the data to design the solution, while \emph{simultaneously} providing rigorous performance guarantees. 
\medskip

\begin{pullquote}
There is an urgent need for methods that use data to simultaneously enable design and provide safety or performance guarantees.
\end{pullquote}

\noindent
\textbf{Pick-to-Learn for Systems and Control.} This paper presents  \emph{Pick-to-Learn (P2L) for Systems and Control}, a recently developed framework designed to equip any data-driven control method with rigorous safety and performance guarantees. P2L was proposed in \cite{P2L2023} in a machine learning context, developing an idea initially introduced in an embryonic form in \cite{CGR2015} into a full-fledged framework. It builds on a recent breakthrough in compression theory, \cite{2023compression}, and, crucially, enables the use of all available data to jointly synthesize and certify a control policy or a system design. As a result, P2L is well-placed to deliver designs and certificates that surpass the state-of-the-art across a wide range of core problems in systems and control. This is demonstrated in the paper through extensive numerical experiments on established benchmarks in reachability analysis, optimal control, safe synthesis, and robust control.
The key features of P2L can be summarized as follows: 

\smallskip
\begin{itemize}[itemsep=5pt]
\item\textbf{Distribution-free.} P2L provides formal guarantees of safety/performance without needing access to the data-generating distribution. This is important because, while data may be available, the data-generating distribution is often unknown or only partially known.\footnote{Importantly, distribution-free does not advocate for forgoing  prior information relevant for the \emph{design} task at hand, but merely refers to the fact that the data-generating distribution need not be known in order to generate formal \emph{guarantees}.} 

\item\textbf{Calibration/test-free.} P2L does not require to set aside  a portion of the dataset for calibration or testing. This is crucial in applications where data are a limited or costly resource since withholding a calibration/test set may worsen the quality of the resulting design. 

\item\textbf{Accurate bounds.} P2L produces accurate probabilistic guarantees, that is, guarantees that take a small margin from the actual system behavior. This is important for accurately assessing performance and safety requirements. 

\item\textbf{Wide applicability.} P2L is a framework, not a single specific design technique. As a result, it can be applied to a variety of problems, ranging from reachability analysis to neural network-based control problems. 
\end{itemize}
\medskip

\begin{figure}[H]
\centering
\hspace*{5mm}
\includegraphics[width=.93\linewidth]{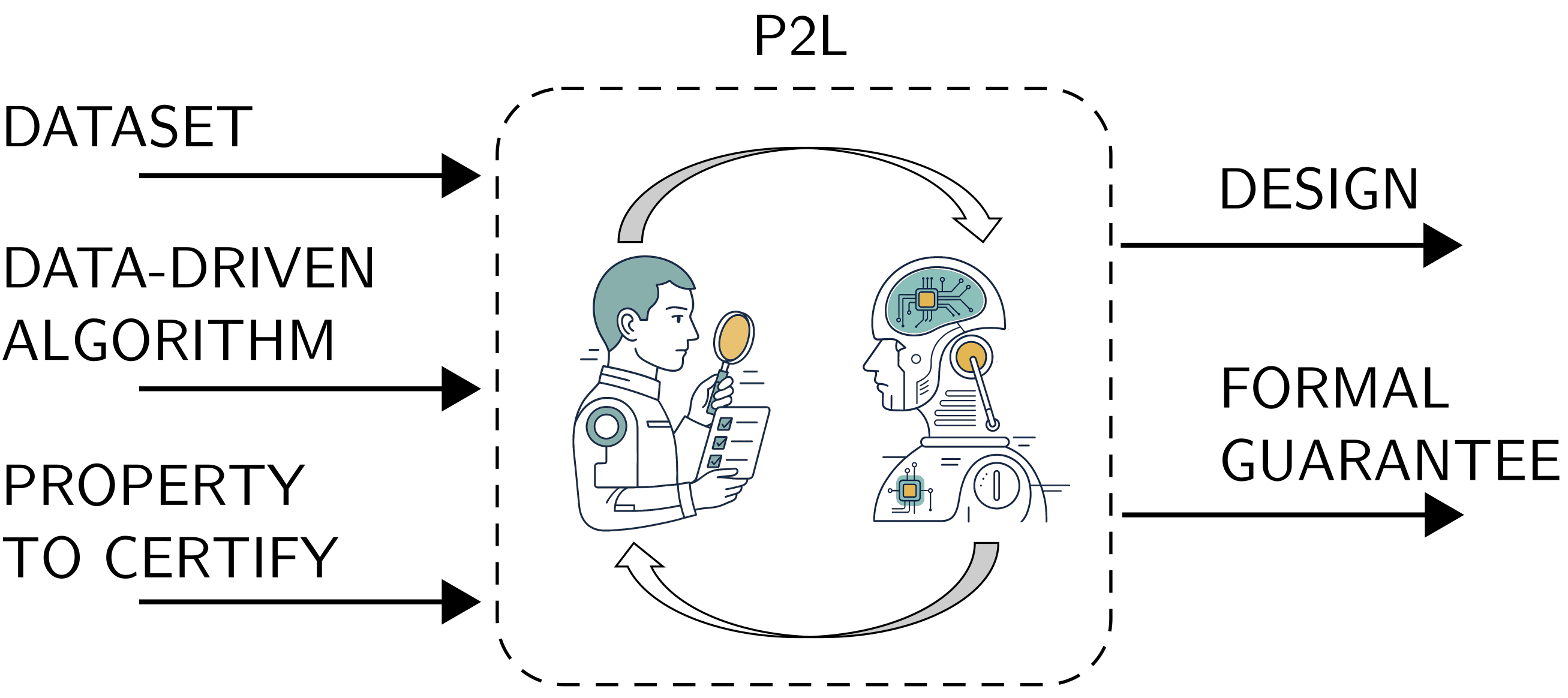}	
\vspace*{1mm}
\caption{Pick-to-Learn equips any data-driven algorithm with formal guarantees. Towards this goal, it receives in input three fundamental ingredients: a dataset, a data-driven algorithm, and a property to be certified. By incrementally feeding the data-driven algorithm with a carefully selected subset of the data, Pick-to-Learn generates a design and a formal guarantee on the property of interest.}
\label{figure:P2L-generic}
\end{figure}

\noindent
At its core, P2L transforms any data-driven algorithm into a \emph{preferent compression scheme}. This enables to leverage recent results in sample compression theory, \cite{2023compression}, to obtain probabilistic guarantees on any property of interest, e.g., constraint satisfaction, or complex safety specifications.  
In more specific terms, P2L (see \Cref{figure:P2L-generic}) consists of a \emph{meta-algorithm} constructing a loop around the original data-driven algorithm and feeding it with increasingly larger subsets of the available data, where the data points that least satisfy the property of interest are iteratively added. The degree of compression achieved, that is, the size of the subset obtained at termination, determines the probabilistic bound on the satisfaction of the property.

\medskip

\noindent
\textbf{P2L as an enabler.} The compression result proved in \cite{2023compression}, on which this work builds, is part of the scenario approach literature. Paper \cite{2023compression} establishes a general methodology for providing tight guarantees in data-driven design schemes that exhibit a preferent compression property. While this property naturally holds in many cases, \cite{2018waitandjudge,2022riskandcomplexity,2025noncvx}, other learning algorithms commonly employed in practice do not possess it, thereby preventing the application of the results in \cite{2023compression}. By transforming any given algorithm into a preferent compression scheme, P2L substantially expands the original scope of \cite{2023compression}. In this light, P2L can be interpreted as an \emph{enabler} that significantly broadens the use of the scenario approach results, particularly in relation to modern machine learning pipelines that have somewhat opaque structures, for which guarantees are otherwise difficult to obtain.

\medskip

\noindent
\textbf{P2L in action across four pillars.}
This paper not only presents P2L as originally introduced in \cite{P2L2023}, it also provides significant new extensions that have not appeared in previous works. These extensions, presented in Section~\ref{sec:extended_framework}, enable the use of P2L across a variety of problems, and a core contribution of this manuscript is to demonstrate its broad applicability in four pillars of data-driven control and system design. For each pillar, the designs and guarantees obtained by P2L are compared with state-of-the-art methodologies for that specific task. The four pillars are the following (see also \Cref{fig:pillars}): 

\begin{figure}[t]
\centering
\includegraphics[width=.8\linewidth]{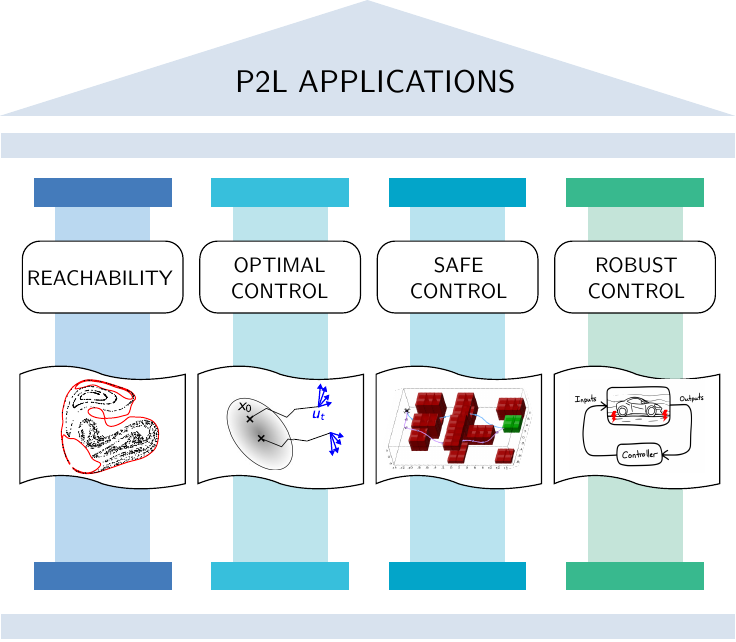}	
\vspace*{2mm}
\caption{In this manuscript, the broad applicability of P2L is demonstrated by deploying it across four pillars in data-driven systems and control: reachability analysis, optimal control, safe control and robust control.}
\label{fig:pillars}
\end{figure}

\vspace*{2mm}
\begin{itemize}[itemsep=5pt]
\item\textbf{P1: Reachability analysis.} Given a sample of terminal states of a stochastic system, determine an approximate reachable set and guarantee that,  with high probability, future realizations of the terminal state will fall within this set. 
\item\textbf{P2: Optimal control.} In stochastic optimal control problems, leverage a dataset of noise trajectories to design a control policy that minimizes the cost while providing probabilistic guarantees on performance or constraint satisfaction. 
\item\textbf{P3: Safe control synthesis.} Given a stochastic system and a complex mission to accomplish, for example specified by temporal logic, synthesize a control policy that maximizes the probability of completing the mission, while providing out-of-sample guarantees.
\item\textbf{P4: Robust Control.} For a linear time invariant system with probabilistic parametric uncertainty, design a state feedback controller that robustly minimizes the $H_2$ norm of the resulting closed-loop system, while guaranteeing the performance against unseen realizations.
\end{itemize}

\subsection{Related work}
\medskip

\begin{sidebar}{Test-set approach}
\label{test-set floating_box}
\noindent
Test-set is a commonly-employed approach to equip data-driven methods with probabilistic guarantees on a property of interest. Rooted in classical hypothesis testing, the method works by splitting the available dataset into two disjoint subsets referred to as the training set and the test set. The training set is used for design, while the test set is employed to assess the extent to which the determined decision empirically verifies the property of interest. This empirical evaluation is turned into formal guarantees by means of probabilistic results, the tightest of which relies on the binomial tail inversion formula, \cite{langford2005tutorial}.
\medskip

\noindent
More formally, assume availability of a dataset $D=\{z_1,\dots,z_N\}$, modeled as a realization of $\bD=\{\bz_1,\dots,\bz_N\}$, where $\bz_1,\dots,\bz_N$ are i.i.d. with distribution $\mb{P}_z$. A subset of $D$ called training set and denoted $T$ is used to formulate a design by means of an algorithm $L$. Given a property of interest, expressed via a function $\varphi(h,z)$ that takes value $1$ if decision $h$ satisfies the property when applied to $z$ and value zero otherwise, the quality of the design $L(T)$ is then evaluated empirically on the test set $V=D \setminus T$. This is just obtained by counting how many times $\varphi(L(T),z_i) = 0$ over the test set $V$: 
\[
k=\left|\{z_i \in V~\text{s.t.}~\varphi(L(T),z_i) = 0\}\right|.
\] 
\noindent
This empirical evaluation is turned into a formal statement by proving that relation 
\[
\mb{P}_z[z: \varphi(L(T),z) = 0]\le \bar \varepsilon(k,|V|),
\]
holds, for suitable expressions of function $\bar \varepsilon(\cdot,\cdot)$, with high confidence $1-\delta$ with respect to the realizations of $\bD$. The tightest evaluation of $\bar \varepsilon(k,|V|)$ formulated in \cite[Thm. 3.3]{langford2005tutorial} is obtaining by solving the following equation 
\[
\sum_{j=0}^k\binom{|V|}{j} \bar\varepsilon^j (1-\bar\varepsilon)^{|V|-j}=\delta.
\]

\noindent
Crucially, the data employed in the test phase cannot be used for training, and vice versa. This results in a critical tradeoff: reducing the number of data points used for training worsens the quality of the obtained design in favor of a tighter bound. Conversely, increasing the size of the training set improves the quality of the decision, but this comes at the price of loosening the probabilistic guarantees. \Cref{fig:sidebar_testset} illustrates this point. 

\newlength\boundswidth
\newlength\boundsheight
\setlength{\boundswidth}{.85\linewidth}
\setlength{\boundsheight}{4cm}

\sdbarfig{\begin{tikzpicture}
\definecolor{darkgray176}{RGB}{210,210,210}

\begin{axis}[
    width=\boundswidth,
    height=\boundsheight,
    tick align=outside,
    tick pos=left,
    x grid style={darkgray176},
    xlabel={Fraction of $D$ used for training},
    xmin=0, xmax=1,
    xtick style={color=black},
    y grid style={darkgray176},
    ylabel={\textcolor{blue}{Risk}},
    ymin=0.05, ymax=0.22,
    ytick style={color=black},
    xtick={0,0.2,0.4,0.6,0.8,1},
    ytick={0, 0.05, 0.1, 0.15, 0.20},
    yticklabel style={
        /pgf/number format/fixed,
        /pgf/number format/precision=5
    },
    legend cell align={left},
    ticklabel style = {font=\footnotesize},
    label style={font=\footnotesize},
    legend pos=outer north east,
    ymajorgrids=true,
    xmajorgrids=true,
    y label style={at={(axis description cs:+0.08,.5)},anchor=south},
]

\addplot [semithick, blue, mark=*, mark size=2, mark options={solid}]
table {%
0.1 0.20647
0.2 0.155705
0.3 0.12654
0.4 0.110221666666667
0.5 0.0989883333333333
0.6 0.09008
0.7 0.0839366666666667
0.8 0.0786366666666667
0.9 0.0744583333333333
};
\end{axis}

\begin{axis}[
    width=\boundswidth,
    height=\boundsheight,
    axis y line*=right,
    axis x line=none,
    ymin=0.04, ymax=0.13,
    ylabel={\textcolor{red}{Bound $-$ Risk}},
    ytick style={color=black},
    yticklabel style={
        /pgf/number format/fixed,
        /pgf/number format/precision=5
    },
    y label style={at={(axis description cs:1.5,.5)},anchor=south},
    ytick={0, 0.05, 0.075, 0.1, 0.125},
    ticklabel style = {font=\footnotesize},
    label style={font=\footnotesize},
    xtick=\empty,
    ymajorgrids=false,
    xmajorgrids=false,
]

\addplot [semithick, red, mark=*, mark size=2, mark options={solid}]
table {
0.1 0.050002801
0.2 0.048809376
0.3 0.050570023
0.4 0.053260355
0.5 0.055827839
0.6 0.061597531
0.7 0.071296213
0.8 0.083786917
0.9 0.124877201
};
\end{axis}
\end{tikzpicture}}{
(adapted from \cite[Fig. 2]{P2L2023}). Misclassification rate (Risk) and difference between the test-set bound and the true misclassification rate (Bound$-$Risk) on a binary version of the MNIST digit classification problem. Note how the quality of the decision, as measured by the risk, worsen as the fraction of $D$ used for training is decreased. Similarly, the bound returned by test-set deteriorates as the latter fraction increases. \label{fig:sidebar_testset}}

\end{sidebar}

\begin{sidebar}{Conformal prediction}
\label{conformal-prediction floating_box}
\noindent 
Calibration-based conformal prediction (also known as split conformal prediction) has recently gained momentum within the control community as a tool to equip data-driven designs with formal guarantees, \cite{chee2024uncertainty,Lindemann2024}. The approach shares common aspects with test-set, from which we borrow the availability of a data-driven algorithm $L$ and of a datasets $D=\{z_1,\dots,z_N\}$ modeled as a realization of $\bD=\{\bz_1,\dots,\bz_N\}$, where $\bz_1,\dots,\bz_N$ are i.i.d. with distribution $\mb{P}_z$. The method also requires access to a so-called \emph{non-conformity score} $s(h,z)$ mapping a decision $h$ and a value of $z$ into a real value. The latter can be thought of as a way to assess the extent to which a desired property of interest is satisfied. 
\medskip

\noindent
Given these ingredients, similarly to the test-set approach, calibration-based conformal prediction requires to split the dataset $D$ into two predefined disjoint sets: a training set $T$, and a calibration set $C$. The training set is used to synthesize the decision $L(T)$, while the calibration set is used to empirically assess its quality. To this end, one computes the scores $s_k=s(L(T),z_k)$, $z_k\in C$, and the conformal prediction theory provides guarantees on the probability that a new data point exceeds any one of the values $s_k$.
\medskip

\noindent
Specifically, let $s_{(1)},\ldots,s_{(k)},\ldots,s_{(|C|)}$ be the $s_k$'s arranged in ascending order: $s_{(1)} \le \dots \le s_{(|C|)}$. Then (see \cite[Prop. 2b]{vovk2012conditional}, \cite[Sec. 3.2]{angelopoulos2021gentlearxiv}), for any a-priori fixed choice of $1 \le k \le |C|$, relation 
\be
\mb{P}_z[z: s(L(T),z) > s_{(k)}] \le \bar\varepsilon(k,|C|),
\label{eq:conformal_bound}
\ee 
holds with high confidence $1-\delta$ with respect to the realizations of $\bD$ when $\bar\varepsilon(\cdot,\cdot)$ is determined by solving the following equation  
\be\sum_{j=0}^{|C|-k}\binom{|C|}{j} \bar\varepsilon^j (1-\bar\varepsilon)^{|C|-j}=\delta.
\label{eq:conformal_bound_eps}
\ee
In calibration-based conformal prediction, data points used for calibration cannot be used for training, leading to a tradeoff similar to that in the test-set approach. On the other hand, using a calibration set adds an important degree of freedom: it allows one to evaluate the probability with which the design satisfies various score values, those obtained from the evaluation of $L(T)$ over $C$. 
\end{sidebar}

\begin{sidebar}{PAC-Bayes bounds}
\label{PAC-Bayes floating_box}
\noindent 
PAC-Bayes bounds offer an attractive framework for deriving probabilistic guarantees on key metrics of interest. Unlike methods such as test-set or calibration-based conformal prediction, PAC-Bayes bounds do not require to split the dataset into separate parts. However, in their native form, they apply to \emph{randomized} decisions, e.g., distributions over control policies. For this reason, the probabilistic guarantees are provided for distributions $\mc{Q}$ defined over the domain of decisions, rather than for individual decisions. 

The core idea of this approach is as follows. One wants to come up with evaluations of the performance achieved by each $\mc{Q}$, while ensuring that these evaluations are simultaneously valid for all $\mc{Q}$. These assessments are of interest \emph{per se}, but they can also guide the selection of a specific $\mc{Q}$, one that optimizes the performance evaluation (at this abstract level, the approach is not dissimilar to the uniform evaluation of hypotheses, as done for instance in the Vapnik-Chervonenkis theory). However, a single dataset does not carry enough information to achieve uniformly accurate evaluations. Nonetheless, and this is the key idea, the PAC-Bayes method suggests picking a distribution $\mc{P}$ that acts as an ``anchor'', and shows that the performance of each $\mathcal{Q}$ can be bounded by its empirical performance plus a margin depending on a ``distance'' between $\mc{Q}$ and $\mc{P}$.

In more precise terms, consider a dataset $D=\{z_1,\dots,z_N\}$ modeled as a realization of $\bD=\{\bz_1,\dots,\bz_N\}$, where $\bz_1,\dots,\bz_N$ are i.i.d. with distribution $\mb{P}_z$. Next, consider a prior distribution $\mc{P}$ over the set of decisions $h$, and let $\mc{Q}$ be any distribution over the same domain. Finally, given any decision $h$ and any value of $z$, let $c(h,z)$ be a metric measuring the cost of using $h$ on $z$. 

In its classical form, PAC-Bayes assesses the performance of a distribution $\mc{Q}$ in terms of the expected value of $c$, where the expectation is taken over both $\bz \sim \mb{P}_z$ and $\bh \sim \mc{Q}$. That is, classical PAC-Bayesian results aim to bound the following quantity 
\[
\mc{L}(\mc{Q})=\mb{E}_{\bz\sim\mb{P}_z} \mb{E}_{\bh\sim \mc{Q}} [c(\bh,\bz)]. 
\]
The goal is to obtain a bound that holds simultaneously over all distributions $\mc{Q}$; this bound will depend on the empirical performance of $\mc{Q}$ on the dataset $D$, i.e., $\hat{\mc{L}}(\mc{Q})=\frac{1}{|D|}\sum_{z_i\in D}\mb{E}_{\bh\sim \mc{Q}} [c(\bh,z_i)]$ and the distance between $\mc{P}$ and $\mc{Q}$. 

While many are the PAC-Bayes bounds introduced in the literature (see \cite{PBbounds24} for an overview), we present here only one of the most popular ones. This allows us to highlight key aspects that are shared by other variants. 

McAllester's result \cite[Thm. 1]{mcallester1999pac} (see also \cite[p. 215]{PBbounds24}) shows that, for $[0,1]$-valued functions $c(h,z)$ and with confidence $1-\delta$ with respect to the realizations of $\bD$, the following bound holds simultaneously for \emph{any} choice of $\mc{Q}$ 
\be
\label{eq:McAllester}
\mc{L}(\mc{Q})\le \hat{\mc{L}}(\mc{Q}) +\sqrt{\frac{\text{KL}(\mc{Q}||\mc{P})+\ln\frac{2\sqrt N}{\delta}}{2N}},
\ee
where $\text{KL}(\mc{Q}||\mc{P})$ is the Kullback-Leibler divergence between $\mc{Q}$ and $\mc{P}$. 
Since the bound holds uniformly across all $\mc{Q}$'s, one can select a $\mc{Q}$ that minimizes the right-hand side of \eqref{eq:McAllester}, and the bound will apply to the selected distribution. 

It should be noted that in the PAC-Bayes approach the uniformity of the bound is obtained at the price of introducing the penalization term $\text{KL}(\mc{Q}||\mc{P})$ for distributions $\mc{Q}$ that deviate from $\mc{P}$. Therefore, although all data points are used to learn and evaluate, the accomplishment of this dual goal is externally guided by the introduction of $\mc{P}$. Choosing a suitable $\mc{P}$ is crucial for the effectiveness of the method, as otherwise distributions $\mc{Q}$ that could be good fits for the problem at hand may be unfairly penalized and excluded due to a large KL divergence term. 

Finally, we observe that there also exist results for individual realizations of $h$, albeit introducing additional error terms. Moreover, the PAC-Bayes approach, which involves the evaluation of the expectation over $\bh\sim\mc{Q}$, can be computationally challenging, especially in high dimensional problems. 
\end{sidebar}

Several established approaches convert available data into decisions accompanied by probabilistic guarantees. The most prominent paradigms are outlined below, and a comparative analysis with P2L is provided later in the paper.

\medskip

\noindent
\textbf{Test-set approach.}
Rooted in statistical hypothesis testing, the test-set approach is a classical tool to obtain formal probabilistic guarantees on a property of interest, \cite{langford2005tutorial}. This approach works by splitting the available dataset into two subsets, the training set and the test set. As the name suggests, the training set is used for synthesis purposes, while the test set is exclusively used to evaluate the quality of the resulting design. Statistical tools, such as the binomial tail inversion bound and other concentration inequalities, \cite{langford2005tutorial}, are used to provide formal probabilistic certificates based on empirical evidence. While simple and computationally efficient, this approach requires withholding a portion of the data from training, thus negatively impacting the quality of the final design. See the \hyperref[test-set floating_box]{floating box} for more details. 
\medskip

\noindent
\textbf{Conformal prediction.}
Conformal prediction (see the \hyperref[conformal-prediction floating_box]{floating box}) is a set of techniques introduced in \cite{Vovketalbook2005,ShaferVovk}, commonly applied in regression and classification tasks to derive prediction regions with certified guarantees. In its calibration-based version, \cite{vovk2012conditional}, conformal prediction has recently gathered momentum within the systems and control community, and has shown to be a versatile methodology (see \cite{Lindemann2024} for a recent survey). Also this approach requires withholding a portion of the dataset in order to generate the desired guarantees. 

\medskip

\noindent
\textbf{PAC-Bayes.}
The Probably Approximately Correct (PAC)-Bayes framework builds upon the earlier work of Valiant \cite{valiant1984theory} and later contributions of McAllester \cite{mcallester1999pac}. The central idea is to assume a prior distribution $\mc{P}$ over the object to be designed, e.g., a control policy, and to leverage the available data to select a distribution $\mc{Q}$ over the same domain that provides a good fit for the problem at hand. Probabilistic bounds on properties of interest are obtained based on empirical evidence, along with a measure of how much $\mc{Q}$ deviates from $\mc{P}$ (see the \hyperref[PAC-Bayes floating_box]{floating box} for a more detailed explanation). While PAC-Bayes was among the first approaches to provide non-vacuous bounds for deep learning architectures, \cite{dziugaite2017computing,Ortiz21}, various challenges limit its applicability. First, PAC-Bayes typically works with distributions over objects, rather than individual objects. This can be problematic when one wishes to certify the safety and performance of a single deterministic design or policy. Second, obtaining accurate bounds requires a suitable choice of the prior $\mc{P}$, which is often pursued by setting aside part of the dataset to pre-train the prior distribution. It should also be noted that PAC-Bayes bounds are computationally demanding, as they require expensive optimization over the space of distributions $\mc{Q}$. 
\medskip

\noindent
\textbf{VC-dimension and other complexity measures.} The Vapnik-Chervonenkis (VC) dimension, \cite{Vapnik98}, and related complexity measures such as Radamacher complexity, \cite{Bartlett02}, stem from the pioneering work of Vapnik and Chervonenkis, \cite{Vapnik71}. These methods provide probabilistic guarantees based on the complexity of a hypothesis class from which the design is selected. The core idea is that classes with lower complexity generalize better, while more complex classes are prone to overfitting. While VC-based approaches do not require a test or calibration set, their uniform treatment of the entire hypothesis class often leads to loose or even vacuous bounds, \cite{dziugaite2017computing}, making them unsuitable for safety-critical control and system design.

\section{Pick-to-Learn for Systems \& Control} \label{sec:P2L_for_systems_and_control}
\medskip

In this section, we introduce the meta-algorithm Pick-to-Learn (P2L) in its basic form. We then discuss extensions in \Cref{sec:extended_framework}, thereby enabling a gradual introduction of the fundamental concepts. We begin by introducing three fundamental ingredients: \emph{a dataset; a synthesis algorithm; a property of interest}, along with the setup in which they operate. 

Consider a data-driven system design or control problem, and suppose that a sample of $N$ data points $\{z_i\}_{i=1}^N$, with $z_i\in \mathcal{Z}$, is given for synthesis purposes. These are historical observations of uncertain elements that affect the problem.\footnote{The $z_i$'s have been also called \emph{scenarios} in the literature on the scenario approach.} For example, $z_1,\ldots,z_N$ may represent realizations of a disturbance in a stochastic optimal control problem. The data are collected in the dataset $D=\{z_1,\dots,z_N\}$. Mathematically, each $z_i$ is interpreted as a realization of a random element $\bz_i$, and $\bz_1,\dots,\bz_N$ are assumed to be independent and identically distributed (i.i.d.) with distribution $\mb{P}_z$.\footnote{Throughout, boldface denotes random quantities.} For future use, it is also convenient to introduce a probability space $(\Omega,\mc{F},\mb{P})$ over which all random elements are defined. 

We assume access to a data-driven synthesis algorithm $L$ that maps any collection of data points of arbitrary length to a decision $h$ selected from a feasible set $\mathcal{H}$ (e.g., a control policy chosen from a set of admissible policies).\footnote{Formally, $L: \bigcup_k \mc{Z}^k \to \mathcal{H}$.} The synthesis algorithm should be suited for the design problem at hand, and may incorporate any available prior knowledge about the problem. For example, $L$ could be gradient descent applied to neural-based control design, \cite{miller1995neural}, or the BFGS algorithm used in robust control to solve non-convex Bilinear Matrix Inequalities, \cite{hifoo2003}. No assumptions or restrictions are imposed on $L$ for the results presented in this paper to hold.\footnote{The algorithm $L$ is also allowed to be permutation-dependent, i.e., it may produce different outputs when fed with $\{z_1,\ldots,z_k\}$ versus a permutation of it, $\{z_{i_1},\ldots,z_{i_k}\}$. In this respect, the collection of data points at the input of $L$ should be regarded as a list (in which elements are ordered), rather than a set. Throughout, also $D$ is regarded as a list, although it is referred to as the ``dataset", following standard terminology.} 

The final ingredient is a given property of interest, whose satisfaction depends on the interaction between the decision and the uncertainty. To formalize this, introduce the map $\varphi: \mathcal{H}\times\mathcal{Z}\rightarrow \{0,1\}$, where $\varphi(h,z)=1$ means that the decision $h$ meets the property of interest for the uncertainty instance $z$ (in this case, we will also use the terminology ``property $\varphi$ is satisfied''), and $\varphi(h,z)=0$ otherwise. For example, when $h$ represents a control policy, $\varphi$ may encode the requirement that the trajectory of the closed-loop system, corresponding to a disturbance realization $z$, satisfies a prescribed state constraint. The main objective of P2L is to construct decisions via $L$, for which formal certifications on the satisfaction of $\varphi$ can also be provided. 

To proceed, we assume to have access to a quantifier of the degree of dissatisfaction when $h$ fails to satisfy property $\varphi$. This enables us to identify the data point in the available dataset that exhibits the \emph{highest} degree of dissatisfaction.\footnote{It is assumed that the quantifier is sufficiently fine to assign distinct degrees of dissatisfaction to distinct observations.} For example, given a control policy $h$, a disturbance realization $z$ may be assigned a dissatisfaction level based on the duration for which the corresponding closed-loop trajectory violates a state constraint; in other cases, one can instead use the maximum distance between the trajectory and the constraint as the quantifier. P2L leaves the greatest freedom to the user in the selection of the quantifier of dissatisfaction, allowing it to be adapted to the specific needs for the problem at hand. 

\medskip

\begin{figure}
	\centering
	\begin{tikzpicture}[
  node distance=0.5cm,
  block/.style={rectangle, draw=teal!50!black, fill=teal!15, rounded corners, align=center, 
                minimum width=5cm, minimum height=1cm, font=\small},
  decision/.style={diamond, draw=orange!70!black, fill=yellow!20, align=center, text width=2.5cm, inner sep=-2pt, font=\small},
  arrow/.style={-Latex}
]

\node[block] (init) {
Initialize decision $h$ and \\ empty training set $T$
};

\node[decision, below=of init] (check) {Does $h$ satisfy property $\varphi$ for all elements in $D \setminus T$?};

\draw[arrow] (check.east) -- ++(1,0) node[midway, above] {Yes} node[right] {$h, T$};

\node[block, below=of check] (findz) {Find data point $\bar{z} \in D \setminus T$ that least\\ satisfies $\varphi$ for current decision $h$};

\node[block, below=of findz] (append) {Append $\bar{z}$ to $T$};

\node[block, below=of append] (redesign) {Design $h=L(T)$};

\draw[arrow] (init) -- (check);
\draw[arrow] (check) -- node[left] {No} (findz);
\draw[arrow] (findz) -- (append);
\draw[arrow] (append) -- (redesign);

\draw[arrow] (redesign.west) -- ++(-1,0) 
                           --([xshift=-1.86cm]check.west) 
                           -- (check.west);

\end{tikzpicture}
	\vspace*{3mm}
	\caption{Schematic presentation of Pick-to-Learn.}
	\label{fig:meta-algorithm-informal}
\end{figure}
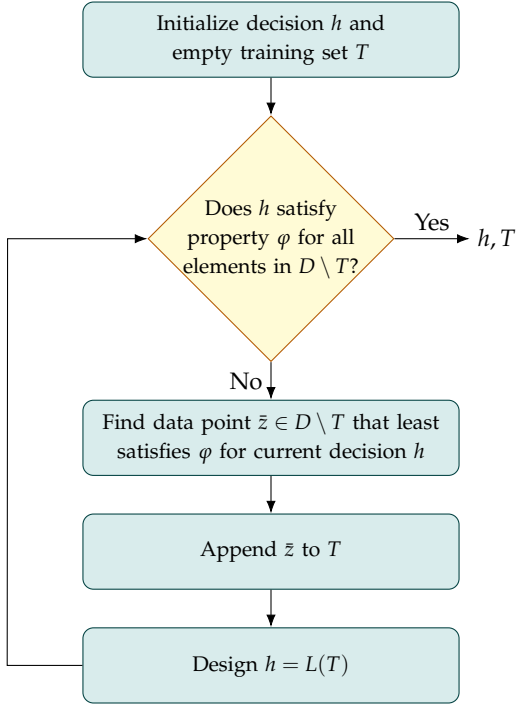

We are now ready to introduce the meta-algorithm P2L. Referring to \Cref{fig:meta-algorithm-informal}, P2L works by iteratively feeding the synthesis algorithm $L$ with a growing list $T$ of data points taken from $D$, while terminating when the current decision satisfies the property of interest for all data points that have \emph{not} yet been used in the synthesis step. Until this condition is met, P2L appends to $T$ the data point, among those not yet used, that exhibits the highest degree of dissatisfaction with the current decision. At termination, P2L returns both the final list $T$ of actively used data points and the final decision $h$. 

\Cref{alg:meta-algo} presents the pseudocode of P2L in more precise terms. Given a synthesis algorithm $L$, a property $\varphi$, and an initial decision $h_0$, P2L maps a dataset $D$ to the pair $(h,T)=\text{P2L}(D)$, where $h\in\mathcal{H}$ is the final decision and $T$ is referred to as the \emph{training set}.\footnote{$T$ is indeed the part of the dataset upon which the final decision is built, which motivates the terminology ``training set''. However, while in test-set and conformal prediction the training set is selected \emph{a priori}, $T$ in P2L is actively constructed using all the available information, and its size is adapted to the specific problem instance at hand.} In the initialization step (Line~1 in \Cref{alg:meta-algo}), the training set $T$ is initialized as empty and the decision is set to $h=h_0$. Lines~2 through 6 contain the main loop. If, with the current $h$, property $\varphi$ is satisfied for all elements in $D \setminus T$, the loop terminates.\footnote{All operations in \Cref{alg:meta-algo} are intended to be operations with lists; for example, $D \setminus T$ denotes the sublist of $D$ obtained by removing from $D$ the elements in $T$ in their order of appearance.} If not, P2L  computes the element $\bar{z}$ in $D \setminus T$ that exhibits the highest degree of dissatisfaction with the current $h$ (Line~3). Then, it enlarges $T$ by appending  $\bar z$ to $T$ (Line~4), and updates the decision based on the enlarged $T$ (Line~5). At termination, P2L returns the current $h$ and $T$.  

\begin{algorithm}[t]
\caption{-- Meta-algorithm P2L}
\label{alg:meta-algo}
\begin{algorithmic}[1]     
    \State \textbf{Initialize:} $T=\emptyset$, $h=h_0$ 
        \While{$\exists \ z_i \in D \setminus T : \; \varphi(h,z_i)=0$}         	
        	\State $\bar{z} \gets$ most unsatisfied point in $D \setminus T$ 
        	\State $T \gets T \cup \{ \bar z\}$ \Comment{augment $T$}
            \State $h \gets L(T)$ \Comment{synthesize $h$}                        
        \EndWhile
        \State \textbf{return} $h$, $T$\Comment{decision $h$ and list $T$}
\end{algorithmic}
\end{algorithm}

\medskip

Two remarks concerning the broad applicability of P2L and its computational complexity are in order. 
\medskip

\noindent
\textbf{Broad-applicability.} The choice of the synthesis algorithm $L$, the property $\varphi$, and the initial decision $h_0$ are left to the user without restriction. Perhaps unexpectedly, strong generalization guarantees can be secured at this high level of abstraction (\Cref{thm:main_result}), thereby enabling the deployment of P2L across significantly different domains in systems and control. As illustrative examples, P2L will be applied to reachability analysis, robust control, stochastic optimal control, and safety control synthesis later in this work. 
\medskip

\noindent
\textbf{Computational complexity.} 
As presented in \Cref{alg:meta-algo}, P2L requires to iteratively synthesize a decision over lists of data of increasing size, which may suggest a significant computational burden. However, during the initial steps, P2L works with small lists, so that, if termination occurs in early stages, its overall runtime is competitive with that obtained by directly applying $L$ to the full dataset $D$. Besides, many synthesis algorithms do not require a complete re-design when a new data point is added. For example, in a Bayesian setting, the posterior can be easily updated when one more data point becomes available, \cite{marks2025pick}. Additionally, previous designs can be used as initializations in subsequent iterations. For example, in synthesizing a neural network, the current network can be used as initialization, and only a few steps of stochastic gradient descent can be performed on the enlarged set of data.\footnote{Using the previous design as initialization falls within the setup of \Cref{alg:meta-algo}. In fact, introduce the notation $L(\bar{h},T)$ to indicate an algorithm that maps an initialization $\bar{h}$ and a list $T$ to a decision. If $\bar{h}$ is chosen as  $L(T')$, where $T = T' \cup \{ \bar z\}$, then $L(\bar{h},T) = L(L(T'),T)$ is still a map from $T$ to a decision, which we can write as $L(T)$.} Finally, the algorithm's runtime can be reduced by appending more than one point to $T$ at each iteration, and probabilistic guarantees similar to those provided in the next \Cref{sec:prob_guarantees} still hold, refer to \cite[Section A.3]{P2L2023} for details. 
  
\subsection{Probabilistic guarantees} \label{sec:prob_guarantees}
\medskip

In this section, we present theoretical results concerning the satisfaction of property $\varphi$ for new, as-yet unseen instances of the uncertainty. This is a \emph{generalization result}, aiming to formally guarantee the effectiveness of the decision in future deployments. Towards this goal, we begin by introducing the notion of risk of a decision. 

\begin{definition}[Risk of decision]
\label{def:stat_risk}
The risk of a decision $h\in\mathcal{H}$ is defined as  
\[
R(h)=\mathbb{P}_z[z: \varphi(h,z) = 0]. 
\] 
\end{definition}
\noindent
In words, the risk of a decision $h$ quantifies the probability that $h$ does not satisfy the property $\varphi$ when facing a new uncertainty instance. For example, if the property encodes the satisfaction of a state constraint, the risk measures the probability with which the state constraint is violated for a new realization of the disturbance~$\bz$. 

Because the dataset $D$ is the realization of a random element $\bD=\{\bz_1,\dots,\bz_N\}$, the decision $h$ returned by P2L is itself a realization of a random element through $(\bh,\bT)=\text{P2L}(\bD)$. As a consequence, one aims to make statements of the form 
\vspace*{2mm}
\begin{quote}
\emph{
``With high confidence $1 \! - \! \delta$ with respect to the $N$ i.i.d. draws in the dataset, the risk of the decision returned by P2L is bounded by $\bar\varepsilon(|\bT|,\delta,N)$}'',
\end{quote}
\vspace*{2mm}
\noindent
for a suitably defined function $\bar\varepsilon(\cdot,\cdot,\cdot)$. Note that, the guarantees obtained will depend on the number of elements in $T$, which we denote throughout with  $|T|$.

In preparation of \Cref{thm:main_result}, we need to introduce the following function ${\Psi}_{\k,\delta}: (0,1) \to \mathbb{R}$ indexed by $\k=0,1,\ldots,N-1$ and by the confidence parameter $\delta\in (0,1)$: 
\be
{\Psi}_{\k,\delta}(\varepsilon) =
\frac{\delta}{N} \sum_{m=k}^{N-1}\frac{\binom{m}{k}}{\binom{N}{k}}(1-\varepsilon)^{-(N-m)}.\footnote{The result similar to \Cref{thm:main_result} presented in \cite{P2L2023} used a slightly different function ${\Psi}_{\k,\delta}$.}
\nonumber
\ee
The following theorem presents the bound on the risk of the decision $h$ returned by P2L. No proof is provided as this theorem will be derived as a special case of a more general theorem stated and proved in \Cref{sec:extended_framework}. 
{
\begin{theorem}
\label{thm:main_result}
Let $(\bh,\bT)=\text{P2L}(\bD)$. For any choice of $\delta \in (0,1)$, it holds that 
\[
\mathbb{P}\{R(\bh)\le \bar\varepsilon(|\bT|,\delta,N)\}\ge 1-\delta,
\]	
where, for $k=0,1,\dots,N-1$,  $\overline\varepsilon(k,\delta, N)$ is defined as the unique solution to the equation ${\Psi}_{\k,\delta}(\varepsilon)=1$ in the interval $[\k/N,1]$, while, for $k=N$, $\overline\varepsilon(N,\delta,N)$ is set to value $1$. 
\end{theorem}}

\begin{figure}[t]
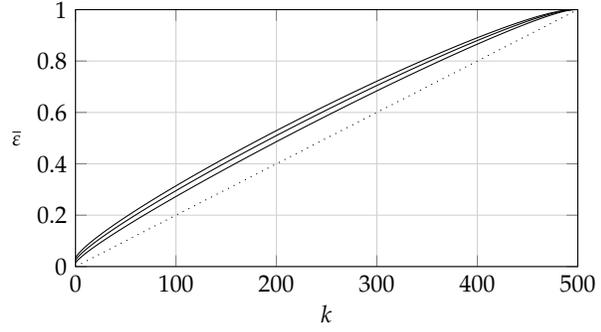

			\centering
			\include{figures_tikz/eps_500}
			\vspace*{-7mm}
			\caption{Function $\bar\varepsilon(k,\delta,N)$ for $N=500$ and $\delta=10^{-6}$, $10^{-4}$, $10^{-2}$ (top to bottom). The dotted line is $k/N$.}	
			\label{fig:epsilon}
\end{figure}

Three important remarks are in order, addressing the distribution-free aspect of the result, its informative content, and the connection with the scenario approach. 
\\[2mm] 
\textbf{Distribution-free and ease of use.} The result in \Cref{thm:main_result} is distribution-free. This means that no knowledge of the data-generating distribution $\mb{P}_z$ is needed for its utilization. In practice, after executing P2L, one is simply required to compute the number of points in $T$ and to evaluate $\bar\varepsilon(|T|,\delta,N)$, a task that, as shown in \cite{2023compression}, can be performed efficiently by means of a bisection procedure (see the Appendix for ready-to-implement \textsc{Matlab} and \textsc{Python} codes). \Cref{thm:main_result} ensures that $\bar\varepsilon(|\bT|,\delta,N)$ is a ($1-\delta$)-confidence bound to the risk of the $\bh$ independently of the distribution of $\bz$. 
\\[2mm]
\textbf{Informativeness.} Higher compressions, i.e., lower values of $|T|$, correspond to better probabilistic bounds, see \Cref{fig:epsilon}. The level of compression depends on the problem at hand and the choices made by the user, particularly the quantifier used to measure the degree of dissatisfaction, which influences how quickly P2L learns from data. In practice, the bound is often informative and it compares favorably with other methods, as demonstrated in later sections. Provably, the dependence of $\bar\varepsilon$ on $\delta$ is logarithmic (see \cite{2023compression}, Proposition 8). 
As a consequence, asking for results that hold with very high confidence, e.g., $1-\delta=0.999999$, corresponding to $\delta=10^{-6}$, only marginally impacts on the value of the bound, see again \Cref{fig:epsilon}. 
\\[2mm]
\noindent
\textbf{Connection with the scenario approach.} Readers familiar with the theory of the scenario approach may have recognized that the probabilistic bound given in \Cref{thm:main_result} is \emph{exactly} the same as that in \cite{2023compression}. This is no coincidence. In fact, at its core, \Cref{thm:main_result} hinges on the statistical results for risk certification within a compression framework appearing in \cite{2023compression}. The remarkable achievement of P2L is that it transforms \emph{any} synthesis algorithm $L$ into a compression scheme for which certificates of reliability can be obtained via the results in \cite{2023compression}. In doing so, P2L uncovers a truly broad domain of utilization of the scenario approach and enables extensive use of its statistical results. 
\begin{pullquote}
Pick-to-Learn equips any data-driven control algorithm with formal guarantees, therefore setting the designer free to choose. 
\end{pullquote}

\section{An extended framework - P2L$^+$} \label{sec:extended_framework}
\medskip

In this section, we present a generalization of the P2L algorithm that adds further flexibility while maintaining probabilistic guarantees in the same spirit as the original P2L. In the second part of the section, we exemplify how this additional flexibility can be concretely leveraged across three different settings. 
 
Thus far, the termination condition of P2L has been tied to the property to be satisfied, that is, P2L terminates when all points in $D\setminus T$ satisfy property $\varphi$. We are now interested in decoupling the termination condition from the property $\varphi$. The original P2L will turn out to be a special case of this extended setup, see Section~\ref{subsec:eventtrig} for a detailed discussion on this point. Towards this goal, begin by observing that the degree of dissatisfaction in Section~\ref{sec:P2L_for_systems_and_control} effectively defines a total order over $\mathcal{Z}$ that depends on $h$, with the most unsatisfied point selected by P2L being the maximal element in $\mathcal{Z}$ according to this total order. Generalizing this standpoint, we introduce here a total order $\le_T$, which may depend on $T$ itself, but is not necessarily tied to $\varphi$, and terminate the procedure when no element $z_i \in D\setminus T$ can be found that is bigger in the $\le_T$ order than a special element called \emph{threshold} and denoted by ${\tt Stop}$. 

In formal terms, the threshold ${\tt Stop}$ is a new, abstract element that is added to $\mathcal{Z}$. The total order $\le_T$ is defined over $\mathcal{Z} \cup \{\tt Stop\}$, and lines up all the elements of this extended set. We allow the order to depend on $T$, which generalizes the situation encountered in P2L where the degree of dissatisfaction depends on $h$ (which, in turn, depends on $T$ via $h = L(T)$).

The extended Pick-to-Learn algorithm P2L$^+$ is essentially P2L with the following modifications: 
\begin{itemize}
	\item the termination condition becomes that $z_i \le_T {\tt Stop}$ holds for all elements in $D \setminus T$;
	\item when entering the loop, the element to be appended to $T$ is selected as $\max_{\leq_T} \{ D \setminus T \}$, the maximal element, according to the order $\leq_T$, among those in $D \setminus T$;
	\item upon termination, the output consists not only of $h$ and $T$, but also of a list $U$ containing the elements in $D \setminus T$ for which $\varphi(h,z) = 0$. Thus, $\text{P2L}^+(D) = (h,T,U)$. 
\end{itemize}
Notice that property $\varphi$ comes into play in the final step only. The pseudocode of P2L$^+$ is given in \Cref{alg:P2L+}. 

\begin{algorithm}[t]
	\caption{-- Meta-algorithm P2L$^+$}
	\label{alg:P2L+}
	\begin{algorithmic}[1] 		
		\State \textbf{Initialize:} $T=\emptyset$, $h=h_0$ 
		\While{$\exists \ z_i \in D \setminus T : \; {\tt Stop} \le_T z_i$} 		
		\State $\bar{z} \gets \max_{\leq_T} \{ D \setminus T \}$ 
		\State $T \gets T \cup \{ \bar z\}$ \Comment{augment $T$}
		\State $h \gets L(T)$ \Comment{synthesize $h$}                        
		\EndWhile
		\State \textbf{compute} $U \subseteq D \setminus T$ such that $z_i \in U$ iff $\varphi(h,z_i) = 0$  
		\State \textbf{return} $h$, $T$, $U$ \Comment{decision $h$ and lists $T$, $U$}
	\end{algorithmic}
\end{algorithm}

In the context of P2L$^+$, one cannot expect that $T$ alone provides valid indications of the risk, as it is in the context of \Cref{thm:main_result}. It is also necessary to account for an \emph{empirical evaluation of the violation} of property $\varphi$ on $D\setminus T$, as done in Line 7 of \Cref{alg:P2L+}. We now have the following theorem. 

\begin{theorem}	\label{thm:P2L+_main_result}
	Let $(\bh,\bT,\boldsymbol{U})=\text{P2L}^+(\bD)$ be the output of P2L$^+$. For any choice of $\delta \in (0,1)$, it holds that 
	\[
	\mathbb{P}\{R(\bh)\le \bar\varepsilon(|\bT|+|\boldsymbol{U}|,\delta,N)\}\ge 1-\delta,
	\]	
	where $\bar\varepsilon(k,\delta,N)$ is defined as in \Cref{thm:main_result}.
\end{theorem}

\noindent
The proof is given in the Appendix. 

As mentioned earlier, the remainder of this section discusses three instances of P2L$^+$ that illustrate the versatility of the framework. First, we show how P2L can be obtained as a special case of P2L$^+$. Second, we enable P2L to terminate after a specified number of steps. Third, we provide more fine-grained guarantees for settings where one is interested in verifying satisfaction at different ``levels'', for example to bound the entire tail of a cost function. 

\subsection{P2L as a special case} 
\label{subsec:eventtrig}
\medskip

We show here that P2L is a special case of P2L$^+$. To this end, it is sufficient to properly define $\le_T$ in P2L$^+$ based on $\varphi$ as follows: 
\begin{itemize}
	\item[1.] for $z \in \mathcal{Z}$ such that $\varphi(L(T),z) = 0$, let $z \geq_T {\tt Stop}$; otherwise, when $\varphi(L(T),z) = 1$, let $z \leq_T {\tt Stop}$; 
    \item[2.] for $z$ such that $\varphi(L(T),z) = 0$, define $\leq_T$ based on the degree of dissatisfaction: elements $z$ with higher degree of dissatisfaction occupy higher positions in the order induced by $\leq_T$; for $z$ such that $\varphi(L(T),z) = 1$, the order is unimportant, and can be defined arbitrarily. 
\end{itemize} 
With this choice, termination of P2L$^+$ occurs when all the elements in $D \setminus T$ satisfy the property $\varphi$, which aligns with algorithm P2L. Moreover, at termination, $U$ is empty, and thus $|U|=0$. This makes the statement of \Cref{thm:P2L+_main_result} identical to that of \Cref{thm:main_result}, showing that the latter is obtained from the former.

\subsection{Time-triggered stop}
\medskip

Sometimes, it is desirable to run a learning algorithm for a desired number of iterations. This section presents such a setup. 

Assume that we have available a total order on $\mathcal{Z}$ that encodes our criterion of choice among elements $z$. This order defines the relation $\leq_T$ over elements in $\mathcal{Z}$. Extend $\leq_T$ to $\mathcal{Z} \cup {\tt Stop}$ with the following position: 
\[
\begin{cases*}
\; {\tt Stop} \leq_T z, \;\ \forall z \in \mathcal{Z}, \quad \text{if } |T| < M \\
\; z \leq_T {\tt Stop}, \;\ \forall z \in \mathcal{Z}, \quad \text{if } |T| \geq M.
\end{cases*}
\]
Since $T$ is constructed incrementally, the P2L$^+$ main loop is repeated $M$ times before exiting (assuming that $D$ contains at least $M$ elements). 

This P2L$^+$ variant is equivalent to \Cref{alg:P2L-TTS} called P2L-TTS (Time-Triggered Stop).
\begin{algorithm}[t]
	\caption{-- P2L-TTS (Time-Triggered Stop)}
	\label{alg:P2L-TTS}
	\begin{algorithmic}[1] 
		\State \textbf{Initialize:} $T=\emptyset$, $h=h_0$ 
		\For{$i = 1,2,\ldots,M$} 
		\State $\bar{z} \gets \max_{\leq_T} \{ D \setminus T \}$ 
		\State $T \gets T \cup \{ \bar z\}$ \Comment{augment $T$}
		\State $h \gets L(T)$ \Comment{synthesize $h$}                        
		\EndFor
		\State \textbf{compute} $U \subseteq D \setminus T$ such that $z_i \in U$ iff $\varphi(h,z_i) = 0$  
		\State \textbf{return} $h$, $T$, $U$ \Comment{decision $h$ and lists $T$, $U$}
	\end{algorithmic}
\end{algorithm}
The interpretation of P2L-TTS is that the user a-priori decides the number $M$ of data points the decision $h$ must be learned from, and utilizes $\leq_T$ as a criterion to select suitable data points from the dataset $D$. \Cref{thm:P2L+_main_result} can then be used to certify the risk of the decision.

\begin{remark} 
The setup described in this section can also be used to enforce a stopping threshold when P2L$^+$ incorporates a criterion of stop that depends on the satisfaction of a specific property. To this end, any ordering $\leq_T$ defined over $\mathcal{Z} \cup \{\tt Stop\}$ that captures the property of interest is used until $|T| < M$, while, for $|T| \geq M$, the order is redefined as $z \leq_T {\tt Stop}, \;\ \forall z \in \mathcal{Z}$. This allows for early termination in case of favorable conditions, while imposing an upper limit to the maximum allowed number of iterations, see \cite{marks2025pick}. 
\end{remark}

\begin{remark} \label{rmk:all_iterations}
The time-triggered setup can be used with different values of $M$, and then, one decision, associated to a specific $M$, can be selected to best fit the user's goals. For example, P2L-TTS can be executed for all values of $M$, from $1$ to $N$, producing outputs denoted by $h_M$, $T_M$, $U_M$.\footnote{Note that implementing this scheme requires to actually run P2L only once without any {\tt Stop} condition until $T = D$, while storing the results at each intermediate step.} For each $M$, \Cref{thm:P2L+_main_result} yields 
\[
\mathbb{P}\{R(\bh_M)\le \bar\varepsilon(|\bT_M|+|\boldsymbol{U}_M|,\delta,N)\}\ge 1-\delta,
\]	
and a simple application of the union bound shows that 
\[
\mathbb{P}\{R(\bh_M)\le \bar\varepsilon(|\bT_M|+|\boldsymbol{U}_M|,\delta,N), \; \forall M\} \ge 1-N\delta.
\]	
This result certifies simultaneously the risk of all decisions $\bh_M$ with confidence $1-N\delta$ (replacing $\delta$ with $N\delta$ has a marginal impact since, as explained in Section \ref{sec:P2L_for_systems_and_control}, $\delta$ can be chosen very small given that the bound on the risk depends logarithmically on $\delta$). Therefore, a posteriori selecting any one of the decisions $h_M$ (including the one that has generated the lowest value of $|T_M|+|U_M|$) retains a valid bound of $\bar\varepsilon(|\bT_M|+|\boldsymbol{U}_M|,\delta,N)$ with confidence $1-N\delta$. 
\end{remark}

\subsection{Multiple levels of satisfaction}
\label{sec:multiple-levels}
\medskip

In certain applications, employing a decision $h$ results in a cost that also depends on the uncertainty instance $z$, as given by a function $c(h,z): \mathcal{H} \times \mathcal{Z} \to \mathbb{R}$.\footnote{This is, for example, the case in an investment problem where the return depends on how we diversify our portfolio and on the evolution of the market during the investment period.}  
In such cases, after the decision $h$ has been obtained according to any design principle, one can be interested in evaluating the probability that the cost exceeds a certain threshold $\gamma$. 

To see that this situation falls under the operation of P2L$^+$, let $L$ and $\leq_T$ be chosen so as to encode the principle used to design $h$, and let $\varphi(h,z) = 1$ when $c(h,z) \leq \gamma$ and $\varphi(h,z) = 0$ otherwise, so that $R(h) = \mathbb{P}_z [z: c(h,z) > \gamma ]$. After P2L$^+$ has come to termination, one computes $|U|$, the number of observations in $D \setminus T$ for which $c(h,z) > \gamma$; \Cref{thm:P2L+_main_result} then ensures that $R(h) = \mathbb{P}_z [z: c(h,z) > \gamma ]$ can be reliably bounded using $\bar\varepsilon(|T|+|U|,\delta,N)$.\footnote{In an investment problem, for example, one can evaluate the probability that the investment will incur a loss bigger than $\gamma$.}

Similarly to  \Cref{rmk:all_iterations}, it is possible to consider multiple thresholds $\gamma^{(1)},\ldots,\gamma^{(r)}$, while keeping $\leq_T$, and therefore the designed $h$, unchanged. In this way, one simultaneously certifies with confidence $1-r\delta$  all the bounds on the risk corresponding to the selected thresholds under a common decision. 
In turn, this enables us to claim that the cumulative distribution function of the cost must remain below given values with high confidence, thus providing a comprehensive assessment of its behavior in the same spirit as \cite{CareGarattiCampi2015}. 

\section{Pillar 1: Data-driven Reachability}
\medskip
Reachability analysis is a commonly employed tool to guarantee safety of a system in the face of uncertainty. Traditionally, given complete knowledge of a dynamical system and the uncertainty acting upon it, the task is to determine a minimal subset of the state space containing all possible terminal states of the system, the so-called \emph{reachable set}. In the context of safety, this information is important as it allows one to assess whether the system may visit a region deemed unsafe or, conversely, whether safety can be certified across all realizations of the uncertainty. 

Here, we follow a growing line of work that, in contrast to the above-mentioned model-based approach, adopts a \emph{data-driven} perspective in which uncertainty is treated probabilistically, \cite{Muratetalreachability2023}. This shift is motivated by the increasing complexity of the modern systems: while a detailed model may be unavailable, one may have access to measurements or may be able to simulate the system's behavior. 

In this setting, the goal is to utilize the available data to  construct a subset of the state-space and to give formal probabilistic guarantees certifying that, with high probability, a newly sampled terminal state will fall inside the constructed set. Naturally, a crucial measure of the quality of such a set is its volume, which should be minimized to reduce conservatism and avoid trivial statements, e.g., identifying the reachable set with the whole state-space.

Data-driven reachability is a rapidly growing field, with several methods proposed in recent years, including \cite{Muratetalreachability2023,dietrich2024nonconvex,devonport2020data,pmlr-v204-tebjou23a,alanwar2023data,hashemi2023data}. Among the approaches providing formal guarantees, conformal prediction has emerged as a method capable of delivering state-of-the-art certificates, \cite{pmlr-v204-tebjou23a}. In this section, we borrow the problem formulation and benchmark proposed in \cite{Muratetalreachability2023,pmlr-v204-tebjou23a}, and compare the performance of P2L with conformal prediction, as well as the test-set approach. 
\begin{pullquote}
	Pick-to-Learn directly quantifies the probability of reaching a target region without requiring any data beyond those used to learn that region.
\end{pullquote}

\subsection{Problem Formulation}
\medskip
 
Our problem formulation follows \cite{Muratetalreachability2023}. Consider a dynamical system described by the state transition function $\Phi(t_1;t_0,x_0,w)$ mapping the initial state $x_0\in\mathbb{R}^{n_x}$ at time $t_0$ to a final state $x_1 = \Phi(t_1;t_0,x_0,w)$ at time $t_1$ under the effect of the disturbance $w \in \mathbb{R}^{n_w}$. For instance, $\Phi(t_1;t_0,x_0,w)$ may return the solution $x(t)$ at time $t_1$ obtained from the integration of an ordinary differential equation $\dot x(t)=f(t,x(t),w(t))$. 
In this context, it is common to assume that both the initial state $x_0$ and the disturbance $w$ are uncertain, and each follow an unknown probability distribution where noise can possibly be correlated in time. This uncertainty is then propagated through the state transition function $\Phi(t_1; t_0, x_0, w)$. As a result, the final state $x_1$ becomes a random variable, which we here denote by $\bz$. We refer to its (unknown) distribution as to $\mathbb{P}_z$.

In the following, we assume access to a dataset $D=\{z_1,\dots,z_N\}$ containing draws of the terminal state. Formally, $D$ is modeled as a realization of $\bD = \{\bz_1,\dots,\bz_N\}$, where $\bz_1,\dots,\bz_N$ are independent and identically distributed, each with distribution $\mathbb{P}_z$. These draws can be obtained in various ways, for instance by simulating the system dynamics, or by measuring the terminal state via experiments. Our task is then to utilize this dataset to construct a subset of the state-space, denoted by $S(D)\subseteq\mathbb{R}^{n_x}$, and certify that a large fraction of the probability mass of $\bz$ falls in $S(D)$. This problem is summarized in the following box: 

 \begin{tcolorbox}[
 colframe=gray!70!white,
 colback=gray!20!white,
 arc=1pt,
 breakable,
left=1pt,right=1pt,top=1pt,bottom=1pt,
 boxrule=0.3pt,
 ]
\textbf{Data-driven reachability.} Given $N$ i.i.d. draws from the distribution of the final state $\bz\sim\mathbb{P}_z$ collected in the dataset $D$, determine a set $S(D)$ and a bound $\varepsilon(D)$ such that, with probability $1-\delta$ with respect to the realizations of $\bD$, it holds that 
\[
R(\bD) \le \varepsilon(\bD), 
\]
where $R(D) = \mathbb{P}_z[z: z \notin S(D)]$.  
\end{tcolorbox}

\begin{remark}
	Three comments are in order. First, the formulation is very general and applies to both continuous-time and discrete-time systems described by arbitrary transition functions. In all cases, the problem of computing a reachable set corresponds to estimating the support of the distribution of the final state, having access to an i.i.d. sample of realizations of this random variable. Second, the approach extends naturally to obtain a reachable set at multiple times $(t_1,t_2,\dots, t_K)$, in which case one considers the joint probability distribution over the states at those times. Third, note that the problem of estimating the support of a random variable with formal probabilistic guarantees has applications beyond reachability analysis. As such, the technique  described can find application in other domains as well. 	
\end{remark}

\subsection{P2L for data-driven reachability}
\medskip

In this section, we show how P2L can be applied to data-driven reachability, while equipping the reachable set with probabilistic guarantees (\Cref{pillar1:cor_boundx}). 

Recall that the meta-algorithm P2L (\Cref{alg:meta-algo}) is fully specified once the following elements are defined: i) the synthesis algorithm $L$, ii) the property $\varphi$ and the quantifier of the degree of dissatisfaction, iii) an initialization $h_0$. We discuss each of them in the following.
\medskip

\noindent
\textbf{Synthesis algorithm.} In the context of data-driven reachability, a synthesis algorithm $L$ maps a  training set $T\subseteq D$ to a set $S(T) \subseteq\mathbb{R}^{n_x}$, with the aim of approximating the support of the distribution $\mathbb{P}_z$.  For this purpose, we use the algorithm proposed by \cite{Lasserreetal2019}, and also employed by \cite{Muratetalreachability2023} and \cite{pmlr-v204-tebjou23a} in the context of reachability analysis via PAC-Bayes and conformal prediction. Such an algorithm utilizes the training set $T$ to construct a so-called \emph{empirical inverse Christoffel function}, from which one determines $S$ as a sublevel set.\footnote{We refer the reader to \cite{Lasserreetal2019} for the theoretical underpinnings of this method, e.g., see \cite[Thm 3.14]{Lasserreetal2019} for the convergence of $S$ to the true support of the distribution as the number of draws grows. Here, we provide a minimal, but self-sufficient exposition.}

Given an integer $d$ that plays the role of a hyper-parameter, denote by $v(x)$ the vector of monomials on $\mathbb{R}^{n_x}$ (i.e., over the state-space) of degree no higher than $d$. 
For example, when $n_x=3$ and $d=2$, one has $v(x)=[1,~x_1,~x_2,~x_3,~x_1^2,~x_2^2,~x_3^2,~x_1x_2,~x_1x_3,~x_2x_3]$. Given the dataset $T$, the empirical inverse Christoffel function is defined as 
\[
\kappa^{-1}(x)=v(x)^\top \hat M^{-1}v(x),
\]
where $\hat M$ is the matrix of empirical moments, defined by 
\[
\hat M = \frac{1}{|T|}\sum_{z_i\in T}v(z_i)v(z_i)^\top.
\]
The set $S$ is obtained as the $\alpha$-sublevel set of this empirical inverse Christoffel function, where $\alpha = \max_{z_i\in T} \kappa^{-1}(z_i)$. This corresponds to ``slicing'' the empirical inverse Christoffel function at its maximum value over the training set $T$, and taking as $S$ the region of the state space where the empirical inverse Chrisoffel function lies below this value: 
\[
S = \{x\in\mathbb{R}^{n_x}~\text{s.t.}~v(x)^\top \hat M^{-1}v(x)\le \alpha\}.
\]
For later use, we let the algorithm $L$ return $h=(S,\kappa^{-1}(x))$, i.e., both the set $S$ and the empirical inverse Christoffel function itself, which will have a role when quantifying the degree of dissatisfaction. 
\medskip

\noindent
\textbf{Property $\varphi$ and quantifier of the degree of dissatisfaction.} Recall that our goal is to bound the probability with which a final state falls outside the set constructed by P2L. Referring to Definition \ref{def:stat_risk}, this corresponds to the risk associated with the following property $\varphi$. Given a decision $h=(S,\kappa^{-1}(x))$ and a $z$, $\varphi(h,z) = 1$ if $z \in S$, and zero otherwise. As for the degree of dissatisfaction, a sensible definition for points $z_i$ in $D\setminus T$ that fall outside $S$ is to assign them a value given by $\kappa^{-1}(z_i)$ (when selecting the point with the highest degree of dissatisfaction, possible ties are broken using any preference criterion over points in $\mathbb{R}^{n_x}$). 
\medskip

\noindent
\textbf{Initialization.} We utilize a small fraction of $D$ to initialize P2L. Specifically, we use $N_i \ll N$ data-points to compute via the above-described synthesis algorithm an initial set and an empirical inverse Christoffel function, which form the initial decision $h_0$ of P2L.
\medskip

\noindent
\textbf{Probabilistic bound.} Having described the ingredients necessary to define the meta-algorithm, we now turn our attention to certifying its output. To bound the probability that a new terminal state falls outside the set $S$ returned by P2L, we can directly apply \Cref{thm:main_result} for the choice of property $\varphi$ introduced above, so obtaining the following corollary. 
\begin{corollary}
\label{pillar1:cor_boundx}
Let $S(D)$ be the reachable set, and $T$ the corresponding compressed set, returned by P2L. Then, with confidence $1-\delta$ with respect to the realizations of $\bD$, it holds that 
\[
R(\bD) \le \bar\varepsilon(|\bT|,\delta,N-N_i).
\]
where $R(D) = \mathbb{P}_z[z: z \notin S(D)]$. 
\end{corollary}

\subsection{Pick-to-Learn in action}
\medskip

We consider the benchmark employed in \cite{Muratetalreachability2023,pmlr-v204-tebjou23a} and compare the performance of P2L  vis-a-vis state-of-the-art methods. 

As in \cite{Muratetalreachability2023,pmlr-v204-tebjou23a}, we consider the Duffing oscillator 

\[
\begin{cases}
\dot x_1 &= x_2\\
\dot x_2 &= x_1(1-x_1^2) -\alpha x_2 + \gamma\cos(\omega t),	
\end{cases}
\]

\noindent
with $\alpha=0.05$, $\gamma=0.4$, $\omega=1.3$ and uncertain initial state at time $t_0=0$. Given access to $N$ i.i.d. realizations of the initial state, we construct the i.i.d. sample $D=\{z_1,\dots, z_N\}$ by exploiting the system dynamics, where each $z_i$ denotes the terminal state at time $t_1=100$ obtained from a corresponding initial condition. Leveraging the dataset $D$, we then generate a set $S$ over $\mathbb{R}^2$ intended to include as much of the probabilistic mass of $\bz$ as possible. Three different methods (P2L, conformal prediction and test-set\footnote{We do not include a direct comparison with PAC-Bayes methods, as recent work, \cite{pmlr-v204-tebjou23a}, has shown that PAC-Bayes approaches yield looser bounds than conformal prediction in this context.}) are employed, and compared using the same confidence level $\delta=0.01$, number of data points $N=2000$, and order $d=10$ for the inverse Christoffel function. For each method, we compute the volume of the resulting set $S$, and evaluate its risk using a Monte-Carlo approach: $500,000$ new terminal states are generated, and the risk is computed as the fraction that falls outside $S$. This entire experiment is repeated $20,000$ times to give statistical significance to our findings. Before presenting the results, we provide details on how each individual method is implemented.
\medskip 

\noindent
\emph{Pick-to-Learn.} We run the meta-algorithm P2L presented in \Cref{alg:meta-algo} with the choice of learning algorithm, property $\varphi$ and quantifier of the degree of dissatisfaction, and initialization previously described. Here we take $N_i=200$, i.e., use $10\%$ of the available dataset to initialize $h_0$. With the common choice of $\delta=0.01$ and $N=2000$ as specified above, P2L returned a bound to the risk of $\varepsilon=0.034$ on average over the $20,000$ repetitions of the experiment, and a standard deviation of $0.0029$. This means that, with confidence $0.99$, the bound on the probability of observing a new terminal state lying outside $S$ was (on average) $3.4\%$.
\medskip

\noindent
\emph{Conformal prediction.} 
We run the conformal prediction approach exactly as presented in \cite{pmlr-v204-tebjou23a},%
\footnote{In other contributions, conformal prediction has also been utilized fixing a priori the ``shape'' of the reachable set, see \cite{Lindemann2024}. We do not include comparisons with these approaches that do not allow the shape of the set $S$ to adapt to the data.}
and give it the benefit of experimenting with different fractions of the dataset (10\%, 20\%, 30\%, 40\%, 50\%, 60\%, 70\%, 80\%, 90\%) to construct the calibration set.\footnote{Testing different configurations and selecting the best one as done here formally requires summing the individual confidences used in each test, resulting in an overall confidence for conformal prediction that is lower (worse) than that obtained by P2L; in this case, we have a confidence of $1-9\cdot \delta = 0.91<0.99$. We allow conformal prediction this small advantage.} In order to ensure a fair comparison with P2L, for each fraction we calibrate (this corresponds to selecting a value $k$ in \Cref{eq:conformal_bound} in the conformal prediction floating box) the conformal risk bound to the smallest $\varepsilon$ equal to or greater than that of P2L, and then select the set with the smallest volume among the nine candidates.\footnote{This procedure was repeated $20,000$ times, once for each realization of $D$.} By doing so, both methods return the same certificate, and can therefore be fairly compared by analyzing the volume of the resulting sets. 
\medskip

\noindent
\emph{Test-set.} 
We run the test-set method, also giving it the benefit of experimenting with different fractions of the dataset for training vs testing (again, 10\%, 20\%, 30\%, 40\%, 50\%, 60\%, 70\%, 80\%, 90\%). We then picked the fraction that, among all those achieving an equal or smaller risk bound than that of P2L, returns the set $S$ with smallest volume. 
\medskip

\noindent
\textbf{Results.} 
\Cref{fig:Empirical_risk_and_volume} (top panel) presents the histogram of the volume of the sets $S$ obtained by the three approaches over $20,000$ repetitions, while \Cref{fig:Empirical_risk_and_volume} (bottom panel) shows the histogram of the risk evaluated over $500,000$ additional draws (Monte-Carlo evaluation). Averages and standard deviations of these values are also reported in \Cref{tab:volumes_risks}.
In \Cref{fig:median_sets}, we present typical sets with volumes close to the median of the histograms to provide a visual impression of their structure. 
\medskip

\setlength\figureheighteps{3.5cm}
\setlength\figurewidtheps{0.85\linewidth} 
\begin{figure}[h!]
	\centering
	\begin{tikzpicture}

\begin{axis}[%
width=\figurewidtheps,
height=\figureheighteps,
at={(1.011in,0.642in)},
scale only axis,
xmin=0.325,
xmax=0.46,
ymin=0,
ymax=180,
scaled x ticks=false, 
xlabel={Volume of reachable set},
ylabel={Frequency},
ylabel style={yshift=-10pt},
legend style={legend cell align=left, align=left, draw=none},
reverse legend
]
\addplot[ybar interval, fill=black, fill opacity=0.6, draw=black, area legend] table[row sep=crcr] {%
x	y\\
0.33075	0\\
0.332433333333333	0\\
0.334116666666667	0\\
0.3358	0\\
0.337483333333333	0\\
0.339166666666667	0\\
0.34085	0\\
0.342533333333333	0\\
0.344216666666667	0\\
0.3459	0\\
0.347583333333333	0\\
0.349266666666667	0\\
0.35095	0\\
0.352633333333333	0\\
0.354316666666667	0\\
0.356	0\\
0.357683333333333	0\\
0.359366666666667	0\\
0.36105	0\\
0.362733333333333	0\\
0.364416666666667	0\\
0.3661	0\\
0.367783333333333	0\\
0.369466666666667	0\\
0.37115	1\\
0.372833333333333	0\\
0.374516666666667	3\\
0.3762	3\\
0.377883333333333	10\\
0.379566666666667	5\\
0.38125	12\\
0.382933333333333	17\\
0.384616666666667	17\\
0.3863	26\\
0.387983333333333	43\\
0.389666666666667	30\\
0.39135	46\\
0.393033333333333	64\\
0.394716666666667	77\\
0.3964	67\\
0.398083333333333	109\\
0.399766666666667	116\\
0.40145	114\\
0.403133333333333	120\\
0.404816666666667	154\\
0.4065	152\\
0.408183333333333	164\\
0.409866666666667	146\\
0.41155	131\\
0.413233333333333	117\\
0.414916666666667	71\\
0.4166	63\\
0.418283333333333	45\\
0.419966666666667	35\\
0.42165	21\\
0.423333333333333	10\\
0.425016666666667	8\\
0.4267	3\\
0.428383333333333	0\\
0.430066666666667	0\\
0.43175	0\\
0.433433333333333	0\\
0.435116666666667	0\\
0.4368	0\\
0.438483333333333	0\\
0.440166666666667	0\\
0.44185	0\\
0.443533333333333	0\\
0.445216666666667	0\\
0.4469	0\\
0.448583333333333	0\\
0.450266666666667	0\\
0.45195	0\\
0.453633333333333	0\\
0.455316666666667	0\\
0.457	0\\
0.458683333333333	0\\
0.460366666666667	0\\
0.46205	0\\
0.463733333333333	0\\
0.465416666666667	0\\
0.4671	0\\
0.468783333333333	0\\
0.470466666666667	0\\
0.47215	0\\
0.473833333333333	0\\
0.475516666666667	0\\
0.4772	0\\
0.478883333333333	0\\
0.480566666666667	0\\
0.48225	0\\
0.483933333333333	0\\
0.485616666666667	0\\
0.4873	0\\
0.488983333333333	0\\
0.490666666666667	0\\
0.49235	0\\
0.494033333333333	0\\
0.495716666666667	0\\
0.4974	0\\
};
\addlegendentry{TS}

\addplot[ybar interval, fill=blue, fill opacity=0.6, draw=black, area legend] table[row sep=crcr] {%
x	y\\
0.33075	0\\
0.332433333333333	0\\
0.334116666666667	0\\
0.3358	0\\
0.337483333333333	0\\
0.339166666666667	0\\
0.34085	0\\
0.342533333333333	0\\
0.344216666666667	0\\
0.3459	0\\
0.347583333333333	0\\
0.349266666666667	0\\
0.35095	0\\
0.352633333333333	0\\
0.354316666666667	0\\
0.356	1\\
0.357683333333333	0\\
0.359366666666667	0\\
0.36105	0\\
0.362733333333333	2\\
0.364416666666667	0\\
0.3661	2\\
0.367783333333333	0\\
0.369466666666667	1\\
0.37115	1\\
0.372833333333333	4\\
0.374516666666667	5\\
0.3762	4\\
0.377883333333333	3\\
0.379566666666667	7\\
0.38125	9\\
0.382933333333333	11\\
0.384616666666667	14\\
0.3863	21\\
0.387983333333333	21\\
0.389666666666667	25\\
0.39135	26\\
0.393033333333333	38\\
0.394716666666667	43\\
0.3964	55\\
0.398083333333333	73\\
0.399766666666667	75\\
0.40145	87\\
0.403133333333333	81\\
0.404816666666667	111\\
0.4065	108\\
0.408183333333333	126\\
0.409866666666667	126\\
0.41155	131\\
0.413233333333333	136\\
0.414916666666667	109\\
0.4166	109\\
0.418283333333333	92\\
0.419966666666667	67\\
0.42165	62\\
0.423333333333333	52\\
0.425016666666667	28\\
0.4267	30\\
0.428383333333333	24\\
0.430066666666667	18\\
0.43175	8\\
0.433433333333333	11\\
0.435116666666667	7\\
0.4368	4\\
0.438483333333333	9\\
0.440166666666667	2\\
0.44185	7\\
0.443533333333333	5\\
0.445216666666667	0\\
0.4469	3\\
0.448583333333333	1\\
0.450266666666667	3\\
0.45195	1\\
0.453633333333333	0\\
0.455316666666667	0\\
0.457	0\\
0.458683333333333	0\\
0.460366666666667	0\\
0.46205	0\\
0.463733333333333	0\\
0.465416666666667	0\\
0.4671	0\\
0.468783333333333	0\\
0.470466666666667	0\\
0.47215	0\\
0.473833333333333	0\\
0.475516666666667	0\\
0.4772	0\\
0.478883333333333	0\\
0.480566666666667	0\\
0.48225	0\\
0.483933333333333	0\\
0.485616666666667	0\\
0.4873	0\\
0.488983333333333	0\\
0.490666666666667	0\\
0.49235	0\\
0.494033333333333	0\\
0.495716666666667	1\\
0.4974	1\\
};
\addlegendentry{Conf}

\addplot[ybar interval, fill=red, fill opacity=0.6, draw=black, area legend] table[row sep=crcr] {%
x	y\\
0.33075	1\\
0.332433333333333	0\\
0.334116666666667	1\\
0.3358	3\\
0.337483333333333	5\\
0.339166666666667	7\\
0.34085	8\\
0.342533333333333	18\\
0.344216666666667	24\\
0.3459	32\\
0.347583333333333	47\\
0.349266666666667	50\\
0.35095	72\\
0.352633333333333	101\\
0.354316666666667	116\\
0.356	127\\
0.357683333333333	140\\
0.359366666666667	151\\
0.36105	157\\
0.362733333333333	144\\
0.364416666666667	144\\
0.3661	134\\
0.367783333333333	131\\
0.369466666666667	118\\
0.37115	60\\
0.372833333333333	66\\
0.374516666666667	35\\
0.3762	36\\
0.377883333333333	23\\
0.379566666666667	29\\
0.38125	10\\
0.382933333333333	6\\
0.384616666666667	3\\
0.3863	1\\
0.387983333333333	0\\
0.389666666666667	0\\
0.39135	0\\
0.393033333333333	0\\
0.394716666666667	0\\
0.3964	0\\
0.398083333333333	0\\
0.399766666666667	0\\
0.40145	0\\
0.403133333333333	0\\
0.404816666666667	0\\
0.4065	0\\
0.408183333333333	0\\
0.409866666666667	0\\
0.41155	0\\
0.413233333333333	0\\
0.414916666666667	0\\
0.4166	0\\
0.418283333333333	0\\
0.419966666666667	0\\
0.42165	0\\
0.423333333333333	0\\
0.425016666666667	0\\
0.4267	0\\
0.428383333333333	0\\
0.430066666666667	0\\
0.43175	0\\
0.433433333333333	0\\
0.435116666666667	0\\
0.4368	0\\
0.438483333333333	0\\
0.440166666666667	0\\
0.44185	0\\
0.443533333333333	0\\
0.445216666666667	0\\
0.4469	0\\
0.448583333333333	0\\
0.450266666666667	0\\
0.45195	0\\
0.453633333333333	0\\
0.455316666666667	0\\
0.457	0\\
0.458683333333333	0\\
0.460366666666667	0\\
0.46205	0\\
0.463733333333333	0\\
0.465416666666667	0\\
0.4671	0\\
0.468783333333333	0\\
0.470466666666667	0\\
0.47215	0\\
0.473833333333333	0\\
0.475516666666667	0\\
0.4772	0\\
0.478883333333333	0\\
0.480566666666667	0\\
0.48225	0\\
0.483933333333333	0\\
0.485616666666667	0\\
0.4873	0\\
0.488983333333333	0\\
0.490666666666667	0\\
0.49235	0\\
0.494033333333333	0\\
0.495716666666667	0\\
0.4974	0\\
};
\addlegendentry{P2L}

\end{axis}

\end{tikzpicture}%
	\vspace*{-5mm}
	\begin{tikzpicture}

\begin{axis}[%
width=\figurewidtheps,
height=\figureheighteps,
at={(1.011in,0.642in)},
scale only axis,
xmin=0,
xmax=0.055,
ymin=0,
ymax=190,
scaled x ticks=false, 
xtick={0.01, 0.02, 0.03, 0.04, 0.05}, 
xticklabels={0.01, 0.02, 0.03, 0.04, 0.05}, 
xlabel={Risk},
ylabel={Frequency},
ylabel style={yshift=-10pt},
legend style={legend cell align=left, align=left, draw=none},
reverse legend
]
\addplot[ybar interval, fill=black, fill opacity=0.6, draw=black, area legend] table[row sep=crcr] {%
x	y\\
0.00087	0\\
0.00138868686868687	0\\
0.00190737373737374	0\\
0.00242606060606061	0\\
0.00294474747474748	0\\
0.00346343434343434	0\\
0.00398212121212121	0\\
0.00450080808080808	0\\
0.00501949494949495	0\\
0.00553818181818182	0\\
0.00605686868686869	0\\
0.00657555555555555	1\\
0.00709424242424242	0\\
0.00761292929292929	0\\
0.00813161616161616	1\\
0.00865030303030303	0\\
0.0091689898989899	0\\
0.00968767676767677	3\\
0.0102063636363636	2\\
0.0107250505050505	4\\
0.0112437373737374	6\\
0.0117624242424242	5\\
0.0122811111111111	9\\
0.012799797979798	10\\
0.0133184848484848	14\\
0.0138371717171717	15\\
0.0143558585858586	11\\
0.0148745454545455	19\\
0.0153932323232323	15\\
0.0159119191919192	22\\
0.0164306060606061	20\\
0.0169492929292929	37\\
0.0174679797979798	32\\
0.0179866666666667	51\\
0.0185053535353535	57\\
0.0190240404040404	46\\
0.0195427272727273	52\\
0.0200614141414141	36\\
0.020580101010101	47\\
0.0210987878787879	48\\
0.0216174747474747	51\\
0.0221361616161616	38\\
0.0226548484848485	51\\
0.0231735353535354	65\\
0.0236922222222222	46\\
0.0242109090909091	49\\
0.024729595959596	47\\
0.0252482828282828	58\\
0.0257669696969697	60\\
0.0262856565656566	60\\
0.0268043434343434	62\\
0.0273230303030303	59\\
0.0278417171717172	44\\
0.028360404040404	45\\
0.0288790909090909	50\\
0.0293977777777778	55\\
0.0299164646464646	46\\
0.0304351515151515	30\\
0.0309538383838384	36\\
0.0314725252525253	30\\
0.0319912121212121	38\\
0.032509898989899	29\\
0.0330285858585859	26\\
0.0335472727272727	31\\
0.0340659595959596	31\\
0.0345846464646465	18\\
0.0351033333333333	33\\
0.0356220202020202	24\\
0.0361407070707071	13\\
0.0366593939393939	21\\
0.0371780808080808	14\\
0.0376967676767677	14\\
0.0382154545454545	19\\
0.0387341414141414	14\\
0.0392528282828283	13\\
0.0397715151515152	11\\
0.040290202020202	18\\
0.0408088888888889	12\\
0.0413275757575758	8\\
0.0418462626262626	3\\
0.0423649494949495	7\\
0.0428836363636364	12\\
0.0434023232323232	4\\
0.0439210101010101	9\\
0.044439696969697	4\\
0.0449583838383838	4\\
0.0454770707070707	3\\
0.0459957575757576	4\\
0.0465144444444444	5\\
0.0470331313131313	5\\
0.0475518181818182	3\\
0.0480705050505051	1\\
0.0485891919191919	1\\
0.0491078787878788	0\\
0.0496265656565657	0\\
0.0501452525252525	0\\
0.0506639393939394	1\\
0.0511826262626263	1\\
0.0517013131313131	1\\
0.05222	1\\
};
\addlegendentry{TS}

\addplot[ybar interval, fill=blue, fill opacity=0.6, draw=black, area legend] table[row sep=crcr] {%
x	y\\
0.00087	0\\
0.00138868686868687	0\\
0.00190737373737374	0\\
0.00242606060606061	0\\
0.00294474747474748	0\\
0.00346343434343434	0\\
0.00398212121212121	0\\
0.00450080808080808	0\\
0.00501949494949495	0\\
0.00553818181818182	0\\
0.00605686868686869	0\\
0.00657555555555555	0\\
0.00709424242424242	0\\
0.00761292929292929	0\\
0.00813161616161616	1\\
0.00865030303030303	0\\
0.0091689898989899	0\\
0.00968767676767677	0\\
0.0102063636363636	3\\
0.0107250505050505	2\\
0.0112437373737374	2\\
0.0117624242424242	6\\
0.0122811111111111	2\\
0.012799797979798	8\\
0.0133184848484848	13\\
0.0138371717171717	16\\
0.0143558585858586	10\\
0.0148745454545455	25\\
0.0153932323232323	20\\
0.0159119191919192	36\\
0.0164306060606061	48\\
0.0169492929292929	51\\
0.0174679797979798	57\\
0.0179866666666667	81\\
0.0185053535353535	77\\
0.0190240404040404	84\\
0.0195427272727273	86\\
0.0200614141414141	92\\
0.020580101010101	80\\
0.0210987878787879	91\\
0.0216174747474747	79\\
0.0221361616161616	89\\
0.0226548484848485	84\\
0.0231735353535354	90\\
0.0236922222222222	68\\
0.0242109090909091	88\\
0.024729595959596	68\\
0.0252482828282828	82\\
0.0257669696969697	61\\
0.0262856565656566	50\\
0.0268043434343434	54\\
0.0273230303030303	53\\
0.0278417171717172	45\\
0.028360404040404	32\\
0.0288790909090909	29\\
0.0293977777777778	20\\
0.0299164646464646	27\\
0.0304351515151515	22\\
0.0309538383838384	13\\
0.0314725252525253	10\\
0.0319912121212121	5\\
0.032509898989899	10\\
0.0330285858585859	8\\
0.0335472727272727	6\\
0.0340659595959596	3\\
0.0345846464646465	3\\
0.0351033333333333	3\\
0.0356220202020202	0\\
0.0361407070707071	2\\
0.0366593939393939	0\\
0.0371780808080808	3\\
0.0376967676767677	0\\
0.0382154545454545	1\\
0.0387341414141414	1\\
0.0392528282828283	0\\
0.0397715151515152	0\\
0.040290202020202	0\\
0.0408088888888889	0\\
0.0413275757575758	0\\
0.0418462626262626	0\\
0.0423649494949495	0\\
0.0428836363636364	0\\
0.0434023232323232	0\\
0.0439210101010101	0\\
0.044439696969697	0\\
0.0449583838383838	0\\
0.0454770707070707	0\\
0.0459957575757576	0\\
0.0465144444444444	0\\
0.0470331313131313	0\\
0.0475518181818182	0\\
0.0480705050505051	0\\
0.0485891919191919	0\\
0.0491078787878788	0\\
0.0496265656565657	0\\
0.0501452525252525	0\\
0.0506639393939394	0\\
0.0511826262626263	0\\
0.0517013131313131	0\\
0.05222	0\\
};
\addlegendentry{Conf}

\addplot[ybar interval, fill=red, fill opacity=0.6, draw=black, area legend] table[row sep=crcr] {%
x	y\\
0.00087	4\\
0.00138868686868687	3\\
0.00190737373737374	20\\
0.00242606060606061	29\\
0.00294474747474748	47\\
0.00346343434343434	77\\
0.00398212121212121	100\\
0.00450080808080808	144\\
0.00501949494949495	141\\
0.00553818181818182	160\\
0.00605686868686869	168\\
0.00657555555555555	172\\
0.00709424242424242	183\\
0.00761292929292929	140\\
0.00813161616161616	111\\
0.00865030303030303	97\\
0.0091689898989899	110\\
0.00968767676767677	73\\
0.0102063636363636	62\\
0.0107250505050505	49\\
0.0112437373737374	38\\
0.0117624242424242	28\\
0.0122811111111111	14\\
0.012799797979798	14\\
0.0133184848484848	3\\
0.0138371717171717	7\\
0.0143558585858586	5\\
0.0148745454545455	0\\
0.0153932323232323	0\\
0.0159119191919192	1\\
0.0164306060606061	0\\
0.0169492929292929	0\\
0.0174679797979798	0\\
0.0179866666666667	0\\
0.0185053535353535	0\\
0.0190240404040404	0\\
0.0195427272727273	0\\
0.0200614141414141	0\\
0.020580101010101	0\\
0.0210987878787879	0\\
0.0216174747474747	0\\
0.0221361616161616	0\\
0.0226548484848485	0\\
0.0231735353535354	0\\
0.0236922222222222	0\\
0.0242109090909091	0\\
0.024729595959596	0\\
0.0252482828282828	0\\
0.0257669696969697	0\\
0.0262856565656566	0\\
0.0268043434343434	0\\
0.0273230303030303	0\\
0.0278417171717172	0\\
0.028360404040404	0\\
0.0288790909090909	0\\
0.0293977777777778	0\\
0.0299164646464646	0\\
0.0304351515151515	0\\
0.0309538383838384	0\\
0.0314725252525253	0\\
0.0319912121212121	0\\
0.032509898989899	0\\
0.0330285858585859	0\\
0.0335472727272727	0\\
0.0340659595959596	0\\
0.0345846464646465	0\\
0.0351033333333333	0\\
0.0356220202020202	0\\
0.0361407070707071	0\\
0.0366593939393939	0\\
0.0371780808080808	0\\
0.0376967676767677	0\\
0.0382154545454545	0\\
0.0387341414141414	0\\
0.0392528282828283	0\\
0.0397715151515152	0\\
0.040290202020202	0\\
0.0408088888888889	0\\
0.0413275757575758	0\\
0.0418462626262626	0\\
0.0423649494949495	0\\
0.0428836363636364	0\\
0.0434023232323232	0\\
0.0439210101010101	0\\
0.044439696969697	0\\
0.0449583838383838	0\\
0.0454770707070707	0\\
0.0459957575757576	0\\
0.0465144444444444	0\\
0.0470331313131313	0\\
0.0475518181818182	0\\
0.0480705050505051	0\\
0.0485891919191919	0\\
0.0491078787878788	0\\
0.0496265656565657	0\\
0.0501452525252525	0\\
0.0506639393939394	0\\
0.0511826262626263	0\\
0.0517013131313131	0\\
0.05222	0\\
};
\addlegendentry{P2L}
\end{axis}

\end{tikzpicture}%
	\vspace*{-5mm}
	\caption{\emph{Top panel} --  P2L returns a set with lower volume compared to both conformal prediction, and test-set. This difference is also notable in \Cref{fig:median_sets}. 
    \emph{Bottom panel} -- P2L has lower, i.e., better, empirical risk compared to both conformal prediction, and test-set.}	
	\label{fig:Empirical_risk_and_volume}
\end{figure}
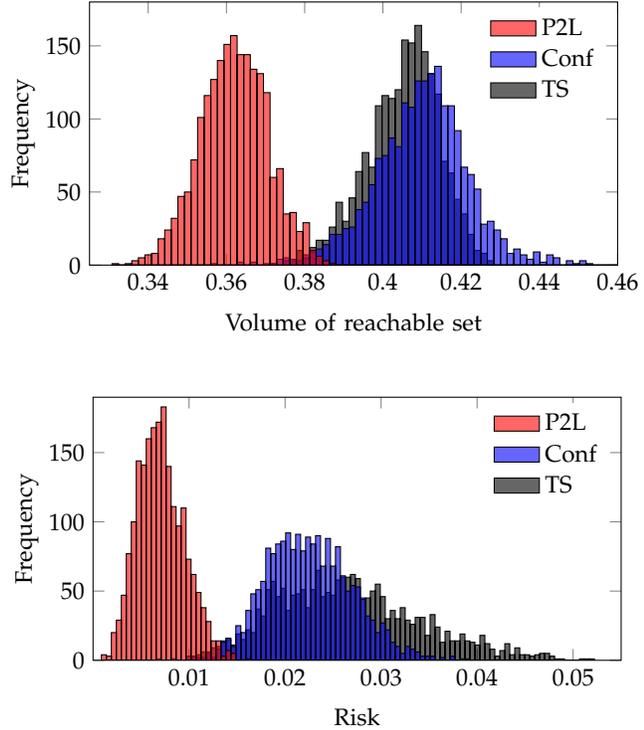

\begin{table}[h]
    \centering
    \begin{tabular}{lccc} %
        \toprule
        & \textbf{P2L} & Conf & TS \\ 
        \midrule
        {Volume } & \textbf{0.362} $\pm$ \textbf{0.009} & 0.410 $\pm$ 0.012 &  0.405 $\pm$ 0.009 \\
        {Risk } & \textbf{0.007} $\pm$ \textbf{0.002} & 0.023 $\pm$ 0.005 & 0.027 $\pm$ 0.008 \\
        \bottomrule
    \end{tabular}
    \vspace*{2mm}
    \caption{Mean $\pm$ standard deviation of volumes and risk of the reachable sets.}
    \label{tab:volumes_risks}
\end{table}

\setlength\figureheighteps{4cm}
\setlength\figurewidtheps{4cm} 
\begin{figure*}
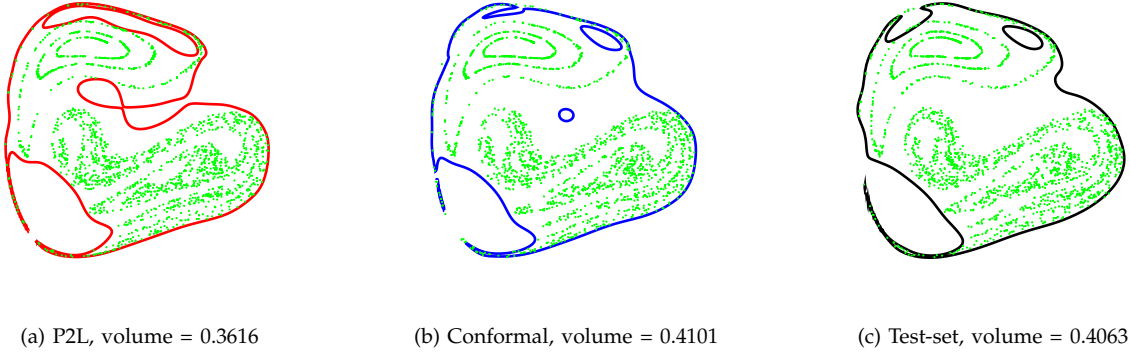

    \centering
    \begin{subfigure}[t]{0.31\textwidth}
    \centering 
       \include{figures_tikz/P2L_median_N=2000} 
        \vspace*{-5mm}
        \caption{P2L, volume = 0.3616}
    \end{subfigure}
    \hspace{2mm}
    \begin{subfigure}[t]{0.31\textwidth}
    \centering 
        	\include{figures_tikz/Conf_median_N=2000}
        	\vspace*{-5mm}
        \caption{Conformal, volume = 0.4101}    \end{subfigure}
    \hspace{2mm}
    \begin{subfigure}[t]{0.31\textwidth}
    \centering 
        \include{figures_tikz/TT_median_N=2000}
        \vspace*{-5mm}
        \caption{Test-set, volume = 0.4063}
    \end{subfigure}
    \vspace*{3mm}
    \caption{Reachable set with median volume obtained by different methods for equal certificate.}
    \label{fig:median_sets}
\end{figure*}

\noindent
\textbf{P2L constructs better sets than conformal prediction and test-set.} \Cref{fig:Empirical_risk_and_volume} shows that the sets built with P2L outperform those obtained with other approaches both in terms of volume and in their ability to include a larger portion of the probabilistic mass of $\bz$. The difference in volume is rather substantial, as can also be visually appreciated in \Cref{fig:median_sets}. This difference, amounting to more than a 10\% reduction, would for example translate into less conservative downstream control designs when the reachable sets are used for safety-critical applications. 

The reason why P2L achieves better performance is that it more effectively exploits the available data. Specifically, P2L is allowed to use the entire dataset to \emph{jointly} learn the set $S$ and provide a risk certificate. This is not the case for the other methods: test-set holds back a portion of the data for testing, while in conformal prediction the data used to construct the inverse Christoffel function cannot be used for certifying the reliability of the set.

\section{Pillar 2: Data-driven Optimal Control}
\medskip

Optimal control deals with designing algorithms that steer the behavior of a dynamical system so as to minimize an objective of interest. Built on the calculus of variations, modern optimal control dates back to the work of Pontryagin and Bellman, who pioneered this field by developing necessary and sufficient conditions of optimality. Their work has since become fundamental, leading to the solution of the linear-quadratic regulator problem \cite{kalman1960contributions}, laying the foundations for reinforcement learning (RL), \cite{watkins1989learning}, and contributing to the development of optimization-based techniques such as model predictive control (MPC), \cite{garcia1989model}. Recently, motivated by the widespread availability of data, there has been an emerging interest in reconsidering optimal control through a \emph{data-driven} lens, i.e., in settings where a complete model of the system dynamics is not available, but one has access to measurements of its behavior and, possibly, additional structural information. Examples range from RL, \cite{watkins1989learning}, to control via neural network, \cite{miller1995neural}, data-driven counterparts of MPC, and many others. 

When control policies are designed from data, it is crucial to equip them with safety and performance guarantees to ensure confidence in their real-world deployment. P2L allows the designer to utilize any data-driven control algorithm, while still providing formal certificates regardless of the specific method used. This pillar demonstrates the use of P2L in the context of optimal control and provides a comparison with alternative state-of-the-art approaches, see e.g. \cite{boroujeni2024pac}. 

\subsection{Problem formulation}
\medskip

Consider a discrete-time dynamical system of the form
\be
\label{eq:pillar_oc_dyn_sys}
x_{t+1}=f_t(x_t,u_t,w_t),
\ee
where $x_t \in \mathbb{R}^{n_x}$ is the state, $u_t \in \mathbb{R}^{n_u}$ is the control input, and $w_t \in \mathbb{R}^{n_w}$ is the disturbance. The system is initialized at time $t = 0$ with $x_0$ and runs for $H$ instants. The initial state $x_0$ and the disturbance signal $w := \{w_0,\ldots,w_{H-1}\}$ are uncertain, and it is assumed that each follows an unknown probability distribution, where the disturbance may exhibit temporal correlation. While these distributions are not known, we assume to have access to $N$ draws of $z = (x_0,w)$, which we collect in the dataset $D=\{z_1,\dots,z_N\}$. This $D$ is modeled as a realization of $\bD=\{\bz_1,\dots,\bz_N\}$, where $\bz_1,\dots,\bz_N$ are i.i.d. with common distribution $\mathbb{P}_z$. 

The input to the system is to be determined by a state-feedback control policy $\pi$:\footnote{$\pi$ is a shorthand to indicate the sequence $\{\pi_t\}_{t=0}^{H-1}$.} 
\[
u_t=\pi_t(\{x_s\}_{s=0}^{t}).
\]
No restrictions are imposed on the structure of this policy, which can amount to any function of $\{x_s\}_{s=0}^{t}$.\footnote{We adopt state-feedback as a working assumption to present the methodology; the P2L framework extends without conceptual modification to output-feedback controllers.} To a given policy $\pi$ and a realization $z$, there is associated a cost 
\begin{equation}
\label{cost optimal control}
J(\pi,z)=\sum_{t=0}^\T \left[ \ell_t(x_t,\pi_t(\{x_s\}_{s=0}^t))\right],
\end{equation}
where $x_t$ denotes the trajectory generated by the policy $\pi$ with $z=(x_0,w)$. 
We want to leverage the dataset $D$ to design a control policy $\pi(D)$ and provide probabilistic guarantees. Specifically, for a given threshold $\bar J$, we want to bound the probability that the cost $J(\pi(D),z)$ incurred under a new realization of the uncertainty exceeds $\bar J$. This goal is formalized as follows. 

\begin{tcolorbox}[
 colframe=gray!70!white,
 colback=gray!20!white,
 arc=1pt,
 breakable,
left=1pt,right=1pt,top=1pt,bottom=1pt,
 boxrule=0.3pt,
 ]
\textbf{Data-driven optimal control.} Given $N$ i.i.d. draws of the initial state $x_0$ and noise trajectory $w$ collected in the dataset $D$, determine a control policy $\pi(D)$ and a bound $\varepsilon(D)$ such that, with confidence $1-\delta$ with respect to the realizations of $\bD$, it holds that 
\[
R(\bD) \le \varepsilon(\bD), 
\]
where $R(D) = \mathbb{P}_z[z: J(\pi(D),z)  >  \bar J]$. 
\end{tcolorbox}

\begin{remark}
\label{rmk:performace_vs_safety}
While the formulation above points to performance guarantees, P2L can also be used to provide other types of guarantees, for instance bounds on the violation of input-state constraints (see also \Cref{sec:pillar-safety} for further discussion). Moreover, as illustrated in the numerical examples of this section, P2L can certify multiple thresholds, thereby bounding the entire cumulative distribution function of the risk.
\end{remark}
\begin{pullquote}
Pick-to-Learn  guarantees the performance of optimal control designs, even when they are obtained through complex or opaque procedures.
\end{pullquote}

\subsection{P2L for Data-driven Optimal Control}
\medskip

In the following, we focus on parameterized control policies $\pi^\theta$, where $\theta \in\mb{R}^{n_\theta}$. While this is not strictly necessary, this choice makes the presentation more concrete and aligns with many commonly encountered settings. For example, $\pi^\theta$ could represent the policy induced by MPC where $\theta$ parametrizes the MPC input and terminal cost matrices, \cite{zuliani2024closed}, or a neural-network-based controller where $\theta$ contains the network's weights and biases, \cite{boroujeni2024pac}. We then consider minimization of the empirical version of the problem obtained by replacing the expected cost $\mathbb{E}_{\bz}[J(\pi,\bz)]$ with its empirical average over the available data. This defines a choice for the algorithm $L$.\footnote{One might consider alternative empirical costs, for example the empirical probability that $J(\pi,z)$ exceeds the threshold $\bar J$. Here, we use the empirical counterpart of $\mathbb{E}_{\bz}[J(\pi,\bz)]$ because this is a common choice in many works.} Finally, guarantees are derived using the P2L framework. The components of the meta-algorithm P2L are discussed in detail below. 
\medskip

\noindent
\textbf{Synthesis Algorithm.} A synthesis algorithm $L$ is a map that associates a training set $T\subseteq D$ to a parameter value $\theta(T)$. In the examples presented below, this $\theta(T)$ is obtained via minimization of the empirical cost 
\[
\frac{1}{|T|}\sum_{z_i\in T} J(\pi^\theta,z_i) 
\]
with respect to $\theta$. P2L returns the policy $\pi(D) = \pi^{\theta(T)}$, where $T$ is the training set at termination. 
\medskip 

\noindent
\textbf{Property $\varphi$ and quantifier of the degree of dissatisfaction.} We are interested in bounding the probability that a newly drawn $z$ incurs a cost higher than $\bar J$. This corresponds to defining the property of interest $\varphi$ through $\varphi(h,z) = \varphi(\theta,z) = 1$ if $J(\pi^\theta,z)\le \bar J$, and zero otherwise. The degree of dissatisfaction of a data point $z$ is measured through the cost $J(\pi^\theta, z)$, with higher cost corresponding to higher dissatisfaction (ties are broken according to any pre-specified rule).
\medskip

\noindent
\textbf{Initialization.} A small number of data points $N_i \ll N$ are used to determine an initial policy $h_0 = \pi^{\theta_0}$.
\medskip

\noindent
\textbf{Performance guarantees for data-driven optimal control.} We now turn attention to equipping the policy $\pi(D)$ returned by P2L with formal guarantees. The result is formalized in the next corollary, which follows readily from \Cref{thm:main_result}.

\begin{corollary}
\label{pillar_SNOC:cor_bound}
Let $\pi(D)$ be the control policy, and $T$ the corresponding compressed set, returned by P2L. Then, with confidence $1-\delta$ with respect to the realizations of $\bD$, it holds that 
\[
R(\bD) \le \bar\varepsilon(|\bT|,\delta,N-N_i), 
\]
where $R(D) = \mathbb{P}_z[z: J(\pi(D),z)  >  \bar J]$. 
\end{corollary}
\begin{remark}
	To highlight the versatility of P2L, we remark that P2L can also be used to certify the probability of incurring a cost greater than a data-dependent threshold, rather than a threshold fixed a priori. As an example, one can define $\bar J(T)$ as the worst-case cost incurred by the designed policy $\theta(T)$ over the elements in $T$: $\bar J(T) = \max_{z_i\in T} J(\pi^{\theta(T)},z_i)$. In this setting, the algorithm $L$ returns $L(T) = (\theta(T),\bar J(T))$, and property $\varphi$ is defined in relation to the data-tuned threshold $\bar J(T)$. Specifically, letting $\bar{\bar{J}}(D) = \bar{J}(T)$, where $T$ is the training set at termination of P2L, property $\varphi$ is satisfied for a given $z$ by the decision $h(D) = (\pi(D),\bar{\bar{J}}(D))$ returned by P2L if $J(\pi(D),z) \leq \bar{\bar{J}}(D)$. Notice that the condition $J(\pi(D),z) \leq \bar{\bar{J}}(D)$ can here be written as $\varphi(h(D),z) = 1$ because we have included in the decision $h(D)$ the value $\bar{\bar{J}}(D)$. 
\end{remark}

\subsection{P2L in action}
\medskip

We first apply P2L to a simple benchmark introduced in \cite{boroujeni2024pac}. In \cite{boroujeni2024pac}, PAC-Bayes bounds are employed to generate performance certificates. Later in the section, we present a more complex robotic example also inspired by \cite{boroujeni2024pac}. We first present details on the benchmark and then provide numerical results. 
\medskip

\noindent
The benchmark considers a simple linear system with a single state variable: $x_{t+1} = a x_{t} +b u_t+w_t$, with $a=0.8$, $b=0.1$. The initial state $x_0$ is distributed as $\mathcal{N}(2.3, 0.009)$, and the noise is i.i.d. with distribution $\mathcal{N}(0.3, 0.009)$. These distributions are disclosed here for completeness, but are not available for policy selection. The objective is to tune the parameters of a simple, static state-feedback policy given by $\pi^\theta(x_t)=\theta_1 x_t+\theta_2$ with the goal of minimizing the cost in \eqref{cost optimal control} over a time horizon $H=9$ with a time-invariant stage cost $\ell_t(x,u)=5x^2+0.003u^2$. Following \cite{boroujeni2024pac}, the values for $\theta_1$ and $\theta_2$ are discretized. Precisely, $(\theta_1,\theta_2)$ are confined to the square $[-18,2] \times [-5,5]$ and $100$ equally-spaced values within their respective domains are considered for each parameter. \footnote{Note that the domain enforced on $\theta_1$ corresponds to (simple) stability of the closed-loop system.}
\medskip

\noindent
We compare the empirical performances and guarantees obtained by P2L and PAC-Bayes for various values of $N=4,8,\dots, 128$. For each $N$, the experiment is repeated $10,000$ times to obtain statistical significance. The value of $\bar J$ is set to $4$. The actual risk is estimated via a Monte-Carlo approach with $10,000$ draws. Before presenting the results, we provide details on how each individual method is implemented. 
\medskip

\noindent
\emph{Pick-to-Learn.} P2L is executed with the choices described in the previous section. The policy is initialized using a single data point, i.e., $N_i=1$. 
\medskip 

\noindent
\emph{PAC-Bayes.} Referring to the floating box describing PAC-Bayes methods, we recall that this approach assigns a bound to the performance of competing distributions $\mc{Q}$, and the distribution yielding the smallest bound is selected. In other words, the type of bound provided by the method is directly tied to the adopted selection criterion. For this reason, PAC-Bayes is implemented with the choice $c(h,z) = {\bf 1}(J(\pi^\theta,z) > \bar J)$, where ${\bf 1}(\cdot)$ is the indicator function (refer to the floating box for the notation $c(h,z)$). In this way, the resulting bound aligns with the bound provided by P2L, thus providing a common ground for comparison.\footnote{Note that this choice departs from the approach in \cite{boroujeni2024pac}, where an averaging strategy is used.} The distribution $\mc{P}$ (refer again to the floating box) was taken to be uniform over the grid of values for $(\theta_1,\theta_2)$: since each parameter can take one of $100$ values, each point in the grid was assigned probability $1/10,000$. 
\medskip

\newlength\figureheightoc
\newlength\figurewidthoc
\setlength\figureheightoc{4.5cm}
\setlength\figurewidthoc{0.8\linewidth} 

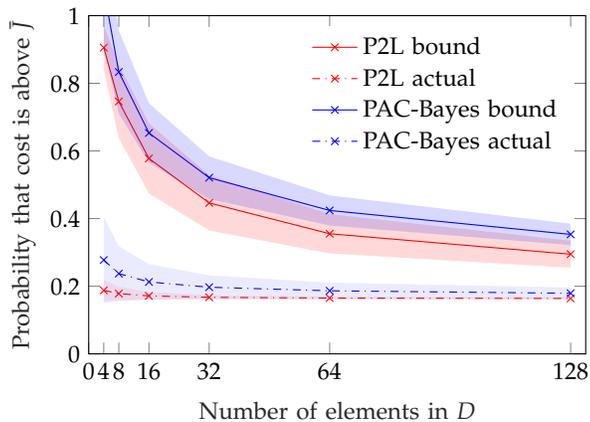
\begin{figure}
	\centering	
	\begin{tikzpicture}
\begin{axis}[%
width=\figurewidthoc,
height=\figureheightoc,
at={(1.011in,0.642in)},
scale only axis,
xmin=0,
xmax=132,
xlabel style={font=\color{white!15!black}},
xlabel={Number of elements in $D$},
ymin=0,
ymax=1,
ylabel style={font=\color{white!15!black}},
ylabel={Probability that cost is above $\bar J$},
xtick={0,4,8,16,32, 64, 128},
title style={font=\bfseries},
every axis plot post/.append style={-},
legend style={legend cell align=left, align=left, draw=none},
legend pos=north east,
ylabel style={yshift=-10pt},
]

\addplot[area legend, draw=none, fill=red, fill opacity=0.15, forget plot]
table[row sep=crcr] {%
x	y\\
4	0.970834089028793\\
8	0.854837308211498\\
16	0.681384583289744\\
32	0.52797991567603\\
64	0.413370825892234\\
128	0.334652082255631\\
128	0.254731950903224\\
64	0.297070066933697\\
32	0.36468910913853\\
16	0.474345413320039\\
8	0.637532729328918\\
4	0.840237375035081\\
}--cycle;
\addplot [color=red, mark=x, mark options={solid, red}]
  table[row sep=crcr]{%
4	0.905535732031937\\
8	0.746185018770208\\
16	0.577864998304892\\
32	0.44633451240728\\
64	0.355220446412966\\
128	0.294692016579428\\
};
\addlegendentry{P2L bound}

\addplot[area legend, draw=none, fill=red, fill opacity=0.08, forget plot]
table[row sep=crcr] {%
x	y\\
4	0.219265295731275\\
8	0.19845068645909\\
16	0.182487552377779\\
32	0.173642405057456\\
64	0.1695044426109\\
128	0.167844400897178\\
128	0.160029447102822\\
64	0.1606007533891\\
32	0.160954684942545\\
16	0.160301043622221\\
8	0.15800809954091\\
4	0.156141096268725\\
}--cycle;
\addplot [color=red, dashdotted, mark=x, mark options={solid, red}]
  table[row sep=crcr]{%
4	0.187703196\\
8	0.178229393\\
16	0.171394298\\
32	0.167298545\\
64	0.165052598\\
128	0.163936924\\
};
\addlegendentry{P2L actual}

\addplot[area legend, draw=none, fill=blue, fill opacity=0.15, forget plot]
table[row sep=crcr] {%
x	y\\
4	1.24808275069277\\
8	0.95646360091061\\
16	0.741125788201136\\
32	0.584104795452671\\
64	0.468616833780397\\
128	0.384959103110792\\
128	0.321689416010591\\
64	0.379446006635366\\
32	0.457907905153088\\
16	0.565174172129855\\
8	0.710322261188638\\
4	0.907254811527179\\
}--cycle;
\addplot [color=blue, mark=x, mark options={solid, blue}]
  table[row sep=crcr]{%
4	1.07766878110998\\
8	0.833392931049624\\
16	0.653149980165496\\
32	0.52100635030288\\
64	0.424031420207882\\
128	0.353324259560692\\
};
\addlegendentry{PAC-Bayes bound}

\addplot[area legend, draw=none, fill=blue, fill opacity=0.08, forget plot]
table[row sep=crcr] {%
x	y\\
4	0.40452536102407\\
8	0.318444781245185\\
16	0.265287857418149\\
32	0.232317800060669\\
64	0.210832292114569\\
128	0.196275495076622\\
128	0.162222768923378\\
64	0.162201113885431\\
32	0.161816709939331\\
16	0.16015620058185\\
8	0.156278988754815\\
4	0.15032644297593\\
}--cycle;
\addplot [color=blue, dashdotted, mark=x, mark options={solid, blue}]
  table[row sep=crcr]{%
4	0.277425902\\
8	0.237361885\\
16	0.212722029\\
32	0.197067255\\
64	0.186516703\\
128	0.179249132\\
};
\addlegendentry{PAC-Bayes actual}

\end{axis}
\end{tikzpicture}%
	\vspace*{-8mm}
	\caption{Linear system benchmark: average risk bounds (solid line) and average actual risks (dashed line) for P2L and PAC-Bayes over $10,000$ trials. The shaded area represents one standard deviation over the trials.}
	\label{fig:Cost_UB_vs_risk}
\end{figure}

\setlength\figureheighteps{3.5cm}
\setlength\figurewidtheps{0.8\linewidth} 
\begin{figure}[h!]
	\centering
	\begin{tikzpicture}

\begin{axis}[%
width=\figurewidtheps,
height=\figureheighteps,
at={(1.011in,0.642in)},
scale only axis,
xmin=0.1,
xmax=0.5,
ymin=0,
ymax=22000,
scaled x ticks=false, 
xlabel={Risk bound},
ylabel={Frequency},
ylabel style={yshift=-2pt},
legend style={legend cell align=left, align=left, draw=none},
reverse legend
]

\addplot[ybar interval, fill=blue, fill opacity=0.6, draw=black, area legend] table[row sep=crcr] {%
x	y\\
0.137823275814299	0\\
0.150031594339655	0\\
0.162239912865011	0\\
0.174448231390367	0\\
0.186656549915723	0\\
0.198864868441079	0\\
0.211073186966435	0\\
0.223281505491791	2\\
0.235489824017148	10\\
0.247698142542504	48\\
0.25990646106786	229\\
0.272114779593216	779\\
0.284323098118572	2146\\
0.296531416643928	4362\\
0.308739735169284	7758\\
0.320948053694641	11358\\
0.333156372219997	14311\\
0.345364690745353	15206\\
0.357573009270709	14124\\
0.369781327796065	11523\\
0.381989646321421	8071\\
0.394197964846777	4986\\
0.406406283372134	2756\\
0.41861460189749	1384\\
0.430822920422846	596\\
0.443031238948202	231\\
0.455239557473558	80\\
0.467447875998914	0\\
0.467447875998914	28\\
0.482513933347411	9\\
0.497579990695908	3\\
0.512646048044406	3\\
};
\addlegendentry{PAC-Bayes}

\addplot[ybar interval, fill=red, fill opacity=0.6, draw=black, area legend] table[row sep=crcr] {%
x	y\\
0.137823275814299	11\\
0.150031594339655	30\\
0.162239912865011	99\\
0.174448231390367	213\\
0.186656549915723	424\\
0.198864868441079	886\\
0.211073186966435	1594\\
0.223281505491791	2504\\
0.235489824017148	8638\\
0.247698142542504	6714\\
0.25990646106786	8013\\
0.272114779593216	8992\\
0.284323098118572	9495\\
0.296531416643928	18665\\
0.308739735169284	8060\\
0.320948053694641	6881\\
0.333156372219997	9843\\
0.345364690745353	3051\\
0.357573009270709	2214\\
0.369781327796065	2390\\
0.381989646321421	575\\
0.394197964846777	315\\
0.406406283372134	291\\
0.41861460189749	53\\
0.430822920422846	37\\
0.443031238948202	8\\
0.455239557473558	4\\
0.467447875998914	4\\
};
\addlegendentry{P2L}

\end{axis}

\end{tikzpicture}%
	\vspace*{-10mm}
	\begin{tikzpicture}

\begin{axis}[%
width=\figurewidtheps,
height=\figureheighteps,
at={(1.011in,0.642in)},
scale only axis,
xmin=0.1,
xmax=0.3,
ymin=0,
ymax=50000,
scaled x ticks=false, 
xlabel={Actual risk},
ylabel={Frequency},
ylabel style={yshift=-2pt},
legend style={legend cell align=left, align=left, draw=none},
reverse legend
]

\addplot[ybar interval, fill=blue, fill opacity=0.6, draw=black, area legend] table[row sep=crcr] {%
x	y\\
0.147	20\\
0.152850847457627	1489\\
0.158701694915254	11863\\
0.164552542372881	22828\\
0.170403389830508	18795\\
0.176254237288136	13495\\
0.182105084745763	9214\\
0.18795593220339	6616\\
0.193806779661017	4534\\
0.199657627118644	3357\\
0.205508474576271	2261\\
0.211359322033898	1640\\
0.217210169491525	1126\\
0.223061016949153	838\\
0.22891186440678	584\\
0.234762711864407	406\\
0.240613559322034	290\\
0.246464406779661	205\\
0.252315254237288	124\\
0.258166101694915	91\\
0.264016949152542	69\\
0.26986779661017	39\\
0.275718644067797	41\\
0.281569491525424	26\\
0.287420338983051	11\\
0.293271186440678	14\\
0.299122033898305	9\\
0.304972881355932	3\\
0.310823728813559	2\\
0.316674576271186	3\\
0.322525423728814	3\\
0.328376271186441	1\\
0.334227118644068	2\\
0.340077966101695	0\\
0.345928813559322	0\\
0.351779661016949	0\\
0.357630508474576	0\\
0.363481355932203	0\\
0.369332203389831	0\\
0.375183050847458	1\\
0.381033898305085	0\\
0.386884745762712	0\\
0.392735593220339	0\\
0.398586440677966	0\\
0.404437288135593	0\\
0.41028813559322	0\\
0.416138983050848	0\\
0.421989830508475	0\\
0.427840677966102	0\\
0.433691525423729	0\\
0.439542372881356	0\\
0.445393220338983	0\\
0.45124406779661	0\\
0.457094915254237	0\\
0.462945762711864	0\\
0.468796610169492	0\\
0.474647457627119	0\\
0.480498305084746	0\\
0.486349152542373	0\\
0.4922	0\\
};
\addlegendentry{PAC-Bayes}

\addplot[ybar interval, fill=red, fill opacity=0.6, draw=black, area legend] table[row sep=crcr] {%
x	y\\
0.147	147\\
0.152850847457627	8899\\
0.158701694915254	47592\\
0.164552542372881	38428\\
0.170403389830508	4788\\
0.176254237288136	143\\
0.182105084745763	3\\
0.18795593220339	0\\
0.193806779661017	0\\
0.199657627118644	0\\
0.205508474576271	0\\
0.211359322033898	0\\
0.217210169491525	0\\
0.223061016949153	0\\
0.22891186440678	0\\
0.234762711864407	0\\
0.240613559322034	0\\
0.246464406779661	0\\
0.252315254237288	0\\
0.258166101694915	0\\
0.264016949152542	0\\
0.26986779661017	0\\
0.275718644067797	0\\
0.281569491525424	0\\
0.287420338983051	0\\
0.293271186440678	0\\
0.299122033898305	0\\
0.304972881355932	0\\
0.310823728813559	0\\
0.316674576271186	0\\
0.322525423728814	0\\
0.328376271186441	0\\
0.334227118644068	0\\
0.340077966101695	0\\
0.345928813559322	0\\
0.351779661016949	0\\
0.357630508474576	0\\
0.363481355932203	0\\
0.369332203389831	0\\
0.375183050847458	0\\
0.381033898305085	0\\
0.386884745762712	0\\
0.392735593220339	0\\
0.398586440677966	0\\
0.404437288135593	0\\
0.41028813559322	0\\
0.416138983050848	0\\
0.421989830508475	0\\
0.427840677966102	0\\
0.433691525423729	0\\
0.439542372881356	0\\
0.445393220338983	0\\
0.45124406779661	0\\
0.457094915254237	0\\
0.462945762711864	0\\
0.468796610169492	0\\
0.474647457627119	0\\
0.480498305084746	0\\
0.486349152542373	0\\
0.4922	0\\
};
\addlegendentry{P2L}

\end{axis}

\end{tikzpicture}%
	\vspace*{-8mm}
	\caption{Linear system benchmark: distribution of the risk bounds (top) and the actual risks (bottom) obtained by P2L and PAC-Bayes for $N=128$.}
	\label{fig:UB_risk_distributions}
\end{figure}
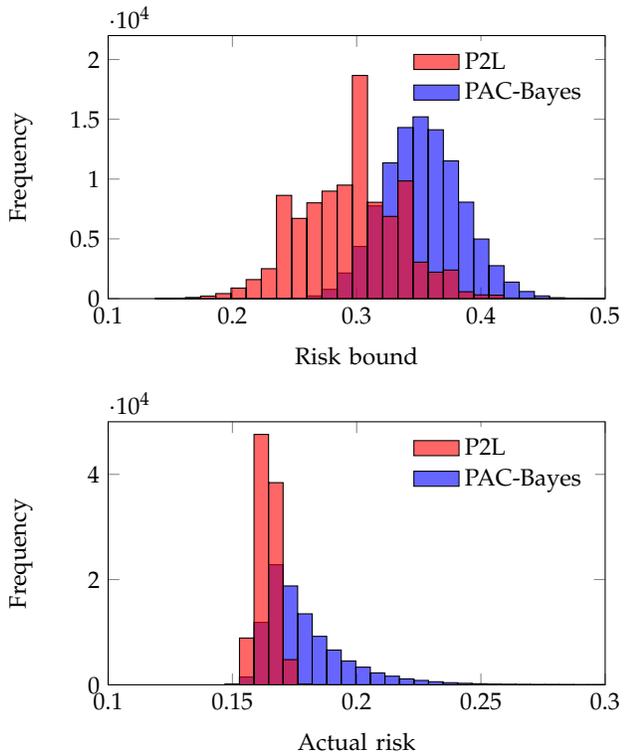

 \begin{table}[h]
	\centering
	\begin{tabular}{lcc} %
		\toprule
		& \textbf{P2L} & PAC-Bayes \\ 
		\midrule
		{Risk bound} & \textbf{0.294} $\pm$ \textbf{0.039} & 0.353 $\pm$ 0.032\\
		{Actual risk } & \textbf{0.164} $\pm$ \textbf{0.004} & 0.179 $\pm$ 0.017\\
		\bottomrule
	\end{tabular}
	\vspace*{2mm}
	\caption{Linear system benchmark: average value $\pm$ one standard deviation of the risk bounds and the actual risks for $N=128$.}
	\label{tab:comparison_N=128}
\end{table}

\noindent
\textbf{Results.} For PAC-Bayes, we use Catoni's bound, as given in Theorem 2.1 of \cite{PBbounds24}, and also equation (9) of \cite{boroujeni2024pac}, instead of the bound based on McAllester's result discussed in the floating box. The reason for this choice is that, after optimizing $\mc{Q}$ (following exactly the same procedure as in the floating box but with Catoni's bound in place of McAllester's), an additional interesting result holds: the performance can be bounded with probability $1 - \delta$ over joint realizations of the datasets and values of $\theta$ drawn according to the optimal distribution $\mc{Q}^\ast$ (refer to formulas (8) and (12) of \cite{boroujeni2024pac} or Theorem 2.7 of \cite{PBbounds24}). In other words, this latter bound holds for the risk associated to a single draw from $\mc{Q}^\ast$ without averaging over $\textbf{\boldmath$\theta$} \sim \mc{Q}^\ast$, a result that better aligns with the bound provided by P2L, where a single $\theta$ is selected and its performance is certified. 

The probabilistic bounds given by P2L and PAC-Bayes are shown in \Cref{fig:Cost_UB_vs_risk} for increasing dataset sizes $N=4,8,\dots, 128$ (solid line), together with the corresponding actual risks (dashed line). Following to the discussion in the last part of the previous paragraph, in PAC-Bayes, for each dataset, we compute $\mc{Q}^\ast$, and draw a single $\theta$ from it; the bound and the actual risk are then evaluated for this specific $\theta$. Both for P2L and PAC-Bayes, the actual risk is evaluated by computing the fraction of times the threshold $\bar J$ is exceeded over $10,000$ test draws of $\bz$ from $\mb{P}_z$ (Monte-Carlo evaluation). The entire experiment is repeated $10,000$ times, and \Cref{fig:Cost_UB_vs_risk} gives the average value and one standard deviation across these repetitions.

We see that, in this experiment, P2L provides the tightest bounds and achieves the lowest actual risks. Our interpretation of this result is that PAC-Bayes employs (possibly conservative) bounds to select a distribution over policies. This mechanism leads the method to depart from a purely data-driven approach, thereby downplaying the importance of the first-hand information contained in the dataset. In contrast, P2L selects a policy by only using the data, and the evaluation is performed a posteriori, without interfering with the selection procedure.\footnote{It can be further noticed that, in PAC-Bayes, we used an empirical cost that aligns with the goal of minimizing the probability of exceeding the threshold $\bar J$, whereas P2L minimizes an average cost. Nevertheless, better results are obtained with P2L despite this apparent advantage given to PAC-Bayes.} 

\Cref{fig:UB_risk_distributions} presents a complete account of the results obtained for $N=128$, while \Cref{tab:comparison_N=128} gives the averages and the corresponding standard deviation values.  
\medskip 

\noindent

\newlength\figureheightcdf
\newlength\figurewidthcdf
\setlength\figureheightcdf{0.4\linewidth}
\setlength\figurewidthcdf{0.4\linewidth} 

\medskip

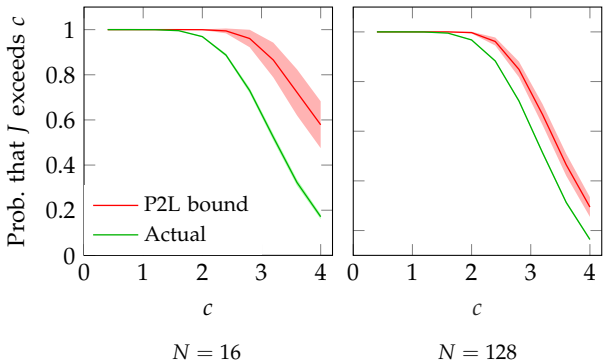
\begin{figure}[b!] 
	\begin{subfigure}[t]{0.5\linewidth}
		\captionsetup{labelformat=empty}
		\centering 
		\begin{tikzpicture}
\begin{axis}[%
width=\figurewidthcdf,
height=\figureheightcdf,
at={(1.011in,0.642in)},
scale only axis,
xmin=0,
xmax=4.2,
xlabel={$c$},
ymin=0,
ymax=1.1,
ylabel style={yshift=-10pt},
ylabel={Prob. that $J$ exceeds $c$},
ytick={0,0.2,...,1},
xtick={0,1,2,3,4},
tick label style={/pgf/number format/fixed},
legend style={at={(0,0)}, 
				anchor=south west,
				legend cell align=left,
				align=left, draw=none}
]%
\addplot[area legend, draw=none, fill=red, fill opacity=0.3, forget plot]
table[row sep=crcr] {%
x	y\\
0.4	1\\
0.8	1.00000043999957\\
1.2	1.00000285861995\\
1.6	1.00011191358959\\
2	1.00124333389372\\
2.4	1.00520040030094\\
2.8	0.999379507970053\\
3.2	0.942280440204222\\
3.6	0.82399859833845\\
4	0.682373575468402\\
4	0.474753882299544\\
3.6	0.619977218199002\\
3.2	0.788226094985575\\
2.8	0.921848303803604\\
2.4	0.983470199796531\\
2	0.998070637853156\\
1.6	0.999870837733938\\
1.2	0.999996714713793\\
0.8	0.999999551111547\\
0.4	1\\
}--cycle;
\addplot [color=red, line width=.5pt]
  table[row sep=crcr]{%
0.4	1\\
0.8	0.99999999555556\\
1.2	0.999999786666874\\
1.6	0.999991375661764\\
2	0.999656985873438\\
2.4	0.994335300048738\\
2.8	0.960613905886828\\
3.2	0.865253267594898\\
3.6	0.721987908268726\\
4	0.578563728883973\\
};
\addlegendentry{\footnotesize P2L bound}

\addplot[area legend, draw=none, fill=green, fill opacity=0.3, forget plot]
table[row sep=crcr] {%
x	y\\
0.4	1\\
0.8	1.00001647382573\\
1.2	0.999879371508889\\
1.6	0.996079700019235\\
2	0.971658679568384\\
2.4	0.893879028690352\\
2.8	0.741806178102285\\
3.2	0.538212516859103\\
3.6	0.337089115739481\\
4	0.182441056388336\\
4	0.160241423611663\\
3.6	0.308393524260519\\
3.2	0.508490403140896\\
2.8	0.718335121897718\\
2.4	0.880480471309649\\
2	0.966366300431616\\
1.6	0.994508559980759\\
1.2	0.999531548491126\\
0.8	0.999975086174267\\
0.4	1\\
}--cycle;
\addplot [color=black!30!green, line width=.5pt]
  table[row sep=crcr]{%
0.4	1\\
0.8	0.999995779999999\\
1.2	0.999705460000008\\
1.6	0.995294129999997\\
2	0.96901249\\
2.4	0.887179750000001\\
2.8	0.730070650000002\\
3.2	0.52335146\\
3.6	0.32274132\\
4	0.17134124\\
};
\addlegendentry{\footnotesize Actual}

\end{axis}%
\end{tikzpicture}%
		\caption{\hspace*{14mm}$N=16$}
	\end{subfigure}
	\begin{subfigure}[t]{0.5\linewidth}
		\captionsetup{labelformat=empty}
		\centering 
		\begin{tikzpicture}
\begin{axis}[%
width=\figurewidthcdf,
height=\figureheightcdf,
at={(1.011in,0.642in)},
scale only axis,
xmin=0,
xmax=4.2,
xlabel style={font=\color{white!15!black}},
xlabel={$c$},
ymin=0.1,
ymax=1.1,
ylabel style={font=\color{white!15!black}},
ytick={0,0.2,...,1},
xtick={0,1,2,3,4},
tick label style={/pgf/number format/fixed},  
yticklabels={},    %
legend style={at={(0,0)}, 
				anchor=south west,
				legend cell align=left,
				align=left, draw=none}
]%
\addplot[area legend, draw=none, fill=red, fill opacity=0.3, forget plot]
table[row sep=crcr] {%
x	y\\
0.4	1\\
0.8	1.00000001214955\\
1.2	1.00000236259454\\
1.6	1.00012618178679\\
2	1.00046835770729\\
2.4	0.977548927732524\\
2.8	0.880700618195379\\
3.2	0.708997086961706\\
3.6	0.508867868147739\\
4	0.334394999559085\\
4	0.254875075563008\\
3.6	0.420267998156073\\
3.2	0.624512131147262\\
2.8	0.817533703929445\\
2.4	0.945098622406074\\
2	0.993710845892323\\
1.6	0.999810097079619\\
1.2	0.999997468637599\\
0.8	0.999999987354476\\
0.4	1\\
}--cycle;
\addplot [color=red, line width=.5pt]
  table[row sep=crcr]{%
0.4	1\\
0.8	0.999999999752012\\
1.2	0.999999915616069\\
1.6	0.999968139433203\\
2	0.997089601799805\\
2.4	0.961323775069299\\
2.8	0.849117161062412\\
3.2	0.666754609054484\\
3.6	0.464567933151906\\
4	0.294635037561046\\
};
\addplot[area legend, draw=none, fill=green, fill opacity=0.3, forget plot]
table[row sep=crcr] {%
x	y\\
0.4	1\\
0.8	1.00001688782289\\
1.2	0.999863261492254\\
1.6	0.995747128852615\\
2	0.969375370632305\\
2.4	0.886187321012449\\
2.8	0.726853821068403\\
3.2	0.518521180217863\\
3.6	0.317848970250816\\
4	0.167855360851307\\
4	0.159974719148693\\
3.6	0.307918849749184\\
3.2	0.507708939782138\\
2.8	0.717281218931598\\
2.4	0.879514558987553\\
2	0.965737289367692\\
1.6	0.994334811147376\\
1.2	0.999506438507759\\
0.8	0.999974292177112\\
0.4	1\\
}--cycle;
\addplot [color=black!30!green, line width=.5pt]
  table[row sep=crcr]{%
0.4	1\\
0.8	0.999995589999999\\
1.2	0.999684850000007\\
1.6	0.995040969999995\\
2	0.967556329999998\\
2.4	0.882850940000001\\
2.8	0.722067520000001\\
3.2	0.51311506\\
3.6	0.31288391\\
4	0.16391504\\
};
\end{axis}
\end{tikzpicture}%
		\caption{~~$N=128$}
	\end{subfigure}	
	\caption{Linear system benchmark: bound on the probability of incurring a cost above $c$ (red) and corresponding actual probability (green) -- results over $100$ realizations of the experiment.}	\label{fig:CDF-bound}
\end{figure}
\textbf{Using P2L$^+$ to bound the cumulative distribution function of the cost.} Following the approach presented in \Cref{sec:multiple-levels}, we consider multiple levels for the cost, so providing an upper bound to its entire cumulative distribution function. These levels are: $\{0.4, 0.8, 1.2, 1.6, 2.0, 2.4, 2.8, 3.2, 3.6, 4.0\}$, and the joint guarantee holds with probability $1-10\cdot \delta = 0.9$, where $10$ is the number of levels. The results are shown in \Cref{fig:CDF-bound} for $N=16$ and $N=128$. 
\medskip

\begin{figure}
	\begin{minipage}{0.3\linewidth}
		\includegraphics[width=\linewidth]{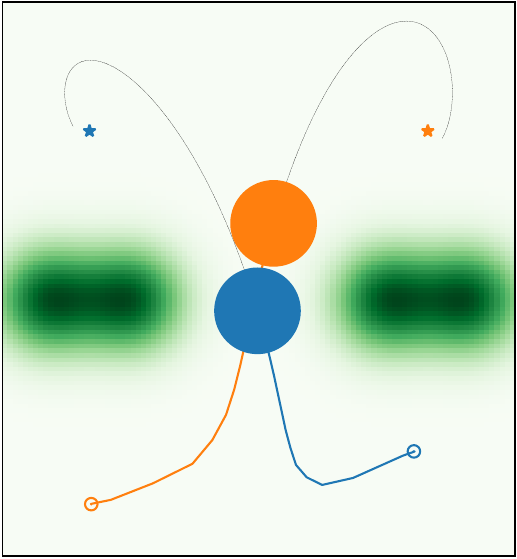}
	\end{minipage}%
	\quad
	\begin{minipage}{0.3\linewidth}
		\includegraphics[width=\linewidth]{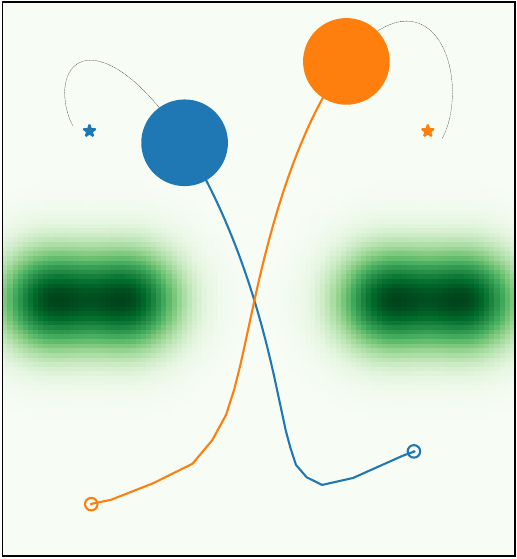}
	\end{minipage}%
	\quad
	\begin{minipage}{0.3\linewidth}
		\includegraphics[width=\linewidth]{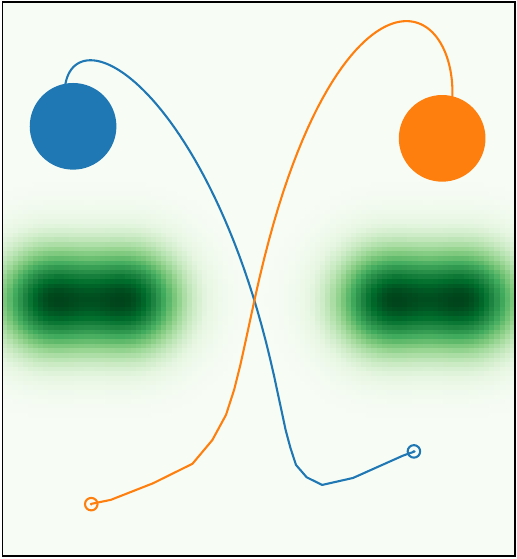}
	\end{minipage}%
	\caption{Three instants of a typical trajectory of the robots controlled with the law designed by P2L. Snapshots are taken at instants $t=15$ (left), $t=30$ (center), $t=99$ (right). Colored balls represent the robots, with their radius indicating their size for collision avoidance.}
	\label{fig:snapshots} 
\end{figure}

\noindent
\textbf{P2L tackles higher dimensional problems -- a robotic example.} The second benchmark presented in \cite{boroujeni2024pac} involves two robots that must move from their initial positions to their destinations going through a narrow opening while avoiding obstacles, as illustrated in \Cref{fig:snapshots}. Each robot is modeled as a double integrator subject to nonlinear friction forces. Uncertainty is present only in the initial state: the initial positions of the robots are drawn from $\mathcal{N}([-2, -2], 0.2\cdot I)$ and $\mathcal{N}([2, -2], 0.2\cdot I)$ respectively, with zero initial velocity, while the system is not subject to any disturbances. The stage cost is quadratic and includes three terms that penalize control effort, deviations from the target, and collisions. The policy $\pi^\theta$ is implemented as a neural network specifically crafted to ensure stability of the closed-loop system. Further details can be found in the original manuscript \cite{boroujeni2024pac}. We set $\bar J =50$, and $\delta = 0.01$. 

The paper \cite{boroujeni2024pac} does not provide risk bounds for this example, likely because of the computational overhead involved in PAC-Bayes. Since P2L does not suffer from this scalability issue, we are in a position to provide both bounds and actual risks, as shown in \Cref{fig:Cost_UB_vs_risk_robotic}. Three snapshots of the behavior of the system controlled with the law designed by P2L are instead reproduced in \Cref{fig:snapshots}. 

\setlength\figureheightoc{4.5cm}
\setlength\figurewidthoc{0.82\linewidth} 
\begin{figure}[tb!]
	\centering	
	\begin{tikzpicture}
\begin{axis}[%
width=\figurewidthoc,
height=\figureheightoc,
at={(1.011in,0.642in)},
scale only axis,
xmin=0,
xmax=256,
xlabel style={font=\color{white!15!black}},
xlabel={Number of elements in $D$},
ymin=0,
ymax=1,
ylabel style={font=\color{white!15!black}},
ylabel={Probability that cost is above $\bar J$},
xtick={0, 4, 8, 16, 32, 64, 128, 256},
xticklabels={0,,,16,32,64,128,256},  
title style={font=\bfseries},
every axis plot post/.append style={-},
legend style={legend cell align=left, align=left, draw=none},
legend pos=north east,
ylabel style={yshift=-10pt},
]

\addplot[area legend, draw=none, fill=red, fill opacity=0.15, forget plot]
table[row sep=crcr] {%
x	y\\
4	0.737109244136033\\
8	0.483546265535989\\
16	0.512558622469785\\
32	0.4162965581648\\
64	0.278695206500946\\
128	0.158093585269344\\
256	0.0756417127552964\\
256	0.108429765992932\\
128	0.200157219127468\\
64	0.327170835394987\\
32	0.601200798408785\\
16	0.828803137622183\\
8	0.864456778245511\\
4	0.921071681253719\\
}--cycle;
\addplot [color=red, mark=x, mark options={solid, red}]
  table[row sep=crcr]{%
4	0.829090462694876\\
8	0.67400152189075\\
16	0.670680880045984\\
32	0.508748678286793\\
64	0.302933020947967\\
128	0.179125402198406\\
256	0.0920357393741142\\
};
\addlegendentry{P2L bound}

\addplot[area legend, draw=none, fill=red, fill opacity=0.08, forget plot]
table[row sep=crcr] {%
x	y\\
4	0.0665973003318641\\
8	0.0739410027029136\\
16	0.0241783582082831\\
32	0.0191865581364662\\
64	0.00533122889566508\\
128	0.000391351759542065\\
256	-0.000104134832843905\\
256	0.0048041348328439\\
128	0.0247286482404579\\
64	0.0248687711043349\\
32	0.0995534418635338\\
16	0.208321641791717\\
8	0.199278997297086\\
4	0.213782699668136\\
}--cycle;
\addplot [color=red, dashdotted, mark=x, mark options={solid, red}]
  table[row sep=crcr]{%
4	0.14019\\
8	0.13661\\
16	0.11625\\
32	0.05937\\
64	0.0151\\
128	0.01256\\
256	0.00235\\
};
\addlegendentry{P2L actual}

\end{axis}
\end{tikzpicture}%
	\vspace*{-8mm}
	\caption{Robotic system benchmark: average risk bounds (solid line) and average actual risk (dashed line) for P2L over $100$ trials. The shaded area represents one standard deviation over the trials.}
	\label{fig:Cost_UB_vs_risk_robotic}
\end{figure}
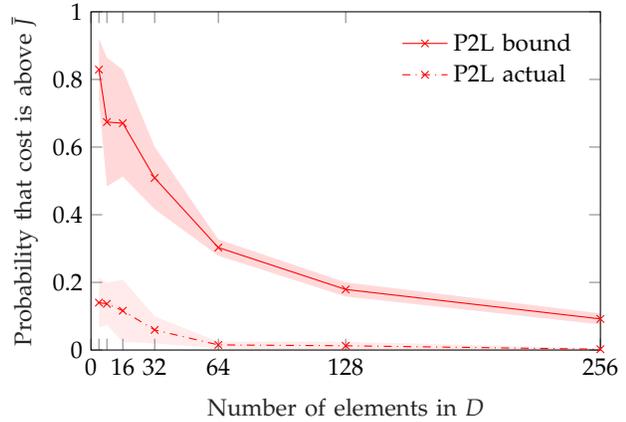

\section{Pillar 3: data-driven Control with safety specifications}
\label{sec:pillar-safety}
\medskip

Control algorithms with formal specifications have received increasing attention in recent years as a crucial component in safety-critical applications such as autonomous driving, where failing to meet the requirements can have severe consequences. Beyond standard control problems with state or input constraints, these algorithms also aim to address temporal properties, \cite{wolff2012robust}. For instance, reach-avoid problems require a dynamical system to reach a target region within a given time horizon while avoiding potentially dangerous conditions. 

While a variety of approaches have been proposed, \cite{wolff2012robust,badings2023robust,abate2024stochastic}, the state-of-the-art for stochastic systems remains that based on abstraction techniques, \cite{nazeri2025data}. In these methods, the available observations are used to construct a \emph{set} of models -- the so-called ``abstraction'' -- such that, with high probability, the performance of the true system lies between the worst-case and best-case performances evaluated over the models in the set. Control policies are then synthesized to robustly satisfy given specifications over the abstraction. In this way, the final design is also guaranteed to satisfy the specifications with high probability when applied to the true system. 

The main challenge with abstraction-based methods lies in the generation of the model set, which is data-intensive and often leads to conservative designs with loose performance bounds, \cite{badings2023robust,nazeri2025data}. In this section, we show that Pick-to-Learn can address this difficulty by directly mapping the available data into a control policy, while simultaneously providing formal guarantees. 
\begin{pullquote}
Pick-to-Learn certifies the safe use of control strategies without first constructing a guaranteed model of the full underlying system.
\end{pullquote}

\subsection{Problem formulation}
\medskip

The system description is the same as in the previous pillar and is repeated here for completeness. Consider a discrete-time dynamical system of the form
\be
\label{eq:pillar_oc_dyn_sys}
x_{t+1}=f_t(x_t,u_t,w_t),
\ee
where $x_t \in \mathbb{R}^{n_x}$ is the state, $u_t \in \mathbb{R}^{n_u}$ is the control input, and $w_t \in \mathbb{R}^{n_w}$ is the disturbance. The system is initialized at time $t = 0$ with $x_0$ and runs for $H$ instants. The initial state $x_0$ and the disturbance trajectory $w := \{w_0,\ldots,w_{H-1}\}$ are uncertain, and it is assumed that each follows an unknown probability distribution, where the disturbance may exhibit temporal correlation. While these distributions are not known, we assume to have access to $N$ draws of $z = (x_0,w)$, which we collect in the dataset $D=\{z_1,\dots,z_N\}$. This $D$ is modeled as a realization of $\bD=\{\bz_1,\dots,\bz_N\}$, where $\bz_1,\dots,\bz_N$ are i.i.d. with common distribution $\mathbb{P}_z$. The input to the system is to be determined by a state-feedback control policy $\pi$:   
\[
u_t=\pi_t(\{x_s\}_{s=0}^{t}).
\]

In this section, we aim to synthesize a policy to satisfy a property of interest $\varphi$ that encodes the given specifications. We let $\varphi(\pi,z)=1$ if the policy $\pi$ satisfies the specifications for $z=(x_0,w)$, and $\varphi(\pi,z)=0$ otherwise. For example, in reach-avoid problems, $\varphi(\pi,z)=1$ if the trajectory generated by policy $\pi$ from the initial condition $x_0$ under noise $w$ reaches the target within the allotted time horizon while avoiding unsafe regions. Our goal is to leverage the dataset $D$ to design a control policy $\pi$ ensuring that property $\varphi$ is satisfied with high probability. 

\begin{tcolorbox}[
 colframe=gray!70!white,
 colback=gray!20!white,
 arc=1pt,
 breakable,
left=1pt,right=1pt,top=1pt,bottom=1pt,
 boxrule=0.3pt,
 ]
\textbf{Data-driven safe synthesis.} Given $N$ i.i.d. draws of the initial state $x_0$ and noise trajectory $w$ collected in the dataset $D$, determine a control policy $\pi(D)$ and a bound $\varepsilon(D)$ such that, with confidence $1-\delta$ with respect to the realizations of $\bD$, it holds that 
\[
R(\bD) \le \varepsilon(\bD), 
\]
where $R(D) = \mathbb{P}_z[z: \varphi(\pi(D),z) = 0]$. 
\end{tcolorbox}

\subsection{P2L for data-driven control with safety specifications}
\medskip

P2L relies on the use of an algorithm $L$ to synthesize a control policy. The algorithm $L$ used in this section is based on abstraction techniques, those proposed in \cite{badings2023robust,nazeri2025data}. Although alternative algorithms could have been employed, we adopt this design methodology to facilitate a more grounded comparison with existing work. Specifically, algorithm $L$ leverages the available data to construct an uncertain Markov decision process in the form of an interval MDP (IMDP), \cite{givan2000bounded}. As is standard in abstraction-based methods, the control policy is synthesized by solving a robust linear programming problem over the IMDP, with the objective of maximizing the probability of satisfying $\varphi$. Crucially, however, unlike the abstraction-based literature we do \emph{not} require the IMDP to capture the true dynamics with high probability: within P2L, the IMDP becomes a mere heuristic for policy design, while P2L generalization results are employed to directly certify the synthesized control policy. We better formalize this idea next by introducing the three elements required to define the meta-algorithm. 
\medskip

\noindent
\textbf{Synthesis algorithm.} A synthesis algorithm $L$ is a map that associates a set $T\subseteq D$ to a control policy $\pi$. Following \cite{badings2023robust}, we discretize the state-control space and construct an IMDP by estimating the transition probabilities from each state $s$ to any other state $s'$ using the data in $T$. This is simply obtained by counting the fraction of occurrences where the system transits from $s$ to $s'$ under a given control action, and then constructing probability intervals around these values to obtain an IMDP.\footnote{\label{ftnt:P2Lfree}In contrast with standard abstraction-based techniques, where the intervals are carefully constructed to ensure that they contain the actual probabilities with high confidence (a requirement needed to obtain overall guarantees for the method), in P2L no such containment condition is enforced when implementing an abstraction-based technique as the $L$ algorithm. In fact, probabilistic guarantees are obtained a posteriori from the theoretical results of P2L, without the need to impose any a priori containment constraint. This feature is fundamental to relax the conservatism inherent in abstraction-based techniques, and allows the data to speak for what the actual risk really is.} The control policy is then obtained by maximizing the probability of satisfying $\varphi$ against the worst-case MDP within the found IMDP. This latter problem is solved via robust Bellman iterations using the off-the-shelf solver {\tt PRISM} (\cite{kwiatkowska2011prism}), as in \cite{badings2023robust}. The policy returned by P2L with the training set $T$ at termination is $\pi(D)$.
\medskip

\noindent
\textbf{Quantifier of the degree of dissatisfaction.} P2L terminates when all trajectories obtained from the pairs $z=(x_0,w)$ contained in $D\setminus T$ satisfy $\varphi$. Until this condition is met, an element from $D\setminus T$ is selected using a lexicographic tie-break rule and appended to $T$.\footnote{The lexicographic tie-break rule does not embody any particular logic of wisdom, and there remains room for further research on improved rules.}
\medskip

\noindent
\textbf{Initialization.} An initial policy is computed based on $N_i$ data points from $D$, with $N_i \ll N$. 
\medskip

\noindent
\textbf{Guarantees on the specifications.} Having fully defined the ingredients of the meta-algorithm, we are now ready to certify the policy obtained by P2L. Specifically, we are interested in certifying the probability that the designed $\pi(D)$ meets the requirements specified by $\varphi$. This follows readily from \Cref{thm:main_result}.

\begin{corollary}
\label{pillar_safe:cor_bound}
Let $\pi(D)$ be the control policy, and $T$ the corresponding compressed set, returned by P2L. Then, with confidence $1-\delta$ with respect to the realizations of $\bD$, it holds that 
\[
R(\bD) \le \bar\varepsilon(|\bT|,\delta,N-N_i), 
\]
where $R(D) = \mathbb{P}_z[z: \varphi(\pi(D),z) = 0]$. 
\end{corollary}

\begin{remark}
Below, as it is common in the literature on abstraction-based methods, we will refer to the probability of satisfying $\varphi$ rather than $R(D)$. This corresponds to considering $\mathbb{P}_z[z: \varphi(\pi(D),z) = 1] = 1-R(D)$, for which a valid lower bound is $1 - \bar\varepsilon(|T|,\delta,N-N_i)$.
\end{remark}

\subsection{Pick-to-Learn in action} 
\medskip

We consider the unmanned aerial vehicle (UAV) benchmark of \cite{badings2023robust}. The UAV is required to travel from an initial state to a target region while avoiding unsafe areas of the state space, corresponding to a reach-avoid specification, see \Cref{fig:UAV_problem}. The system has $6$ state variables, the UAV's 3D position and velocity, and a $3$-dimensional control input, where each component represents a one-directional force. The system's dynamics is modeled as a double integrator with additive disturbance obtained from a Dryden gust model. The time horizon is $H = 64$, and the disturbance is independent through time. In the simulations, the initial condition is deterministic and set to position $(-14,6,-2)$ with zero initial velocity. The reader is referred to  \cite{badings2023robust} for a more complete presentation of this benchmark. 

Unlike \cite{badings2023robust}, in the abstraction-based approach the probabilistic limits for the IMDP are computed using Clopper-Pearson bounds, as suggested in a later paper \cite{meggendorfer2024odds} (see also \cite{nazeri2025data}) by the same research group of \cite{badings2023robust}). The value of $\delta$ is set to $0.008$ both in P2L and in the abstraction-based approach when this approach is executed to provide a comparison with P2L.\footnote{When executed to provide a comparison with P2L, the abstraction-based approach is given access to all the available data in $D$, and the probability of satisfying $\varphi$ against the worst-case MDP within the found IMDP becomes a valid assessment of the probability of satisfying $\varphi$ with the real system because the IMDP is constructed so as to capture the true system's dynamics with confidence $1-0.008$.} However, the abstraction-based approach is also used as algorithm $L$ inside P2L. In this role, the abstraction-based approach is run with $\delta$ set to the value $0.5$. This large value of $\delta$ frees the algorithm $L$ from rigid a priori requirements that may lead to conservatism. Importantly, the guarantees provided by P2L retain their rigorous validity with confidence $1 - 0.008$ even with such a large $\delta$ in $L$, since the P2L guarantees hold irrespective of the algorithm $L$ used. In other words, as already mentioned in \Cref{ftnt:P2Lfree}, in P2L the abstraction-based approach is only used as a means to obtain a policy, while P2L itself is responsible for providing guarantees. P2L is initialized with a single disturbance trajectory, i.e., $N_i=1$. 

P2L and the abstraction-based approach are applied to an increasing number $N$ of disturbance trajectories, ranging from $N=5$ to $N=200$. Since each trajectory contains $64$ disturbance draws (one per time step), the abstraction-based approach utilizes $64 \times N$ disturbance draws, while algorithm $L$ in P2L utilizes progressively increasing number of draws given by $64 \times |T|$. For each value of $N$, the simulations are repeated $100$ times. 
\medskip

\begin{figure}
\centering
\includegraphics[width=\linewidth]{figures/UAV_v3}
\caption{Reach-avoid benchmark: a UAV is required to reach its goal region (green box) from its initial position, while avoiding unsafe states (red boxes). Figure adapted from \cite{badings2023robust}.}
\label{fig:UAV_problem}
\end{figure}

\newlength\figureheight
\newlength\figurewidth
\setlength\figureheight{4.5cm}
\setlength\figurewidth{0.8\linewidth} 
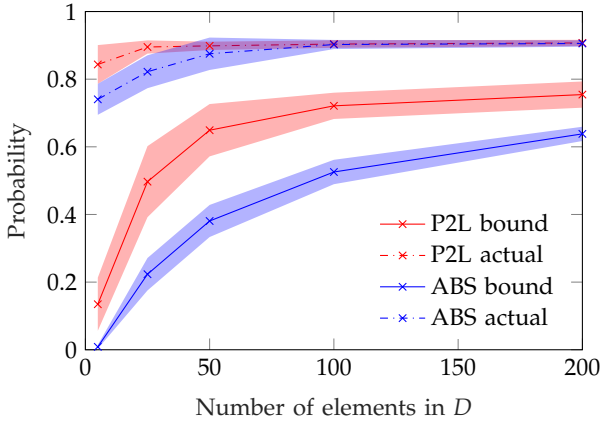
\begin{figure}
\begin{tikzpicture}
\begin{axis}[%
width=\figurewidth,
height=\figureheight,
at={(1.011in,0.642in)},
scale only axis,
xmin=0,
xmax=200,
xlabel style={font=\color{white!15!black}},
xlabel={Number of elements in $D$},
ymin=0,
ymax=1,
ylabel style={font=\color{white!15!black}},
ylabel={Probability},
title style={font=\bfseries},
every axis plot post/.append style={-},
legend style={legend cell align=left, align=left, draw=none},
legend pos=south east,
ylabel style={yshift=-10pt},
]

\addplot[area legend, draw=none, fill=red, fill opacity=0.3, forget plot]
table[row sep=crcr] {%
x	y\\
5	0.213203738708866\\
25	0.602053643427481\\
50	0.726489506067947\\
100	0.760027864777191\\
200	0.793224085859769\\
200	0.715687401324463\\
100	0.68231398309191\\
50	0.571738413941575\\
25	0.391924177827622\\
5	0.0560048914515626\\
}--cycle;
\addplot [color=red, mark=x, mark options={solid, red}]
  table[row sep=crcr]{%
5	0.134604315080214\\
25	0.496988910627551\\
50	0.649113960004761\\
100	0.721170923934551\\
200	0.754455743592116\\
};
\addlegendentry{P2L bound}

\addplot[area legend, draw=none, fill=red, fill opacity=0.3, forget plot]
table[row sep=crcr] {%
x	y\\
5	0.900829400635009\\
25	0.914662751568542\\
50	0.910873938928998\\
100	0.915002038266702\\
200	0.916415832721558\\
200	0.897944167278442\\
100	0.892637961733298\\
50	0.886186061071002\\
25	0.875597248431458\\
5	0.786230599364991\\
}--cycle;
\addplot [color=red, dashdotted, mark=x, mark options={solid, red}]
  table[row sep=crcr]{%
5	0.84353\\
25	0.89513\\
50	0.89853\\
100	0.90382\\
200	0.90718\\
};
\addlegendentry{P2L actual}

\addplot[area legend, draw=none, fill=blue, fill opacity=0.3, forget plot]
table[row sep=crcr] {%
x	y\\
5	0.0142658882747653\\
25	0.271368127617457\\
50	0.428197588510991\\
100	0.561663193972489\\
200	0.659371589038786\\
200	0.617169243961214\\
100	0.489479771027511\\
50	0.333267895089009\\
25	0.175381676782543\\
5	0.0029743007252347\\
}--cycle;
\addplot [color=blue, mark=x, mark options={solid, blue}]
  table[row sep=crcr]{%
5	0.0086200945\\
25	0.2233749022\\
50	0.3807327418\\
100	0.5255714825\\
200	0.6382704165\\
};
\addlegendentry{ABS bound}

\addplot[area legend, draw=none, fill=blue, fill opacity=0.3, forget plot]
table[row sep=crcr] {%
x	y\\
5	0.78619621086174\\
25	0.870634207620636\\
50	0.923048418676798\\
100	0.916048325184851\\
200	0.91491545473443\\
200	0.895964545265569\\
100	0.888931674815148\\
50	0.827211581323203\\
25	0.773325792379364\\
5	0.69460378913826\\
}--cycle;
\addplot [color=blue, dashdotted, mark=x, mark options={solid, blue}]
  table[row sep=crcr]{%
5	0.7404\\
25	0.82198\\
50	0.87513\\
100	0.90249\\
200	0.90544\\
};
\addlegendentry{ABS actual}

\end{axis}
\end{tikzpicture}%
\vspace*{-5mm}
\caption{Average probability bounds (solid line) and average actual probability (dashed line) of successfully completing the reach-avoid mission for P2L and the abstraction-based approach (ABS) over $100$ simulations. The shaded area represents one standard deviation over the simulations.}
\label{fig:reach-avoid}
\end{figure}

\newlength\figureheightheatmap
\newlength\figurewidthheatmap
\setlength\figureheightheatmap{4.5cm}
\setlength\figurewidthheatmap{0.32\linewidth}

\begin{figure}
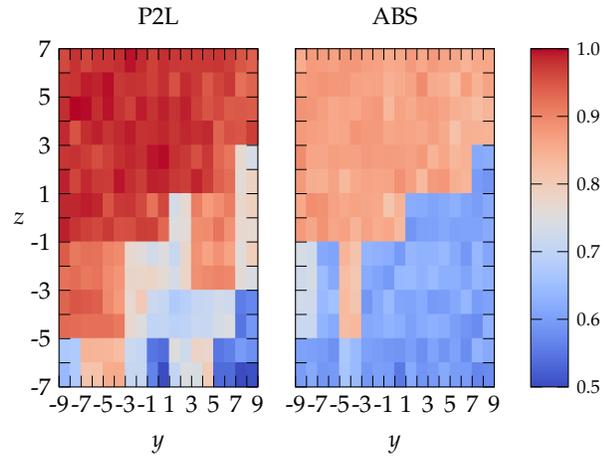

\centering
\include{figures_tikz/heatmap_vanilla_vs_P2L}
\vspace*{-5mm}
\caption{Actual probability of successfully completing the reach-avoid mission for P2L and the abstraction-based approach (ABS) with the UAV starting from various positions in the plane $x=-13$ and zero initial velocity. The number of disturbance trajectories used for policy synthesis is $N=5$. The probability values refer to a single typical design.}
\label{fig:empirical_reach_avoid}
\end{figure}

\noindent
\textbf{Results.} \Cref{fig:reach-avoid} provides a comparison of the results obtained by P2L and the abstraction-based approach. P2L delivers tighter probabilistic certificates (solid line). For example with $N=50$, the average bound on the probability of successfully completing the reach-avoid mission is above $64\%$ for P2L, compared to just $38\%$ for the abstraction-based approach. P2L also synthesizes a better policy, particularly when only few trajectories are used (dashed line).\footnote{The actual probabilities are evaluated via a Monte-Carlo approach with $20,000$ draws.} This results from relaxing the conservatism caused by the rigid a priori probabilistic constraints of the abstraction-based approach. 

\Cref{fig:empirical_reach_avoid} further illustrates this point for $N=5$. Here, the initial condition is no longer fixed at $(-14,6,-2)$; instead, initial conditions in the $(y,z)$ plane with $x=-13$ (and zero initial velocity) are extensively tested. The heatmap in the figure shows the actual probability of successfully completing the reach-avoid mission for P2L and the abstraction-based technique. \\

\noindent
\textbf{A broader evaluation of P2L's operation.} In closing this section, it is useful to take a broader perspective on P2L and highlight some fundamental aspects of the philosophy underlying the method. When selecting a synthesis algorithm $L$  in P2L, the user's only concern is that the algorithm should produce satisfactory designs; at this stage, formal guarantees are not required. In fact, P2L provides rigorous probabilistic guarantees a posteriori for any choice of $L$. This flexibility frees the designer from restrictive assumptions, in particular the need to choose based on uniform evaluations across sets of admissible designs, an approach that often leads to conservatism. Typically, the choice of $L$ is informed by multiple sources, including pre-judgments or even conjectures (for example, in this benchmark, the idea that the disturbance is independent over time). The fundamental fact is that the probabilistic guarantees provided by P2L maintain intact their validity whether these judgments are correct or not. Therefore, the user can inspect the results and, if they are not satisfactory, consider trying an alternative approach based on a different synthesis algorithm $L$.\footnote{When multiple algorithms $L$ are used, methodological rigor requires that the value of $\delta$ is multiplied by the number of attempts made; however, this does not pose any specific problem because, as already noted, selecting a small value of $\delta$ affects very marginally the bounds.}  
\medskip

\noindent

\section{Pillar 4: Data-driven robust control}
\medskip

Robust control is concerned with the synthesis of controllers that maintain desired performances in the presence of structural uncertainty and unmodeled dynamics. Classical lines of work focus on guaranteeing both robust stability and robust performance, drawing on tools such as the small-gain theorem \cite{zames2003input}, Lyapunov-based methods \cite{khalil2002nonlinear}, and optimization-based synthesis techniques for $H_2$, $H_\infty$, or mixed $H_2/H_\infty$ control, \cite{robustbook1996}. Many of these approaches can be formulated or approximated using linear matrix inequalities (LMIs) \cite{caverly2019lmi,boyd1994linear}. 

In this pillar, we focus on $H_2$ optimal control problems, however the presented methodology naturally extends to other frameworks such as $H_\infty$, mixed $H_2 / H_\infty$, and also to nonlinear counterparts of these problems. The goal is to synthesize a controller that achieves desired performances with high probability.\footnote{This probabilistic viewpoint departs from the classical deterministic treatment of robust control, and aligns with a rich stream of literature developed over the past two decades following pioneering works such as \cite{calafiore2006scenario}.} This is achieved by leveraging a set of draws from the uncertainty distribution, which in applications may come from historical data or from realizations generated by a stochastic model of the uncertainty. 
\begin{pullquote}
Pick-to-Learn enables $H_2$ and $H_\infty$ control design beyond the limits of quadratic stability, while providing formal robustness guarantees.
\end{pullquote}

\subsection{Problem Formulation} 
\medskip

Consider a linear time-invariant dynamical system described as 
\[
\begin{split}
\dot x &= A(z)x+B(z)u+E(z)w\\
     y &= C(z)x+D(z)u, 
\end{split}
\]
where $x\in\mathbb{R}^{n_x}$ is the state, $u\in\mathbb{R}^{n_u}$ is the control input, $w\in\mathbb{R}^{n_w}$ is the disturbance, and $y\in\mathbb{R}^{n_y}$ is the output. Matrices $A(z), B(z), C(z), D(z), E(z)$ are real-valued of appropriate size and depend on the uncertainty parameter $z\in \mathbb{R}^{n_z}$. For a given value of $z$ and a given state-feedback controller $u=Kx$, denote with $\Sigma_{z,K}$ the resulting closed-loop system and with $G_{z,K}(s)$ the corresponding transfer function between $w$ and $y$. For stable closed-loop systems, the $H_2$ norm of this transfer function is defined as
\[
H_2(\Sigma_{z,K}) = \sqrt{\frac{1}{2\pi}\int_{-\infty}^\infty \text{trace}(G_{z,K}(j\omega)G_{z,K}^*(j\omega)) d\omega},
\]
where $^*$ denotes the conjugate transpose, while $H_2(\Sigma_{z,K})$ is set to $\infty$ if the closed-loop system is unstable. When $w$ is a white noise, the $H_2$ norm equals the trace of the stationary output covariance matrix, \cite{robustbook1996}. Therefore, minimizing the $H_2$ norm corresponds to minimizing the average effect of white noise disturbances on the system output. 

Towards the goal of minimizing the $H_2$ norm, we assume to have access to a dataset $D=\{z_1,\dots,z_N\}$ containing draws of the uncertainty parameter. We model $D$ as a realization of $\bD=\{\bz_1,\dots,\bz_N\}$, where $\bz_1,\dots,\bz_N$ are i.i.d. with distribution $\mathbb{P}_z$. By leveraging this information, we want to design a state-feedback controller $u=Kx$ and provide probabilistic guarantees on the resulting $H_2$ norm, as formalized below.

\begin{tcolorbox}[
 colframe=gray!70!white,
 colback=gray!20!white,
 arc=1pt,
 breakable,
left=1pt,right=1pt,top=1pt,bottom=1pt,
 boxrule=0.3pt,
 ]
\textbf{Data-driven $\boldsymbol{H}_2$ robust control.} Given $N$ i.i.d. draws of the uncertainty parameter $z$ collected in the dataset $D$, determine a state-feedback matrix $K(D)$, a threshold $\bar H_2(D)$, and a bound $\varepsilon(D)$ such that, with confidence $1-\delta$ with respect to the realizations of $\bD$, it holds that 
\[
R(\bD) \le \varepsilon(\bD), 
\]
where $R(D) = \mathbb{P}_z[z: H_2(\Sigma_{z,K(D)}) > \bar H_2(D)]$.
\end{tcolorbox}

\subsection{P2L for data-driven $\boldsymbol{H}_2$ control}
\medskip

Following a classical approach, the $H_2$ control problem is reformulated as an optimization program with positive-semidefiniteness conditions, which can be solved using an off-the-shelf solver.
Interestingly, the performance certificate provided by P2L does not require that the solver attains exact optimality upon termination. 
\medskip

\noindent
\textbf{Synthesis Algorithm.} In the current context, a synthesis algorithm $L$ is a map that associates a set $T\subseteq D$ to a pair consisting of a state-feedback matrix $K$ and a threshold $\bar H_2$. 

Following the formalization in \cite{caverly2019lmi} and \cite{SchererWeiland2015}, the worst-case minimization over $T$ of the $H_2$ norm of the transfer function from $w$ to $y$ is obtained by solving the following program in the variables $\gamma\in\mathbb{R}$, $P_i\in\mathbb{S}^{n_x}$, $Z_i \in\mathbb{S}^{n_w}$, $K\in\mathbb{R}^{n_u \times n_x}$: 
\be
\begin{split}
	&\min~
	\gamma\\
	&\text{subject to}~ \\
    &
	\begin{bmatrix}
		(A_i+B_iK)P_i+P_i(A_i+B_iK)^\top & P_i(C_i+D_iK)^\top\\
		(C_i+D_iK)P_i & -I	
	\end{bmatrix}\prec 0\\
	&
	\begin{bmatrix}
		Z_i & E_i\\
		E_i^\top & P_i	
	\end{bmatrix}\succ 0
	\\
	&~\text{trace}(Z_i)<\gamma\\ 
	&~P_i\succ 0,~Z_i\succ 0,~\gamma>0\\
	&\mbox{for all } i \mbox{ such that } z_i \in T, 
\end{split}
\label{eq:BMI-H2} 
\ee
where $\mathbb{S}^{n}$ is the set of $n \! \times \! n$ symmetric matrices, and we let $A_i=A(z_i)$, and similarly for $B_i$, $C_i$, $D_i$, $E_i$, to simplify the notation. 

Owing to the presence of the bilinear terms $KP_i$, program \eqref{eq:BMI-H2} is \emph{not} convex. To (approximately) solve it, we use {\tt hifoo}, a BFGS-based off-the-shelf solver specifically designed for $H_2$ and $H_{\infty}$ control problems, \cite{hifoo2009}. The algorithm $L$ returns both the synthesized controller $K$ and the value $\bar H_2$ obtained as the largest $H_2$ norm induced by $K$ over the systems corresponding to $z_i$ values in $T$.\footnote{Even when problem \eqref{eq:BMI-H2} is feasible, the numerical solver {\tt hifoo} may converge to a sub-optimal solution, let alone the fact that no $K$ is returned in the infeasible case.  In the latter, we conventionally set $K=0$ and, in all situations, $\bar H_2$ is computed as $\max_{z_i \in T} H_2(\Sigma_{z_i,K})$ after that $K$ has been determined.} That is, the algorithm output is $h=(K,\bar H_2)$. 
\begin{remark}
\label{Remark quadratic stability}
\textbf{P2L: beyond quadratic stability.}
It is known that convex reformulations of the robust $H_2$ control problem can be obtained by considering a single Lyapunov matrix $P$ and redefining the bilinear term $KP$ as a new variable $F$. Upon termination of the program, the controller is recovered as $K = FP^{-1}$.\footnote{This trick does not extend to the case of several $P_i$'s with the definition $F_i = KP_i$ since $K_i = F_i P_i^{-1}$ does not yield a single state-feedback matrix, as is required for implementation purposes.} This approach is widely used in robust control to make the problem tractable. On the other hand, restricting to a single $P$ corresponds to enforcing quadratic stability, as opposed to robust stability, and often results in overly conservative solutions. Interestingly, P2L only requires considering finitely many systems (those in $T$), which allows the use of distinct $P_i$'s while still keeping the problem tractable. 
\end{remark}

\noindent
\textbf{Property $\varphi$ and quantifier of the degree of dissatisfaction.} 
Given a decision $h=(K,\bar H_2)$, and a realization of the uncertainty parameter $z$, the property we wish to certify is that the $H_2$ norm of the closed-loop system $\Sigma_{z,K}$ is no more than $\bar H_2$. This corresponds to defining $\varphi$ as $\varphi(h,z)=1$ if $H_2(\Sigma_{z,K})\le \bar H_2$, and zero otherwise. The degree of dissatisfaction for a given $z$ is measured by the value of $H_2(\Sigma_{z,K})$, with higher values corresponding to higher dissatisfaction (ties are broken according to any pre-specified rule).
\medskip

\noindent
\textbf{Initialization.} We initialize the controller filling all the entries of $K\in\mathbb{R}^{{n_x}\times{n_u}}$ with zeros and let $\bar H_2 = 0$. 
\medskip

\noindent
\textbf{$\mathbf{H_2}$ robustness guarantees.} We aim to endow the controller $K(D)$ designed by P2L with formal robustness guarantees. Specifically, we want to quantify the probability that $K(D)$ achieves an $H_2$ norm of $\Sigma_{z,K}$ below $\bar H_2(D)$. The following corollary, which directly follows from \Cref{thm:main_result}, provides this result. 

\begin{corollary}
\label{pillar1:cor_bound}
Let $K(D)$ and $\bar H_2(D)$ be the controller and the threshold on the $H_2$ norm, and $T$ the corresponding compressed set, returned by P2L. Then, with confidence $1-\delta$ with respect to the realizations of $\bD$, it holds that 
\[
R(\bD) \le \bar\varepsilon(|\bT|,\delta,N), 
\]
where $R(D) = \mathbb{P}_z[z: H_2(\Sigma_{z,K(D)}) > \bar H_2(D)]$. 
\end{corollary}

\subsection{Pick-to-Learn in action}
\medskip

We demonstrate the performance of P2L on systems taken from {\tt COMPLib}, a publicly available library for robust control synthesis and matrix optimization, \cite{COMPleib2004}. A comparison is provided with the classical LMI-based quadratic stability approach. More precisely, we consider the following LMI-based program, see \cite[Sec 5.2.1]{caverly2019lmi}, 
\be
\begin{split}
&\min~
      \gamma\\
	   & \text{subject to} \\
	   &\begin{bmatrix}
	   A_iP + B_iF + PA_i^\top + F^\top B_i^\top & PC_i^\top+F^\top D_i^\top\\
	   C_iP+ D_iF & -I	
	   \end{bmatrix}\prec 0\\
	   &
	   \begin{bmatrix}
	   Z_i & E_i\\
	   E_i^\top & P	
	   \end{bmatrix}\succ 0
	   \\
	   &~\text{trace}(Z_i)<\gamma\\ 
	   &~P\succ 0,~Z_i\succ 0,~\gamma>0\\
	&\mbox{for all } i \in D, 
\end{split}
\label{eq:LMI-quadstab}
\ee
where the optimization variables are $\gamma\in\mathbb{R}$, $P\in\mathbb{S}^{n_x}$, $Z_i \in\mathbb{S}^{n_w}$, $F\in \mathbb{R}^{n_u\times n_x}$.\footnote{Unlike \eqref{eq:BMI-H2}, this program directly enforces the satisfaction of all constraints for $i \in D$ since, following the classical robust scenario optimization formulation, this is the form in which it is solved in the simulations.} Note that \eqref{eq:LMI-quadstab} is obtained from the nonconvex program \eqref{eq:BMI-H2} by enforcing $P_i=P$, i.e., by restricting the search to a common Lyapunov function, as explained in \Cref{Remark quadratic stability}. To solve \eqref{eq:LMI-quadstab}, we use the off-the-shelf solver {\tt MOSEK}, and the controller is then recovered via the inverse transformation $K=FP^{-1}$. 
\medskip

\noindent
Our experiments are conducted on two systems taken from {\tt COMPlib}: {\tt DIS5} and {\tt HE1}. The first describes a distributed, interconnected system, while the second represents the longitudinal motion of a VOTL helicopter under typical loading and flight conditions, \cite{COMPleib2004}. Some entries of the matrices describing these systems are perturbed (independently of each other) within an interval with uniform distribution. Precisely, these entries are: the upper triangular elements (excluding the diagonal elements) of the nominal state matrices and all the elements of the matrices $B$ and $E$. We do not perturb the elements of $C$, and $D(z)$ is kept fixed at zero throughout. The same perturbation range is applied across all perturbed entries, and simulations are conducted for increasing widths of the perturbation range. Specifically, if $e$ denotes a given entry, the perturbation assigns a uniform distribution over the interval $[e/s, es]$, where $s \geq 1$ is a scaling factor, common across all entries, that enlarges or shrinks the intervals.\footnote{When $s=1$, no perturbation occurs as the interval reduces to $[e,e]$; larger values of $s$ correspond to stronger perturbations. This perturbation scheme is designed to stress-test the LMI-based quadratic stability approach and thereby evaluate P2L in a setup that is difficult to handle with more traditional tools.} All experiments are performed with a dataset containing $N=10,000$ draws. 
\medskip

\noindent
\textbf{Results.} We first compare P2L with the LMI-based approach of \eqref{eq:LMI-quadstab}. They are executed on the same dataset $D$, and the obtained controller $K$ is tested against an additional draw $z$ of the uncertainty. \Cref{fig:LMI_fails} (left: {\tt DIS5}, right: {\tt HE1}) shows the profile of the corresponding norm $H_2(\Sigma_{\bz,K(D)})$ across increasing values of the the uncertainty scaling factor $s$. The result in this figure is the typical behavior also observed for other draws of $D$ and $z$. The LMI-based design underperforms compared to the design produced by P2L, showing the advantage of using Lyapunov matrices adapted to the realization of the uncertainty parameter. This effect becomes increasingly evident as the uncertainty scaling factor $s$ grows, eventually reaching a breaking point beyond which the LMI-based design fails to quadratically stabilize the systems associated with all draws in $D$, which yields an unbounded $H_2$ norm. In contrast, the P2L-based design maintains strong performance, consistently achieving an $H_2$ norm well below unity across the considered values of $s$. 
\medskip

\noindent
We next consider a value of $s$ beyond the breaking point where the LMI-based approach fails, and test P2L more thoroughly. Specifically, we focus on {\tt DIS5} with $s=1.35$. The top panel of \Cref{fig:tightness-bounds} shows the distribution of the threshold $\bar H_2(D)$ obtained by P2L over $1000$ trials. The bottom panel of the same figure offers an evaluation of the tightness of the P2L bounds: each of the $1000$ realizations of the dataset $D$ is represented by a $+$ symbol, whose $x$-coordinate is the bound on the risk and the $y$-coordinate is the actual risk, evaluated via the Monte-Carlo approach with $20,000$ draws (i.e., the proportion of draws that exceed the threshold $\bar H_2(D)$ in 20,000 tests). The closer these $+$ symbols to the diagonal, the tighter the bound. Overall, the results indicate that P2L enables both the synthesis of good controllers (\Cref{fig:tightness-bounds}, top) and the computation of informative bounds on the risk (\Cref{fig:tightness-bounds}, bottom). 

\setlength\figureheighteps{3cm}
\setlength\figurewidtheps{3.2cm} 
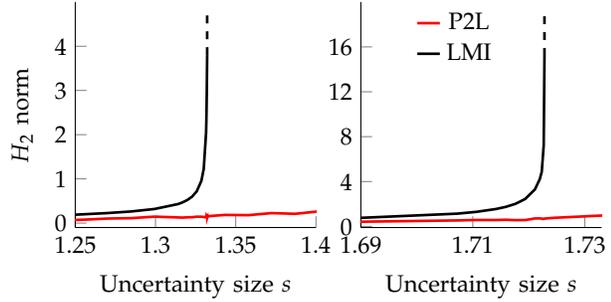
\begin{figure}
    \centering
    \begin{subfigure}[t]{0.24\textwidth}
    	\begin{tikzpicture}

\begin{axis}[%
width=\figurewidtheps,
height=\figureheighteps,
at={(1.011in,0.642in)},
scale only axis,
every axis plot post/.append style={-},
unbounded coords=jump,
xmin=1.25,
xmax=1.4,
ymin=-0.1,
ymax=5,
axis background/.style={fill=white},
axis x line*=bottom,
axis y line*=left,
xlabel={Uncertainty size $s$},
ylabel={$H_2$ norm},
legend style={draw=none},
ylabel style={yshift=-15pt},
ytick={0, 1, 2, 3, 4},
]
\addplot [color=red, line width=1pt, forget plot]
  table[row sep=crcr]{%
1.168	0.000697066914390816\\
1.1826	0.00064890601241681\\
1.1972	0.000607337758976174\\
1.2118	0.000557792423317687\\
1.2264	0.00046708609718973\\
1.241	0.0626937931626371\\
1.2556	0.0771299395436254\\
1.2702	0.10273026234275\\
1.2848	0.111579732447865\\
1.2994	0.145660013795566\\
1.314	0.127958591257404\\
1.31692	0.123759056810985\\
1.31984	0.130693227127486\\
1.32276	0.132666861668748\\
1.32568	0.144244349622416\\
1.3286	0.143996338385199\\
1.33006	0.13701270959892\\
1.33152	0.130090631096062\\
1.331666	0.160164515378397\\
1.331812	0.148124916286014\\
1.331958	0.141381180255815\\
1.33225	0.15542849505851\\
1.3322646	0.140106995613249\\
1.3322792	0.154411948600189\\
1.3322865	0.142969721726029\\
1.33228942	0.149237422734239\\
1.33229088	0.15039549165089\\
1.33229234	0.156154260956781\\
1.3322938	0.147477847063659\\
1.3323084	0.138505979203778\\
1.332323	0.144376864664157\\
1.3323376	0.138078753994842\\
1.3323522	0.166275391210586\\
1.3323668	0.138666870342556\\
1.3323814	0.140354661246953\\
1.33298	0.156471358028149\\
1.3432	0.184726589534995\\
1.3578	0.1790045131592\\
1.3724	0.226046034604258\\
1.387	0.209598405845148\\
1.4016	0.264545790710438\\
1.4162	0.437430057025654\\
1.4308	0.295647311107081\\
1.4454	0.368036698232438\\
1.46	0.511247495065685\\
};
\addplot [color=black, line width=1pt, mark=none, solid, forget plot]
  table[row sep=crcr]{%
1.168	0.00313325737493487\\
1.1826	0.00292422705237117\\
1.1972	0.00243559003218379\\
1.2118	0.000462331486554841\\
1.2264	0.00887650523018249\\
1.241	0.173242790541238\\
1.2556	0.199694814123838\\
1.2702	0.226634701627085\\
1.2848	0.262562564909109\\
1.2994	0.318763214957313\\
1.314	0.429706047470644\\
1.31692	0.469066929121673\\
1.31984	0.521430127583189\\
1.32276	0.596571414221028\\
1.32568	0.716196238357838\\
1.3286	0.955686241987678\\
1.33006	1.22352960366043\\
1.33152	2.04005003965773\\
1.331666	2.24347709732391\\
1.331812	2.54757486839017\\
1.331958	2.93289282816393\\
1.3322 3.8\\
};
\addplot [color=black, line width=1pt, mark=none, dashed, forget plot]
  table[row sep=crcr]{%
1.3322 3.8\\
1.3322 4.8\\
};
\end{axis}
\end{tikzpicture}%
%
   \end{subfigure}%
   ~~%
   \begin{subfigure}[t]{0.24\textwidth}
    	\begin{tikzpicture}

\begin{axis}[%
width=\figurewidtheps,
height=\figureheighteps,
at={(1.011in,0.642in)},
scale only axis,
every axis plot post/.append style={-},
unbounded coords=jump,
xmin=1.69,
xmax=1.733,
ymin=-0.1,
ymax=20,
axis background/.style={fill=white},
axis x line*=bottom,
axis y line*=left,
xlabel={Uncertainty size $s$},
legend style={draw=none},
ylabel style={yshift=-10pt},
ytick={0, 4, 8, 12, 16},
xtick={1.69, 1.71, 1.73},
legend style={at={(0.6,1)}},
legend image post style={xscale=0.5},%
]
\addplot [color=red, line width=1pt]
  table[row sep=crcr]{%
1.6549	0.298382755137526\\
1.67232	0.323391874350803\\
1.68974	0.415853664770275\\
1.70716	0.531611271734947\\
1.710644	0.572222374320364\\
1.714128	0.567826185965619\\
1.71587	0.594265397837857\\
1.717612	0.55595534371019\\
1.719354	0.566162214837506\\
1.721096	0.707584432222162\\
1.721967	0.71334061240667\\
1.7223154	0.690938738047227\\
1.7226638	0.66516897711425\\
1.722838	0.698513326615015\\
1.72458	0.766594485771442\\
1.73329	0.972803427826544\\
1.742	4.75771248543476\\
};
\addlegendentry{P2L}
\addplot [color=black, line width=1pt, mark=none, solid]
  table[row sep=crcr]{%
1.6549	0.512492655090818\\
1.67232	0.60124152605913\\
1.68974	0.756317574753387\\
1.70716	1.13360775539106\\
1.710644	1.29692540962729\\
1.714128	1.54921930365428\\
1.71587	1.73877795449498\\
1.717612	2.01013367188856\\
1.719354	2.44263497665684\\
1.721096	3.29763267612886\\
1.721967	4.22071978976027\\
1.7223154	4.86842463189039\\
1.7226638	7.21980278179448\\
1.7228		15.2\\
};
\addlegendentry{LMI}
\addplot [color=black, line width=1pt, mark=none, dashed]
  table[row sep=crcr]{%
1.7228 15.2\\
1.7228 19.2\\
};
\end{axis}
\end{tikzpicture}%
%
   \end{subfigure}
   \vspace*{-3mm}
   \caption{The $H_2$ norm of $\Sigma_{\bz,K(D)}$ obtained with P2L (red) and the LMI-based quadratic-stability approach (black). Left panel: {\tt DSI5}, right panel: {\tt HE1}.}
   \label{fig:LMI_fails}
\end{figure}

\setlength\figureheight{5cm}
\setlength\figurewidth{0.85\linewidth} 
\setlength\figureheighteps{3.5cm}
\setlength\figurewidtheps{0.67\linewidth}
\begin{figure}[h!]
	\centering
   	\begin{tikzpicture}

\begin{axis}[%
width=\figurewidtheps,
height=\figureheighteps,
at={(1.011in,0.642in)},
scale only axis,
xmin=-0.1,
xmax=3,%
ymin=0,
ymax=300,
scaled x ticks=false, 
xlabel={Threshold $\bar H_2(D)$},
ylabel={Frequency},
ylabel style={yshift=-10pt},
legend style={legend cell align=left, align=left, draw=none},
reverse legend
]
\addplot[ybar interval, fill=red, fill opacity=1, draw=black, area legend] table[row sep=crcr] {%
x	y\\
0	6\\
0.0639	150\\
0.1278	277\\
0.1917	155\\
0.2556	129\\
0.3195	53\\
0.3834	57\\
0.4473	39\\
0.5112	24\\
0.5751	17\\
0.639	13\\
0.7029	13\\
0.7668	8\\
0.8307	8\\
0.8946	4\\
0.9585	3\\
1.0224	4\\
1.0863	2\\
1.1502	3\\
1.2141	0\\
1.278	1\\
1.3419	2\\
1.4058	0\\
1.4697	0\\
1.5336	0\\
1.5975	3\\
1.6614	0\\
1.7253	0\\
1.7892	0\\
1.8531	2\\
1.917	0\\
1.9809	0\\
2.0448	0\\
2.1087	1\\
2.1726	0\\
2.2365	1\\
2.3004	0\\
2.3643	0\\
2.4282	0\\
2.4921	0\\
2.556	0\\
2.6199	0\\
2.6838	0\\
2.7477	0\\
2.8116	0\\
2.8755	1\\
2.9394	0\\
3.0033	0\\
3.0672	0\\
3.1311	0\\
3.195	0\\
3.2589	0\\
3.3228	0\\
3.3867	0\\
3.4506	0\\
3.5145	0\\
3.5784	0\\
3.6423	0\\
3.7062	0\\
3.7701	0\\
3.834	0\\
3.8979	0\\
3.9618	0\\
4.0257	1\\
4.0896	0\\
4.1535	0\\
4.2174	0\\
4.2813	0\\
4.3452	0\\
4.4091	0\\
4.473	0\\
4.5369	0\\
4.6008	0\\
4.6647	0\\
4.7286	0\\
4.7925	0\\
4.8564	0\\
4.9203	0\\
4.9842	0\\
5.0481	0\\
5.112	0\\
5.1759	0\\
5.2398	0\\
5.3037	0\\
5.3676	0\\
5.4315	0\\
5.4954	0\\
5.5593	0\\
5.6232	1\\
5.6871	1\\
5.751	0\\
5.8149	0\\
5.8788	0\\
5.9427	0\\
6.0066	0\\
6.0705	0\\
6.1344	0\\
6.1983	0\\
6.2622	0\\
6.3261	1\\
6.39	1\\
};
\addlegendentry{P2L}
\end{axis}

\end{tikzpicture}%
   	\include{figures_tikz/UB_vis_misclass_DSI5}%
    \caption{Performance of P2L over $1000$ datasets for {\tt DIS5} with $s = 1.35$. Top: Distribution of the bounds; bottom: risk bounds against actual risks. 
    }
    \label{fig:tightness-bounds}
\end{figure}
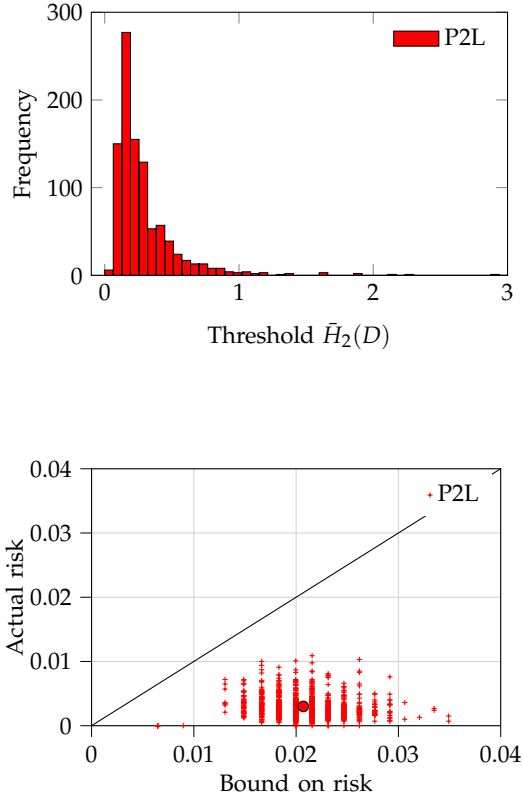

\section{Acknowledgments}
\medskip

This work was partially supported by the EPSRC grant EP/Y001001/1, funded by the International Science Partnerships Fund (ISPF) and UKRI, by the PRIN 2022 project 2022RRNAEX ``The Scenario Approach for Control and Non-Convex Design'' (CUP: D53D23001440006), funded by the NextGeneration EU program (Mission 4, Component 2, Investment 1.1), and by the PRIN PNRR project  P2022NB77E “A data-driven cooperative framework for the management of distributed energy and water resources” (CUP: D53D23016100001), funded by the NextGeneration EU program (Mission 4, Component 2, Investment 1.1).

\section{Author Information}
\medskip

\begin{IEEEbiography}		
{\textbf{Dario Paccagnan}} is an Associate Professor at the Department of Computing, Imperial College London where he joined in the Fall 2020. Before that, he was a postdoctoral fellow at the University of California, Santa Barbara. He obtained his PhD from the Automatic Control Laboratory, ETH Zurich, Switzerland, in 2018. He received a B.Sc. and M.Sc. in Aerospace Engineering from the University of Padova, Italy, in 2011 and 2014, and a M.Sc. in Mathematical Modelling and Computation from the Technical University of Denmark in 2014; all with Honors. Dario's interests are at the interface of game theory, control theory, and learning theory with a focus on tackling societal-scale challenges. Dario was a finalist for the 2019 EECI best PhD thesis award and was recognized with the SNSF Early Postdoc Mobility Fellowship, the SNSF Doc Mobility Fellowship, and the ETH medal for his doctoral work. He is the recipient of the best student paper award (as advisor) at AISTATS 2025.
\spacing \\
\emph{\textbf{Daniel Marks}} is a DPhil candidate in Statistical Machine Learning at the University of Oxford, funded by the G-Research Graduate Scholarship. He also serves as a Lecturer in Computer Science at Christ Church, Oxford. Daniel previously completed a four-year MEng in Computing at Imperial College London specialising in Artificial Intelligence and Machine Learning. His Master’s thesis on the Pick-to-Learn algorithm and self-certified Gaussian Process approximations received the Splunk Prize and a Distinguished Project Award from the Department of Computing, and led to a first-author paper that won the Best Student Paper Award at AISTATS 2025. His research interests include Statistical Learning Theory, Approximate Bayesian Inference, and Gaussian Processes.
\spacing \\
\emph{\textbf{Marco C. Campi}} is Professor of Control and Data-Driven Methods at the University of Brescia, Italy. He has held visiting and teaching appointments with Australian, USA, Japanese, Indian, Turkish and various European universities, besides the NASA Langley Research Center in Hampton, Virginia. He has chaired the IFAC TCs on Modeling, Identification and Signal Processing (MISP) and that on Stochastic Systems (SS), and has served in diverse capacities on the Editorial Board of top-tier control journals such as Automatica and Systems and Control Letters. In 2008, he received the IEEE CSS George S. Axelby award for the article “The Scenario Approach to Robust Control Design”. He has delivered plenary and semi-plenary addresses at major conferences including CDC, SYSID, and OPTIMIZATION, besides presenting numerous talks as IEEE distinguished lecturers for the Control Systems Society. Professor Campi is a Fellow of IEEE and a Fellow of IFAC. His research interests include: data-driven decision-making, stochastic optimization, inductive methods, and the fundamentals of probability theory. 
\spacing \\
\emph{\textbf{Simone Garatti}} is Associate Professor at the Dipartimento di Elettronica ed Informazione of the Politecnico di Milano, Milan, Italy. He received the Laurea degree cum laude and the Ph.D. cum laude in Information Technology Engineering in 2000 and 2004, respectively, both from the Politecnico di Milano. He held visiting positions at the Lund University of Technology, at the University of California San Diego, at the Massachusetts Institute of Technology, and at the University of Oxford. Simone Garatti is currently member of the IEEE-CSS Conference Editorial Board and Associate Editor of the International Journal of Adaptive Control and Signal Processing and of the Machine Learning and Knowledge Extraction journal. In 2024, he served as Tutorial Chair in the organizing committee of the 6th Learning for Dynamics and Control Conference (L4DC), while he was a member of the EUCA Conference Editorial Board from 2013 to 2019. With his co-authors, Simone Garatti has pioneered the theory of the scenario approach for which he was keynote speaker at the IEEE 3rd Conference on Norbert Wiener in the 21st Century in 2021, and semi-plenary speaker at the 2022 European Conference on Stochastic Optimization and Computational Management Science (ECSO-CMS). His current research interests include data-driven optimization and decision-making, stochastic optimization, system identification, uncertainty quantification, and statistical learning theory.
\end{IEEEbiography}

\appendix[Numerical computation of $\boldsymbol{\overline\varepsilon(k,\delta, N)}$]
\medskip

\noindent $\diamond$ \underline{\textsc{Matlab} code}

{\small
\begin{verbatim}
function eps = find_eps(k,N,delta)
 
    if k==N 
        eps = 1;
    else
        t1 = 0;
        t2 = 1;
        while t2-t1 > 1e-10
            t = (t1+t2)/2;
            left = delta*betainc(t,k+1,N-k);
            right = t*N*(betainc(t,k,N-k+1) ...
                    -betainc(t,k+1,N-k));
            if left > right
                t2=t;
            else
                t1=t;
            end
        end
        eps = t2;
    end 
end
\end{verbatim}
}

\noindent $\diamond$ \underline{\textsc{Python} code}

{\small
\begin{verbatim}
from scipy.special import betainc

def find_eps(k,N,delta):

    if k == N:
        return 1.0
    else:
        t1 = 0.0
        t2 = 1.0
        while t2 - t1 > 1e-10:
            t = (t1 + t2) / 2
            left = delta*betainc(k+1,N-k,t)
            right = t*N*(betainc(k,N-k+1,t) \
                    -betainc(k+1,N-k,t))
            if left > right:
                t2 = t
            else:   
                t1 = t
    return t2
\end{verbatim}
}

\appendix[Proof of Theorem 2] 
\medskip

To start with, we note that the P2L$^+$ algorithm is, by construction, permutation invariant: feeding P2L$^+$ with $D$ or with a permutation of it produces the same output. This invariance holds even though the synthesis algorithm $L$ may itself be permutation variant. Thus, while we use ordered lists since this is relevant for the internal synthesis algorithm $L$, from the external perspective of P2L$^+$, the input can be viewed as a multiset: a set with repetitions but without a concept of position of its elements.\footnote{This means, for example, that $\{1, 1, 2\}$ is the same as $\{1, 2, 1\}$, but both are different from $\{1, 2\}$.} For future use, we introduce the notation ${\sf ms}(H)$ to indicate a multiset obtained from a list $H$, where the operator ${\sf ms}$ removes any information on the position of the elements while maintaining repetitions. 
	
Given these premises, we recognize that P2L$^+$ defines a so-called compression function, \cite{2023compression}, that is, a map ${\sf c}$ from ${\sf ms}(D)$ to ${\sf ms}(T)$, where the latter is a sub-multiset of ${\sf ms}(D)$ -- which we denote as ${\sf ms}(T) \subseteq {\sf ms}(D)$ -- since, by construction, the elements of ${\sf ms}(T)$ are also elements of ${\sf ms}(D)$ with a number of repetitions in ${\sf ms}(T)$ that is no bigger than that in ${\sf ms}(D)$. If we now consider any sub-multiset $S$ such that ${\sf ms}(T) \subseteq S \subseteq {\sf ms}(D)$, one can easily prove that ${\sf c}(S) = {\sf ms}(T)$. As a matter of fact, all the maximal elements computed during the execution of \Cref{alg:P2L+} with ${\sf ms}(D)$ as input are all in $S$ (because ${\sf ms}(T) \subseteq S$), and they will be identified as maximal elements -- exactly in the same order -- during the execution of \Cref{alg:P2L+} fed with $S$ as input (because $S \subseteq {\sf ms}(D)$ and, moreover, the algorithm is not presented with additional choices). This implies that $T$ is identically constructed in the two executions of P2L$^+$, one with ${\sf ms}(D)$ as input and one with $S$ as input, and when the termination condition is met with ${\sf ms}(D)$, it is also met with $S$ because $S \setminus {\sf ms}(T) \subseteq D \setminus {\sf ms}(T)$. Therefore, the returned $T$ coincides in the two cases, and: ${\sf c}(S) = {\sf ms}(T) = {\sf c}({\sf ms}(D))$. In view of Lemma 3 in \cite{2023compression}, this proves that the compression function ${\sf c}$ satisfies the \emph{preference} property (Property 2 in \cite{2023compression}). 
	
Turn now to consider the map $\mathcal{A}$ from ${\sf ms}(D)$ to the decision $h$ defined by P2L$^+$. The same argument used before to prove that ${\sf c}$ is \emph{preferent} shows that, for any sub-multiset $S$ such that ${\sf ms}(T) \subseteq S \subseteq {\sf ms}(D)$, it holds that $\mathcal{A}(S) = \mathcal{A}({\sf ms}(D))$ (in fact, the returned $h$ is given by $L(T)$, and we have proven that $T$ coincides in the two executions of P2L$^+$, with ${\sf ms}(D)$ and with $S$ as inputs). In particular, this holds true for $S = {\sf ms}(T) = {\sf c}({\sf ms}(D))$, which gives $\mathcal{A}({\sf c}({\sf ms}(D))) = \mathcal{A}({\sf ms}(D))$. This implies that $\mathcal{A}$ itself acts as \emph{reconstruction function}, i.e. a map from multisets to decisions through which the $h$ constructed from ${\sf ms}(D)$ can be obtained starting from the compression of ${\sf ms}(D)$.
	
The result in \Cref{thm:P2L+_main_result} now follows by a direct application of Theorem 17 in \cite{2023compression}, by: 1. taking $\ell(h,z)=0$ when $\varphi(h,z)=1$ and $\ell(h,z)=1$ when $\varphi(h,z)=0$ in the definition of $R(\mathcal{A}(\bz_1,\ldots,\bz_N))$ appearing in Theorem 17 of \cite{2023compression}; 2. noticing that $|\tilde{{\sf c}}(\bz_1,\ldots,\bz_N)|$ appearing in Theorem 17 of \cite{2023compression} is equal to $|\bT|+|\boldsymbol{U}|$ in the present context. \qed 

\medskip

\bibliographystyle{unsrtnat}
\bibliography{bib_works}

\endarticle
\end{document}